\documentclass[12pt]{article}

\usepackage{graphicx}
\usepackage{epstopdf}
\usepackage{amsmath}
\usepackage{amssymb}
\usepackage{amsthm}
\usepackage{ccaption}
\usepackage[usenames]{color}

\usepackage[
      colorlinks=true,
      linkcolor=blue,
      citecolor=blue,
      urlcolor=black,
      filecolor=blue,
      pdfstartview=FitV,
      pdftitle={Covariant entanglement entropy},
        pdfauthor={Hubeny, Rangamani, Takayanagi},
        pdfsubject={entanglement entropy},
        pdfkeywords={min surface, covariant, entanglement},
        pdfpagemode=None,
        bookmarksopen=true
      ]{hyperref}

\makeatletter
\@addtoreset{equation}{section}

\makeatletter
\renewcommand\section{\@startsection {section}{1}{\z@}%
                                   {-3.5ex \@plus -1ex \@minus -.2ex}
                                   {2.3ex \@plus.2ex}%
                                   {\normalfont\large\bfseries}}
\renewcommand\subsection{\@startsection{subsection}{2}{\z@}%
                                     {-3.25ex\@plus -1ex \@minus -.2ex}%
                                     {1.5ex \@plus .2ex}%
                                     {\normalfont\bfseries}}

  \captionnamefont{\bfseries}
  \captiontitlefont{\small\sffamily}
  \captiondelim{: }
  \hangcaption

\def\baselinestretch{1.2}
\newcommand{\be}{\begin{equation}}
\newcommand{\ee}{\end{equation}}
\newcommand{\beq}{\begin{eqnarray}}
\newcommand{\eeq}{\end{eqnarray}}

\def\sec#1{\S \;\ref{#1}}
\def\fig#1{Fig.\,\ref{#1}}
\def\req#1{(\ref{#1})}
\def\App#1{Appendix \ref{#1}}

\def\[{\left [}
\def\]{\right ]}
\def\({\left (}
\def\){\right )}

\def\eg{{\it e.g.}}

\def\cf{{\it cf.}}
\def\ie{{\it i.e.}}

\def\btab{\begin{table}[h] \begin{center} \begin{tabular}{l lp{3in}}}
\def\etab{\end{tabular} \end{center} \end{table}}
\def\btabm{\begin{center} \begin{tabular}}
\def\etabm{\end{tabular} \end{center}}

\def\a{\alpha}

\def\d{\delta}
\def\eps{\epsilon}
\def\veps{\varepsilon}

\def\vp{\varphi}
\def\ph{\varphi}
\def\p{\partial}

\def\D{\Delta}

\def\s{\sigma}
\def\l{\ell}

\def\CA{{\cal A}}
\def\CB{{\cal B}}

\def\CH{{\cal H}}

\def\CM{{\cal M}}
\def\CN{{\cal N}}
\def\CO{{\cal O}}

\def\CS{{\cal S}}

\def\CW{{\cal W}}
\def\CX{{\cal X}}
\def\CY{{\cal Y}}
\def\CZ{{\cal Z}}
\def\CSig{{\Sigma}}

\def\R{{\bf R}}
\def\Sp{{\bf S}}

\def\A5S5{{\rm AdS}_5 \times \S^5}

\def\l{\ell}

\def\p{\partial}

\def\f#1#2{{\frac{#1}{#2}}}

\def\f#1#2{{\frac{#1}{#2}}}

\def\p{\partial}

\def\Tr#1{{\rm Tr}\(#1\)}

\def\ket#1{\mid  \! \! #1   \rangle}
\def\bra#1{\langle   #1 \! \! \mid}

\newcommand{\bbibitem}[1]{\bibitem{#1}\marginpar{#1}}

\def\Label#1{\label{#1}%
{ \color{blue}{\smash{\hbox to0pt{\raise2ex\hbox{\tiny[#1]}\hss}}}}}
\def\noLabels{\let\Label=\label}
\def\nobbibitem{\let\bbibitem=\bibitem}

\def\bulk{{\cal M}}
\def\bdy{\p{\cal M}}
\def\bdys{\p{\cal N}}
\def\ms{\CS}
\def\Gms{\CW}
\def\Lms{\CY}
\def\Xms{\CX}
\def\Cms{\CZ}
\def\Sms{\CSig}
\def\rA{\CA}
\def\rB{\CB}
\def\brA{\p \CA}

\def\area#1{{\rm Area}(#1)}

\def\curt#1{\gamma_{#1}}
\def\curf#1{\gamma_{#1}^+}
\def\curp#1{\gamma_{#1}^-}
\def\set#1{\{ \, #1 \, \}}
\def\st{ \, \mid \, }
\def\pho{\phi_0}
\def\rmin{r_{\rm min}}
\def\rh{r_+}
\def\rd{{\dot r}}
\def\rb{{\bar r}}
\def\ro{r_0}
\def\Veff{V_{\rm eff}}

\def\ptm{\({\p \over \p t}\)^\mu}
\def\ads#1{AdS$_{#1}$}
\def\cof{{\veps}}
\def\Dr{\D}

\def\s{\sqrt}
\def\de{\partial}

\def\f {\frac}
\def\ti{\tilde}
\def\ap{\alpha}

\def\no{\nonumber \\}
 \def\la{\langle}
 \def\lb{\rangle}
 \def\ep{\epsilon}
  \def\vp{\varphi}
 
 \def\ba{\begin{eqnarray}}
\def\ea{\end{eqnarray}}

\renewcommand{\thefootnote}{\fnsymbol{footnote}}
\title{{\bf \large A Covariant Holographic Entanglement Entropy Proposal}}

\author{\normalsize Veronika E. Hubeny\footnote{veronika.hubeny@durham.ac.uk}\ ,
Mukund Rangamani\footnote{mukund.rangamani@durham.ac.uk}\ ,
and Tadashi Takayanagi\footnote{takayana@gauge.scphys.kyoto-u.ac.jp}\\ \\
\small \sl $^{\ast,\dagger}$  Centre for Particle Theory \& Department of
Mathematical Sciences,
\\[-1.5mm]
\small \sl Science Laboratories, South Road, Durham DH1 3LE, United Kingdom. \\
\small \sl $\ddagger$ Department of Physics, Kyoto University, Kyoto, 606-8502, Japan.
}

\begin{document}

\noLabels 
\nobbibitem 

\setlength{\baselineskip}{16pt}
\begin{titlepage}
\maketitle
\begin{picture}(0,0)(0,0)
\put(350, 320){DCPT-07/13} \put(350,303){KUNS-2069}
\end{picture}
\vspace{-36pt}

\begin{abstract}

With an aim towards understanding the time-dependence of
entanglement entropy in generic quantum field theories, we propose a
covariant generalization of the holographic entanglement entropy
proposal of \href{http://arxiv.org/abs/hep-th/0603001}{
hep-th/0603001}. Apart from providing several examples of possible
covariant generalizations, we study a particular construction based
on light-sheets, motivated in similar spirit to the covariant
entropy bound underlying the holographic principle. In particular,
we argue that the entanglement entropy associated with a specified
region on the boundary in the context of the AdS/CFT correspondence
is given by the area of a co-dimension two bulk surface with
vanishing expansions of null geodesics.  We demonstrate our
construction with several examples to illustrate its reduction to
the holographic entanglement entropy proposal in static spacetimes.
We further show how this proposal may be used to understand the time
evolution of entanglement entropy in a time varying QFT state dual
to a collapsing black hole background. Finally, we use our proposal
to argue that the Euclidean wormhole geometries with multiple
boundaries should be regarded as states in a non-interacting but
entangled  set of QFTs, one associated to each boundary.

 \end{abstract}
\thispagestyle{empty}
\setcounter{page}{0}
\end{titlepage}

\renewcommand{\baselinestretch}{1.4}  
\renewcommand{\thefootnote}{\arabic{footnote}}


\tableofcontents

\section{Introduction}
 \label{intro}

\hspace{5mm} One of the important questions in quantum field
theories  is to understand the number of operative degrees of
freedom  in the theory at a given scale. In conventional RG parlance
this is measured by the Zamolodchikov's c-function (in two
dimensions) \cite{Zamolodchikov:1986gt}   which at the critical
points takes on the value of the central charge. One believes this
picture to persist in higher dimensions; in particular, there  ought
to exist some analog of a c-function in higher dimensional quantum
field theories of interest. Clearly there is a well-defined notion
of the central charges  for conformal field theories in $d >2$
\cite{Cardy:1988cw}, which may quantify  the total degrees of
freedom. However, this interpretation in terms of the degrees of
freedom has not yet been rigorously proved  except in two
dimensions.  Measuring degrees of freedom in
time-dependent backgrounds is an especially important open problem. A
detailed understanding of this issue   is very important for making
precise the notion of holography in quantum  gravity. For example,
in the context of string theory in unstable backgrounds  with closed
string tachyons, one expects that as the tachyon condenses,  the
number of degrees of freedom does change \cite{Nishioka:2006gr}; to
verify this expectation it is crucial to have a precise notion of
the time-dependent degrees of  freedom.

A simple way to get a measure of the degrees of freedom is to couple
the system
 to a heat bath and study its thermal properties, in particular
its entropy. However, we could  also ask the equally important
question: suppose we concentrate on a particular region of the
background spacetime on which the QFT is defined and ask what is the
correct measure of the operative degrees of freedom in that region
(even at zero temperature).  One important aspect of this is
captured by the entanglement entropy, which provides a measure of
how the degrees of freedom localized in that region interact (are
``entangled'') with the rest of the theory. In a sense the
entanglement entropy is a measure of the effective operative degrees
of freedom, \ie, those that are active participants in the dynamics,
in a given region of the background geometry.  Refer to
\cite{Calabrese:2005zw} for a short review of entanglement entropy
in QFT.

Consider a QFT defined on a spacetime manifold $\bdy$ (the peculiar
choice of notation for the background  will become clear
momentarily), and assume that $\bdy$ allows the
 foliation by time-slices $\bdys_t$ as $\bdy=\bdys_t \times \R_t$.
We wish to focus on a region $\rA_t \subset \bdys_t$ at a fixed time
$t$.  Denote also the complement of $\rA_t$ with respect to
$\bdys_t$ by $\rB_t$ so that $\rA_t \cup \rB_t= \bdys_t$. This
procedure divides the Hilbert space for the total system ${\cal
{H}}$ into a direct product of two Hilbert spaces ${\cal {H}}_{\rA}$
and ${\cal {H}}_{\rB}$ for the two subsystems, corresponding to the
regions $\rA_t$ and $\rB_t$, respectively, \ie, ${\cal {H}}_{tot}=
{\cal {H}}_{\rA}\otimes {\cal {H}}_{\rB}$. In this setup, one
measure of the number of degrees of freedom associated with region
(or sub-system) $\rA_t$ is given by the entanglement entropy
$S_{\rA_t}$. It is defined as the von Neumann entropy $S_{\rA_t}(t)
=-\mbox{Tr} \, \rho_{\rA_t}(t) \, \log \rho_{\rA_t}(t)$  associated
with the reduced density matrix
$\rho_{\rA_t}(t)=\mbox{Tr}_{\rB}\,\rho_{tot}(t)$,  obtained by
taking a trace of the density matrix $\rho_{tot}(t)$  for the total
system at time $t$  over the Hilbert space $\CH_{\rB}$. Notice that
the entanglement entropy defined in  this way is manifestly
time-dependent. Below, we will suppress the index ${}_t$ which shows
the time-dependence when we consider a static system, where
$S_{\rA_t}(t)$ does not depend on $t$.

In the same way, we can define the entanglement entropy $S_{\rB_t}(t)$ for the other subsystem $\rB_t$. In  general, $S_{\rA_t}(t)$ is different from $S_{\rB_t}(t)$. However, they are equivalent if the total system is described by a pure state $\ket{\Psi(t)} = \ket{\Psi_{\rA_t}} \,\otimes  \ket{\Psi_{\rB_t}}$, where the total and reduced density matrices are given by $\rho_{tot}(t)=\ket{\Psi(t)}\bra{\Psi(t)}$ and $\rho_{\rA_t}(t)=\mbox{Tr}_{\rB_t}\ket{\Psi(t)}\bra{\Psi(t)}$, respectively.

 In a two dimensional CFT, we can analytically calculate the entanglement entropy for arbitrary choice of the subsystem $\rA_t$ as shown recently in \cite{Calabrese:2004eu}, generalizing the previously known result \cite{Holzhey:1994we}. Moreover, an analogue of the Zamolodchikov's c-theorem (called entropic c-theorem) has been shown in  \cite{Casini:2004bw, Casini:uq} (see also \cite{Solodukhin:2006ic}). However, in higher dimensions it is rather difficult to obtain analytical results for generic $\rA_t$. Its state of the art is reviewed in \cite{Ryu:2006ef} from the viewpoint of the QFT.

Recently, entanglement entropies of various $1+1$ and $2+1$ dimensional condensed matter systems have been actively investigated in order to understand zero temperature quantum phase transitions   \cite{Vidal:2002rm, Calabrese:2005zw, Kitaev:2005dm, LevinWen, Fendley:2006gr}. In this context, entanglement entropy plays an important role of an order  parameter of the phase transition. For example, in a  material exhibiting topological ordering, such as the system with anyons in  fractional quantum Hall effect, correlation functions are not useful since the theory is topological. Instead we need a quantity which probes non-local information like fractional statistics of anyons. It turns out that the entanglement entropy can do this job elegantly, because it is defined non-locally~\cite{Kitaev:2005dm, LevinWen, Fendley:2006gr}.

As already mentioned, one of the important reasons to be interested in issues related to measuring degrees of freedom has to do with quantum gravity and the notion of holography.
 Roughly speaking, the holographic principle states that the number of degrees of freedom
 in a quantum theory of gravity scales with the area of the system, in contrast to standard
 QFTs where the entropy is extensive and scales with the
 volume \cite{Hooft:1993gx, Susskind:1994vu, Bigatti:1999dp}. In string theory a natural
 realization of the holographic principle is manifested by the AdS/CFT correspondence
 \cite{Maldacena:1997re, Aharony:1999ti} which gives us a precise map between a quantum
 gravity theory on an asymptotically AdS spacetime $\bulk$
 and an ordinary QFT  on the conformal boundary $\bdy$
 of $\bulk$. In this context we can ask whether there is a gravitational  dual of
 the entanglement entropy associated with a subsystem of the boundary QFT. Refer to
 \cite{Hawking:2000da, Maldacena:2001kr} for earlier pioneering works.

Interestingly, for a long time it has been known that  the leading
ultraviolet divergent contribution to the entanglement entropy
$S_{\rA}$ in QFTs is proportional to the area of the boundary
$\de\rA$ of the subsystem $\rA$ (known as the area law of
entanglement entropy) \cite{Bombelli:1986rw, Srednicki:1993im}. This
means that unlike the thermal entropy, the entanglement entropy is not
an extensive quantity.\footnote{For systems at finite temperature,
the entanglement entropy also includes a finite extensive term
which is proportional to the thermal entropy.} Instead, this
property looks very analogous to the holographic principle and
 the area law of Bekenstein-Hawking
black hole entropy. This fact strongly suggests a simple
gravitational interpretation of entanglement entropy in QFTs via a
holographic relation.

Recently,  a geometric procedure has been discovered to compute the
entanglement entropy of a sub-system $\rA \subset \bdys$ in the
context of the AdS/CFT correspondence~\cite{Ryu:2006bv,Ryu:2006ef}.
The construction which we review in \sec{lightsh} proceeds as
follows: given a region $\rA$ in $\bdys$ (at a fixed time) of a
static asymptotically AdS spacetime, we construct a minimal surface
$\ms$ (\ie, a surface whose area takes the minimum value) in the
bulk spacetime $\bulk$ which is anchored at the boundary $\brA$ of
$\rA$. The area of  this minimal surface in the bulk Planck units
provides an accurate measure of the entanglement of the degrees of
freedom in $\rA$ with those in its spatial complement, $\rB$. This
prescription has been verified by several non-trivial checks
\cite{Ryu:2006bv, Ryu:2006ef, Hirata:2006jx, Nishioka:2006gr,Matt}
as well as a direct proof \cite{Fursaev:2006ih}. This holographic
prescription provides a simple way to calculate the entanglement
entropy in spacetimes with no temporal evolution. Moreover, this
holographic relation is successfully applied to the brane-world
black holes \cite{Emparan:2006ni, Solodukhin:2006xv} and de-Sitter
spaces \cite{Iwashita:2006zj} as well, which enable us to interpret
the horizon entropy with quantum corrections as the entanglement
entropy (see also recent discussions \cite{Casini:2006ws,
Lee:2007zq, Cadoni:2007vf}).

The geometric perspective provided by the minimal surface
construction has many advantages,  especially for QFTs in dimensions
$d > 2$, since there are relatively few techniques to calculate the
entanglement entropy in interacting field theories. Furthermore,
herein lies the hope to address an interesting question related to
entanglement entropy, namely its behaviour as a function of time in
an interacting QFT.  In this context it is important to note that
since the entanglement entropy is not an extensive quantity, unlike
the conventional thermodynamic entropy, {\it a priori}  it does not
have to obey the Second Law. Nevertheless, it seems natural to
expect that when we consider an interacting QFT, the degrees of
freedom in region $\rA$ will interact with those in $\rB$ and
consequently get more entangled, thereby increasing the entanglement
entropy $S_\rA$.   Indeed the following theorem is well-known: let
$\Lambda_t$ $(t\in R^+)$ be a one parameter family of positive
linear  transformations of a Hilbert space ${\cal{H}}$ such that
they constitute a semi-group\footnote{Here a positive matrix is
defined to be a Hermitian matrix whose trace is positive. A positive
linear transformation is the one which maps a positive matrix to
another positive matrix. Also we require that it does not change the
identity and satisfies Tr$\,\Lambda_t(\rho)=\mbox{Tr}\,\rho$ for any
density matrix on ${\cal{H}}$.}; then $S(\Lambda_t(\rho))\geq
S(\rho)$. This ``monotonicity" property essentially comes from the
concavity of $-\rho\log\rho$ as a function of $\rho$. In the setup
of this theorem, we interpret $\Lambda_t$ as the irreversible and
non-unitary time-evolution such as a  quantum analogue of the Markov
process. Also $\rho$ is taken to be the reduced density matrix
$\rho_{\rA}$. On the other hand, in the case of a unitary time
evolution of an excited state, the entropy for the total system
remains the same while the entanglement entropy for a subsystem can
change. Explicit examples are borne out in the analysis
of~\cite{Calabrese:2005in} where the authors analyse the situation
in two dimensional field theories. However, the general story is far
from clear and one would like to get a better handle on the problem.
Hence, instead of examining time-dependence of entanglement entropy
from the QFT point of view, we would like to analyse it using the
holographic prescription mentioned above.

In the context of the AdS/CFT correspondence, the prescription for
calculating the entanglement entropy from the  area of a minimal
surface  suffers from one stumbling block: the minimal surfaces  are
usually associated with Euclidean geometries. In Lorentzian
spacetimes one has trouble defining  a minimal surface, because by
wiggling a spacelike surface in the time direction, one can make its
area arbitrarily small.  For static spacetimes, this problem is
usually avoided by Wick rotating and working  in the Euclidean
set-up, or equivalently by restricting attention to a constant time
slice.  But for the  most interesting, dynamical questions, this
method is not applicable. However, this does not necessarily mean
that the notion of the geometric dual of the entanglement entropy
cannot be defined in general. Indeed, as we have explicitly seen,
entanglement entropy is well-defined in terms of the time-dependent
density matrix, and therefore has to admit a well-defined
holographic dual.  By well-defined we mean generally covariant.
Hence, our strategy for examining the entanglement entropy dual in a
general time-dependent scenario will be to first find a suitable
fully covariant generalization of the minimal-surface proposal, and
then to use this `covariant holographic entanglement entropy'
definition\footnote{ In what follows, to simplify the terminology
somewhat, we will sometimes denote this as simply ``covariant
entanglement entropy"; it should be clear from context when we mean
the gravitational dual and when we are talking about the QFT
quantity. } to find the time-variation in the specific cases of
interest.

 To motivate the possibility of generalizing the dual of entanglement entropy in time-dependent scenarios,
  it is useful to think of the analogy with a spacelike geodesic
 (which in fact describes the minimal surface for a 3-dimensional bulk).  In Euclidean
 spacetimes, spacelike geodesics are local minima of the proper length functional. However,
 in Lorentzian spacetimes they are extrema of the proper length. Likewise, we expect that the
  natural analog of the Euclidean minimal surface to be an {\it extremal surface}, denoted by
   $\Gms$ below, which is a saddle point of the proper area
   functional. This expectation is indeed realized, and forms the primary result of this paper.

For stationary bulk geometries with a timelike Killing field, the entanglement entropy is likewise time independent,  and there exists a canonical foliation of the bulk spacetime $\bulk$ by spacelike surfaces. In a generic time-dependent background there is no preferred canonical foliation in the bulk. In contrast, for a QFT on a fixed background we do have a natural notion of time.  The issue from a gravitational standpoint is then whether a given  spacelike foliation in  $\bdy = \prod_t\, \bdys_t \times \R_t$, extends in a unique fashion into the bulk to provide us with the requisite foliation of $\bulk$. If the answer is in the
 affirmative, then we can use the spacelike slices thus constructed and find minimal surfaces
 localized  within them.

 Indeed, even in time-dependent geometries it is plausible  that there is a natural slicing of the bulk spacetime $\bulk$: we can define ``maximal area"
 co-dimension one spacelike slices $\Sms$, by the vanishing trace of the
  extrinsic curvature on $\Sms$.  Since each $\Sms$ is spacelike, we now have
   a well-defined prescription for finding a minimal-area (bulk co-dimension two)
   surface localized within $\Sms$ and anchored at $\brA$.
   We denote this `minimal surface on maximal slice' by $\Xms$.  The surface $\Xms$ is  covariantly defined, and like $\Gms$ it reduces correctly to the requisite minimal surface for static spacetimes, thereby providing another candidate for the covariant entanglement entropy.
 However, as we will see, to make contact with a holographic perspective we
     will have to elevate the notion of the maximal slice to that of a
     totally geodesic co-dimension one slice.

In this paper we examine the two constructions $\Gms$ and $\Xms$ motivated above, and propose a more appealing covariant generalization $\Lms$ of the geometric construction of \cite{Ryu:2006bv, Ryu:2006ef} to compute entanglement entropy in general asymptotically AdS spacetimes.  The basic idea behind our proposal is to exploit the light-sheet construction of the covariant entropy bounds of Bousso \cite{Bousso:1999xy, Bousso:1999cb, Bousso:2002ju}. Light-sheets are a natural concept in Lorentzian spacetimes and serve to single out a co-dimension two spacelike surface of the  bulk manifold whose area bounds the entropy passing through its light-sheet in the context of the covariant entropy bounds.
 We will denote this surface, whose construction we focus on in what follows, by $\Lms$.
 The minimal surface $\Xms$ construction also singles out a co-dimension two spacelike surface,
 albeit by first picking a spacelike foliation and then finding a co-dimension one surface within
 the leaves of the foliation. It is thus natural to expect that there is an intimate  relation
 between the light-sheet construction and minimal surfaces
 and indeed we will show that they are equivalent if a given time slice
 is totally geodesic. We  hope a similar argument can be applied to more general
spacetimes with boundaries allowing bulk non-trivial minimal surfaces.

A natural way to motivate the light-sheet construction is to consider a cut-off field theory in
 asymptotically AdS spacetimes. The dual description of the bulk is then in terms of a cut-off
 field theory coupled to dynamical gravity on the cut-off surface.  Due to the gravitational
 dynamics in the boundary field theory  the entropy associated with any co-dimension two surface
 bounds the amount of information that passes through the light-sheet associated with that surface.
 The spacelike co-dimension two surface can be taken to be the boundary $\brA$ of the subsystem $\rA$.
  Aided by this construction we can extend the light-sheets that live on the cut-off surface
  into bulk light-sheets and ask what is the spacelike surface in the bulk associated with these?
  Imposing the constraint that the spacelike co-dimension two\footnote{Note that co-dimension two
  surface in the cut-off boundary corresponds to a co-dimension three surface in the full bulk;
  so here the requisite surface has the same dimension
  as $\rA$ rather than $\brA$ -- see Table \ref{dimensionality} in \App{3dspl}. } surface in $\bulk$ be required to   have boundary $\brA$ on $\bdy$ so that light-sheets can end on
  it,  we can find  the bulk surface we were looking for.

While this motivates the proposal for a covariantization of the
geometric prescription for finding the entanglement entropy in terms
of light-sheets in this formulation, it is not very constructive.
There is in fact a simple algorithm for actually constructing the
bulk surface $\Lms$ in question: find the spacelike co-dimension two
surface whose boundary coincides with $\brA$ on $\bdy$ with the
constraint that the trace of the null extrinsic curvatures (\ie\ the
null expansions) associated with the two null normals to $\Lms$
vanish. For smooth surfaces parameterized by two functions this
leads in general to some partial differential equations which can be
solved to obtain a precise construction of the surface. Furthermore,
we can show that this definition of $\Lms$ is  actually equivalent
to the requirement that $\Lms$ is the co-dimension two extremal
surface  in the Lorentzian manifold with the specified boundary
condition. In other words, $\Lms = \Gms$. Thus this construction
naturally reduces to the minimal surface prescription of
\cite{Ryu:2006bv,Ryu:2006ef}.  

As a check in a simple non-static
example, we analytically compute the holographic entanglement
entropy of three dimensional rotating (BTZ) black holes employing
our covariant prescription. The result precisely agrees with the
entropy calculated in the dual two dimensional CFT.
 Furthermore, we also argue that the prescription of finding
the surface $\Lms$ using the vanishing null extrinsic curvatures can
be derived from a bulk-boundary relation {\it a la}., GKP-W relation
\cite{Gubser:1998bc, Witten:1998qj} for the AdS/CFT correspondence.
One can set up a variational problem by exploiting these ideas and
show that the action principle in gravity singles out the extremal
surface.

Once we have a covariant prescription for computing the Lorentzian
extremal surface in $\bulk$ we can ask the basic questions that
motivated the investigation in the first place, such as whether the
entanglement entropy has definite monotonicity
 properties {\it vis a vis} temporal evolution. To address this issue we
  discuss the example of a spacetime background involving a collapse scenario
   leading to black hole formation; the spacetime is modeled by a Vaidya-AdS spacetime.
    Due to the formation of a black hole in the bulk, we expect that the dual field theory
    on the boundary thermalizes. The thermalization is expected to lead to an increase
     in the entanglement entropy: the ergodic mixing of the boundary degrees of freedom
      would suggest that the degrees of freedom localized in region $\rA$ interact more
       with those in $\rB$ and thereby one expects that the entanglement entropy grows
        in time. The  bulk computation using the light-sheet prescription bears out this picture nicely.

Another example where a covariant formulation is necessary is the
case of wormhole spacetimes in AdS with two disconnected boundaries
\cite{Maldacena:2004rf}.
 Even though the two CFTs on the two disconnected boundaries look
 decoupled from each other, there are non-vanishing correlation functions
  between two theories in the dual gravity calculation as pointed out in
   \cite{Maldacena:2004rf}. We would like to present a possible resolution
    to this puzzle  by computing the entanglement entropy between the two CFTs.

The outline of the paper is as follows: we begin in \sec{lightsh}
with a quick review of the minimal surface proposal for static
spacetimes and the reasons to expect a covariant generalization of
this picture. We then proceed in \sec{cholols} to motivate the
light-sheet construction for time-dependent backgrounds. We present a manifestly
covariant holographic entanglement entropy in this section, which is
the most important conclusion of this paper. We explain
how this construction can be naturally motivated from a variational
principle and its connection to the bulk-boundary relation within the
AdS/CFT context in \sec{covother}. 
In \sec{examples} we illustrate the calculations of the
entanglement entropy using our covariant proposal and demonstrate
the consistent agreement with the minimal surface prescription of
\cite{Ryu:2006bv, Ryu:2006ef}. We also examine rotating BTZ black
holes, which are stationary but non-static, and show that our
covariant proposal precisely reproduces the entanglement entropy
computed from the CFT side.  In
\sec{timedep} we discuss the explicit time-dependent situation of
gravitational collapse and argue that the entanglement entropy
increases monotonically in this context. We discuss other
interesting time-dependent backgrounds, such as AdS wormholes and
bubbles of nothing in \sec{wholebub} and end with a discussion in
\sec{discuss}. In \App{3dspl} we present a simpler covariant
construction which whilst not reproducing the correct minimal
surface in general  is nevertheless interesting in that it provides
a bound on the entanglement entropy.  In \App{apmin} we give a proof
of equivalence between the vanishing of null expansions and the
extremal surface. In \App{apvaidya} we presents some details of the
calculations of the time-dependent entanglement entropy in the
Vaidya-AdS background using perturbative methods.

\section{Entanglement entropy and time-dependent QFTs}
\label{lightsh}
 As mentioned in the Introduction, our main aim is to find a covariant
 prescription for calculating the entanglement entropy associated with a
  given region of the boundary conformal field theory. We begin by reviewing the
  minimal surface proposal of \cite{Ryu:2006bv, Ryu:2006ef}, which
  provides the first step of
   geometrization of entanglement entropy in the AdS/CFT context, and which will
   serve to set up the background and notation for the subsequent generalization
   to non-stationary spacetimes.
 We then argue that entanglement entropy remains a well-defined concept in time-varying states in the field theory, and motivate a correspondingly well-defined dual geometric construction which would accommodate any time-dependence in the bulk.  Finally, we remark that there are in fact many such plausible constructions, and give an overview of those we focus on in the present paper.

\subsection{Review of holographic entanglement entropy}
\label{minsurTT}

Consider a $d+1$ dimensional asymptotically AdS spacetime $\bulk$ with conformal
boundary $\bdy$. For the present we will concentrate on the static case when $\bulk$ admits
 a timelike Killing field $\ptm$. On the boundary $\bdy$  of $\bulk$, which serves as the
 background for the dual field theory,  time translations are generated by $\ptm$ which is
 simply the pullback of the bulk Killing field. Thus we can naturally foliate the boundary $\bdy$
 by spacelike surfaces which are normal to this  timelike Killing field so that
 $\bdy = \prod_t\, \bdys_t \times \R_t$. Consider then  a particular leaf $\bdys$ of
 this foliation which we wish to divide into into two regions $\rA$ and $\rB$ so that
  $\bdys = \rA \cup \rB$.  The boundary between these regions is denoted as $\brA(=\de{\cal  B})$
  assuming that $\bdys$ is a compact manifold.  Note that $\rA$ is $(d-1)$-dimensional and $\brA$
  is therefore $(d-2)$-dimensional.

For a QFT on $\bdy$ we can calculate the entanglement entropy
associated with the region $\rA$. Since there are infinitely many
degrees of freedom in an ordinary QFT, it is known that the
entanglement entropy suffers from an ultraviolet divergence. The
standard result is that the leading divergence of entanglement
entropy scales as the area of the boundary $\brA$ between the two
regions (or sub-systems, as they are conventionally referred to in
the entanglement entropy literature)
\cite{Bombelli:1986rw,Srednicki:1993im}. The intuitive reason of
this area law for the divergent part is that the most entangled
degrees of freedom are the high energy ones localized within an
infinitesimal neighbourhood of $\brA$.
  Essentially,
\begin{equation}
S_\rA = \a \, {\area{\brA} \over \cof^{d-2} } + \cdots \ ,
\Label{entexp}
\end{equation}
where we have indicated the leading divergent behaviour ($\ap$ is a
constant factor). The infinitesimally small parameter $\cof$ denotes
the ultraviolet divergence (\ie\ lattice spacing). The subleading
terms contain slower power law or logarithmic divergences apart from
finite terms which are of interest.

Since we work within the AdS/CFT context we can ask whether the entanglement entropy
 for the boundary QFT can be calculated using a purely geometric construction in
 the bulk; this question was answered in the affirmative in
 \cite{Ryu:2006bv, Ryu:2006ef}. The essential
  idea behind the picture of \cite{Ryu:2006bv, Ryu:2006ef}
  is the following: by virtue of time
   translation invariance, the boundary spacelike foliation naturally extends into the bulk
    to provide a canonical spacelike foliation $\prod_{t}{\cal N}_t$ of $\bulk$.
    On a given spacelike
     slice in the $\bulk$ we are instructed to construct a minimal (area) surface which
      ends on $\brA \subset \bdys$. This is a well defined problem and the minimal surface
       which is a spacelike surface of vanishing mean curvature is guaranteed to exist
        due to the Euclidean signature of the bulk spacelike slice. Thus, given the minimal
         surface $\ms_{min}$, the entanglement entropy associated with region $\rA$ is
\begin{equation}
S_\rA = {\area{\ms_{min}}  \over 4 \, G_N^{(d+1)} } \ .
\Label{minTT}
\end{equation}
Note that the minimal surface $\ms$ is a co-dimension two surface in the bulk spacetime
 $\bulk$ by virtue of being a co-dimension one submanifold of a particular leaf of the
  spacelike foliation.

\subsection{Entanglement entropy in time-dependent states in QFT}
\label{tdqft}

For states in QFT with trivial time-dependence, one can calculate the entanglement entropy in a conventional manner by looking at the decomposition of the total Hilbert space on a given time-slice. The holographic perspective of this is captured by the minimal surface prescription indicated in \req{minTT}. It is clear from the outset that in QFT nothing prevents us from considering explicitly time-varying states and computing entanglement entropy for subsystems thereof. It is easy to give a path-integral prescription for computing the entanglement entropy in these circumstances and we outline the basic methodology below.

Consider a quantum field theory in a time-dependent background. Its
evolution in time is described by the time-dependent Hamiltonian
$H(t)$. A state at the time $t=t_1$ is defined by $\ket{\Psi(t_1)}$.
It is
 related to the state at the time $t_0$ via the familiar formula
 \begin{equation}
\ket{\Psi(t_1)}=T\, \exp\left(-i\int^{t_1}_{t_0} dt \,
H(t)\right)\ket{\Psi(t_0)} \ .
\Label{statTevol}
\end{equation}

In the path integral formulation, the ket state $\ket{\Psi(t_1)}$
is equivalently constructed by the path-integral\footnote{By
employing the conventional $i \, \eps$ -- prescription we can project the
asymptotic state $\ket{\Psi(t=\infty)}$ to the ground state.} from
$t=-\infty$ to $t=t_1$
\begin{equation}
\Psi\left(t_1,\phi_0(x)\right)= \int^{t=t_1}_{t=-\infty} \,[D\phi]\;
e^{iS(\phi)}~\delta\left(\phi(t_1,x)-\phi_0(x)\right) \ ,
\Label{pathina}
\end{equation}
 where we represent all fields by $\phi$. On the
other hand, the bra state $\bra{\Psi(t_1)}$ is expressed as
follows:
\begin{equation}
 \overline{\Psi}\left(t_1,\phi_0(x)\right)=
\int^{t=\infty}_{t=t_1} \, [D\phi] ~ e^{iS(\phi)}~
\delta\left(\phi(t_1,x)-\phi_0(x)\right) \ . \Label{pathinb}
\end{equation}
Clearly they satisfy
\begin{equation}
i\, \frac{\p}{\p t}\ket{\Psi(t)}=H(t) \, \ket{\Psi(t)} \ ,\qquad
i\, \frac{\p}{\p t}\bra{\Psi(t)}=-\bra{\Psi(t)} \, H(t) \ .
\Label{schrodeq}
\end{equation}

Let us first assume the total system is described by a pure state
$\ket{\Psi(t)}$ with unit norm  at vanishing
temperature.  Then the total density matrix is given by
\begin{equation}
\rho_{tot}(t)=|\Psi(t)\rangle \langle\Psi(t)| \ ,\Label{pures}
\end{equation}
and its time-evolution is dictated by the von-Neumann equation
\be
i\,\f{\p \rho_{tot}(t)}{\p t}=[H(t),\rho_{tot}(t)] \ . \ee
In the gravitational context we will consider interesting examples
with event horizons. These will be described in the dual CFT by a
mixed state and so  we need to formulate the theory by using only
the density matrix $\rho_{tot}(t)$. However, even in such  cases we
expect to have an equivalent description in terms of a pure state by
assuming another CFT sector hidden inside the horizons as in the
Schwarzschild-AdS case \cite{Maldacena:2001kr}, or other degrees of
freedom in more general circumstances as in the examples of
\cite{Freivogel:2005qh}.  Thus the assumption (\ref{pures}) does not
exclude the choice of density matrices, so long as we can purify the
state by passage to a enlarged Hilbert space (which, in the
geometry, corresponds to another sector behind the horizon).

Divide the total Hilbert space into a direct product of two Hilbert
spaces at time $t$: $\CH_{tot}=\CH_\rA\otimes \CH_\rB$.
 In the quantum field theory, this is realized by dividing the
  total space manifold $\bdys$ at a fixed time into two parts $\rA$ and $\rB$.
Then the entanglement entropy $S_\rA(t)$ at time $t$ is defined as follows
\begin{equation}
S_\rA(t)=-\mbox{Tr}_{\rA}\,\(\rho_\rA(t)\, \log\rho_\rA(t) \) \ ,
\Label{ent}
\end{equation}
where $\rho_\rA(t)$ is the reduced density matrix
\begin{equation}
\rho_\rA(t)=\mbox{Tr}_{\rB}\, \rho_{tot}(t)
=\mbox{Tr}_{\rB}\,\ket{\Psi(t)}\, \bra{\Psi(t)} \ . \Label{reduce}
\end{equation}
We always normalize any (reduced) density matrices $\rho$ such that
their trace is one \ie, $\mbox{Tr}\, \rho=1$. In order to express
$S_\rA(t)$ in the path integral formalism, we need to first describe
the reduced density matrix in that formalism (see also
\cite{Calabrese:2004eu, Calabrese:2005in, Fursaev:2006ng,
Ryu:2006ef}). Taking the trace in the Hilbert space $\CH_\rB$ in
(\ref{reduce}) is equivalent to partially gluing two boundaries in
(\ref{pathina}) and (\ref{pathinb}) along $\rB$. Thus it is
described by the path-integral over the whole spacetime with an
infinitesimally small slit along $\rA$ at a fixed time $t$
\begin{equation}
\[ \rho_\rA(t)\]_{\phi_+ \phi_-} = {1\over Z_1} \cdot \int
^{t=\infty}_{t=-\infty} \,[D\phi]~ e^{iS(\phi)}\, \prod_{x\in
\rA}\, \delta\left(\phi(t+\eps,x)-\phi_+(x)\right)\, \,
\delta\left(\phi(t-\eps,x)-\phi_-(x)\right) \ ,
\Label{pathrho}
\end{equation}
 where
$\eps$ is an infinitesimal positive constant and we also defined
\be Z_1=\int ^{t=\infty}_{t=-\infty} \, [D\phi]~
e^{iS(\phi)} \ .\ee
Given the definition of the trace of the density matrix $\rho_\rA$ in \req{pathrho},
 it is easy to calculate the trace ${\rm Tr }(\rho_\rA)^n$. This
  is  calculated by integrating the products of path-integrals
\be
[\rho_\rA(t)]_{\phi_{1+} \phi_{1-}} \; [\rho_\rA(t)]_{\phi_{2+} \phi_{2-}}\; \cdots
\; [\rho_\rA(t)]_{\phi_{n+} \phi_{n-}},\ee
successively with the identifications:
$\phi_{1-}=\phi_{2+}$, $\phi_{2-}=\phi_{3+},\cdots$ and $\phi_{n-}=\phi_{1+}$.
 In other words, this is essentially the partition function $Z_n(t)$ on the
 (singular) manifold $\bdy_n$ which is
 defined by the $n$ copies of the total manifold ${\cal{M}}$ glued along $\rA$
at the fixed time $t$
\begin{equation}
\mbox{Tr}\(\rho_\rA(t)\)^n=\frac{Z_n(t)}{(Z_1)^n}  \ .
\Label{parex}
\end{equation}
Knowledge of the partition function $Z_n(t)$ on the singular
manifold, then allows us to compute the entanglement entropy using:
\begin{equation}
 S_\rA(t)=-\frac{\p}{\p n} \, \log \mbox{Tr}\(\rho_A(t)\)^n\biggr|_{n=1}= \log
Z_1-\frac{\p \log Z_n(t)}{\p n}\biggr|_{n=1}. \Label{entfor}
\end{equation}
%

\subsection{Towards holographic entanglement entropy in time-dependent states}
\label{holtimedep}

The discussion of entanglement entropy in time-dependent  QFT states
 in the previous section makes it clear that there is no
  {\it a priori} obstruction in thinking about this issue from a
   field theoretic perspective. In the AdS/CFT context we would then like
    to ask whether the holographic entanglement entropy proposal of~\cite{Ryu:2006bv,Ryu:2006ef}  can be generalized to time-dependent scenarios.  In particular, can we find a suitable generalization of the minimal surface which is fully covariant?  The answer is of course {\it yes}, and
    in fact we will propose several covariant constructions in this and the next section, and
    examine the relations between them.

To motivate the existence of a suitable covariantly well-defined
surface, we start\footnote{Those readers who
 would like to know the final conclusion immediately are advised to skip to the covariant
 entanglement entropy proposal (I) and (II) in
 \sec{lsmotive} and \sec{covls}.}
  by indicating the construction of the surface which we will denote as  $\Xms$,
 which is the most naive generalization of the minimal surface in the case of static bulk spacetimes.
  Consider a time-dependent  version of the AdS/CFT correspondence where the boundary
  theory is taken to be in a time-varying state on a  fixed background $\bdy$. The corresponding
  bulk geometry $\bulk$ will have an explicit time-dependence
   and hence no timelike Killing field. Since the metric on $\bdy$ is non-dynamical in the boundary,
   we can choose a foliation by equal time slices, by picking our time coordinate such that it
   implements the natural Hamiltonian evolution of the field theory, so that
 $\bdy =  \bdys_t \times \R_t$. We can choose to consider a region $\rA_t \in \bdys_t$ on a given
  time slice as in \sec{tdqft}  and compute the entanglement entropy using the path integral prescription.
   The question then is what is the analog of this computation from a bulk perspective?

Naively, one would expect that the minimal surface prescription for computing the
 holographic entanglement entropy should go through. However, this cannot quite be
 the case; as mentioned in \sec{intro}, in Lorentzian spacetimes one has to be careful about defining suitable  minimal area surfaces due to the indefinite metric signature.

 The crucial issue in a Lorentzian setting is the fact that generically,  the equal-time foliation on the  boundary $\bdy$ does not necessarily lead to a canonical (\ie, symmetry-motivated) foliation of the bulk $\bulk$.  Supposing for the moment that a natural foliation was singled out; we could then compute the holographic entanglement entropy by first picking the
preferred spacelike slice $\CN_t$ of $\bulk$ given by extending the slice from $\bdy$.
On $\CN_t$ the induced metric is spacelike and  the notion of the minimal surface is well defined. The holographic prescription then amounts to finding a minimal surface $\ms \in \CN$ such that  $\p \ms |_{\bdy}= \brA$.

The above observation suggests that we look for a covariantly
defined spacelike slice of the bulk, $\CN_t$, anchored at $\bdys_t$,
which reduces to the constant-$t$ slice for static bulk.
Generically, while one expects no preferred/natural time slicing of
$\bulk$, it is plausible that for asymptotically AdS spacetimes one
has a preferred foliation by zero\footnote{ In general, any constant
mean curvature slice of the bulk provides a covariantly well-defined
surface; however, when this constant is non-zero it doesn't satisfy
the requirement of reducing to the constant-$t$ slice in static
bulk. } mean curvature slices \ie, slices with vanishing trace of
extrinsic curvature. Physically, each of these slices corresponds to the
{\it maximal area} spacelike slice through the bulk, anchored at the
boundary slice $\bdys$.\footnote{We thank Doug Eardley, Gary
Horowitz, and Don Marolf for discussions on this issue.} We denote
the leaves of this maximal-area foliation by $\Sms_t$.

One might worry that the maximal-area slice is not well defined because an area of a given surface can always be increased by ``crumpling" or wiggling the surface in the spatial directions; however, here the crucial point is that our slice has co-dimension one, extending over all the available spatial directions, and therefore allows no room for wiggling.  Another possible concern is the fact that in asymptotically AdS spacetimes, the area of any spacelike slice is manifestly infinite.  However, this is the familiar problem of regulating the lengths/areas/volumes in AdS, which we know how to deal with.  Below, we will use a simple background subtraction technique, and regulate all quantities by subtracting off the corresponding values in pure AdS.

Provided we have this special, maximal spacelike slice $\Sms_t$
through the bulk, we proceed as outlined above: on this slice, we
construct the minimal-area surface anchored at $\brA_t$. This
amounts to a mini-max algorithm for the holographic entanglement
entropy; find a maximal slice in the bulk which agrees with the
spacelike foliation of $\bdy$ and in that maximal slice find a
minimal surface $\Xms$. In this setup, one may obtain a natural
proposal that the area of $\Xms$ then gives the entanglement
entropy,
\begin{equation}
S_\rA = {\area{\Xms}  \over 4 \, G_N^{(d+1)} } \ .
\Label{minmax}
\end{equation}
Note that the surface $\Xms$ by construction satisfies the three basic pre-requisites for being a candidate dual of the entanglement entropy of $\rA$:  it is covariantly well-defined, it is anchored at $\brA$, and it reduces to the requisite minimal surface when the bulk spacetime is static.

However, we will argue that this prescription doesn't follow naturally from a holographic viewpoint and needs to be finessed slightly to make contact with the holographic perspective.
Thus, rather than stopping at our candidate surface $\Xms$, in the next section we will propose another candidate surface, $\Lms$, as a more natural dual of the entanglement entropy.
In fact, this holographic formulation will provide a more straightforward algorithmic construction of the requisite surface $\Lms$. Our starting point is motivated by the idea of light-sheets introduced in the context  of covariant entropy bounds in gravitational theories. We will argue that the prescription which we find in terms of light-sheets reduces to the intuitive picture presented above, modulo some subtleties, with the added bonus of providing an explicit equation for the extremal surface.

\subsection{Preview of covariant constructions}
\label{covEEprev}

Before delving into the details of these constructions, we briefly
list them with a short summary of our final conclusion, to orient
the reader and fix the notation. In all cases, the
requisite surface is a co-dimension two bulk surface which is
anchored on the boundary at $\brA_t$.
  In addition, all of these constructions are fully covariant -- they
   do not depend on any particular choice of coordinates -- and therefore
    are physically well-defined.  Also, these surfaces are mutually closely
    related; although different symbols are used to indicate different constructions,
     this is not meant to imply that the surfaces thus constructed are necessarily distinct.
      In part of what follows we will examine the specific relations between them.
\begin{itemize}
\item
$\Gms$: extremal surface, given by a saddle point of the area action.
In 3-dimensional bulk, this is simply the spacelike geodesic through the
bulk connecting the points $\brA_t$.  We return to discuss this surface in \sec{gkpmotive}.
\item
$\Xms$: minimal-area surface on maximal-area (co-dimension one)
slice of the bulk.  The construction was motivated in
\sec{holtimedep} and we will discuss some subtleties and
generalizations in \sec{extminmax}. In particular, we will show that
$\Xms$ coincides with the extremal surface $\Gms$ if $\Xms$ is
situated on a totally geodesic spacelike surface.
\item
$\Lms$: surface wherefrom the null expansions along the requisite
future and past light-sheets vanish. This will be the construction
we primarily focus on.  In fact, we propose two constructions,
$\Lms_{\rA_t}^{min}$ in \sec{lsmotive}, and $\Lms_{ext}$ in
\sec{covls}. We will later see that $\Lms_{\rA_t}^{min}$ and
$\Lms_{ext}$ are equivalent to the extremal surface $\Gms$ (we
present a proof in the \App{apmin}).
\item
$\Cms$: `causal construction,' discussed mainly in \App{3dspl}: minimal-area surface on the boundary
of the causal wedge of the boundary domain of dependence of $\rA_t$.\footnote{Note added in v3: In the previous versions we had erroneously identified $\Cms$ as the maximal-area surface on the boundary of the causal wedge. It has recently been proved in \cite{Hubeny:2012wa} that $\Cms$ is instead a minimal surface.\label{v3addition}} \end{itemize}

Our main claim of this paper will be that the covariant holographic
entanglement entropy is obtained from the area of the surface
$\Gms=\Lms$. In a generic time-dependent spacetime, we will find
that another surface $\Xms$ deviates slightly from $\Gms=\Lms$. We
will also confirm that $\Gms$, $\Xms$, and $\Lms$ all reduce to the
minimal surface in static bulk spacetime; however this is not
necessarily the case for $\Cms$. Nevertheless, $\Cms$ will be useful
because, as we motivate in \App{3dspl}, it provides a bound on the
entanglement entropy, and is computationally simpler to
find.

\section{Covariant holographic entanglement entropy and light-sheets}
\label{cholols}
\subsection{Light-sheets and covariant constructions}
\label{lsmotive}

To motivate the natural covariant generalization of a holographic entanglement
 entropy proposal, it is useful to recall the construction of covariant
  entropy bounds in gravitational theories. The main issue in defining covariant entropy bounds
  was to put a bound on the entropy/information passing
   through a given region of spacetime in a fashion that is independent
    of the choice of coordinates or slicing. A clear formulation of
    covariant entropy bounds was achieved by Bousso
     \cite{Bousso:1999xy, Bousso:1999cb, Bousso:2002ju} using the concept of
     light-sheets.
     A discussion of this entropy bound applied to the AdS/CFT,
     which stimulates our arguments below, can be found in \cite{Bousso:2001cf}.

\begin{figure}[htbp]
\begin{center}
\includegraphics[width=5cm]{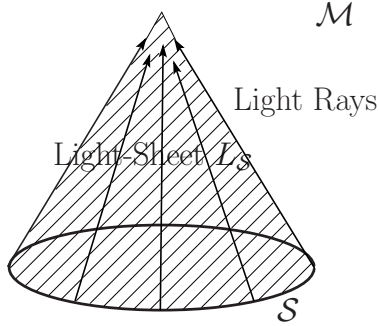}
\caption{A light-sheet $L_{{\cal S}}$ for a co-dimension two
space-like surface ${\cal S}$. The null geodesics on the light-sheet
are converging, \ie, the expansion is non-positive. }
\label{boussofig}
\end{center}
\end{figure}

Before proceeding to discuss the relevance of light-sheets for
calculating entanglement entropy, let us review the concept of a
light-sheet. Given any co-dimension two spacelike surface ${\cal
S}$ in a spacetime manifold $\bulk$, we construct four congruences of
future/past null geodesics from
 the surface in in-going and out-going directions.  A light-sheet
 $L_{{\cal S}}$ for ${\cal S}$ corresponds to those null geodesic
  congruences for which the expansion of the null geodesics is
  non-positive definite (we will explain the definition of the expansion of
  null geodesics in the next subsection; physically, we require that
  the cross sectional area at a constant affine parameter
   along the congruence does not increase).  The null geodesics along the
  light-sheet are converging and will
   eventually develop caustics; at any such point the light-sheet gets cut  off.  According to the
    covariant entropy bound (Bousso bound), the entropy or amount of information
    $S_{L_{{\cal S}}}$ that can pass through
    a light-sheet (\ie\ the integral of the entropy flux on the
    light-sheet \cite{Flanagan:1999jp})
    is bounded by the area of spacelike surface as
    follows:
\be
S_{L_{{\cal S}}}\leq \f{\mbox{Area}({\cal S})}{4\, G_N} \ .
\Label{bousso}
\ee

We would like to propose that the correct generalization of the
holographic entanglement  entropy is in terms of these light-sheets. One can motivate this claim by analyzing the  QFT coupled to gravity as in a brane-world set-up (\ie, RS II model \cite{Randall:1999vf}). Indeed, the Bekenstein-Hawking entropy of brane-world black holes can be interpreted as an entanglement entropy in this setup, as discussed in \cite{Emparan:2006ni, Hawking:2000da}.

 Consider the setup of the AdS$_{d+1}$/CFT$_{d}$ with an explicit UV cut-off in the bulk, $z>\cof$, where $z$ is the AdS radial coordinate, chosen such that the boundary is at  $z=0$. We choose Poincar\'e coordinates for AdS$_{d+1}$ (with the radius of AdS set to unity for simplicity)
\be ds^2=\f{1}{z^2}\, \(-dt^2+dz^2+\sum_{i=1}^{d-1}dx_i^2\) \ .
\Label{poincarem}
\ee
 The UV cut-off $\cof$ is infinitesimally small and is interpreted as a lattice spacing. This setup is  equivalent to the one of the brane-world where a very weak gravity exists on the $d$ dimensional brane located on the cut-off surface. The Newton's constant for the brane will be taken to be $G^{(d)}_N$. By the AdS/CFT
correspondence this set-up is dual to the bulk $d+1$ dimensional AdS
spacetime with the cut-off and a bulk Newton's constant
$G^{(d+1)}_N$ related via the rule
\begin{equation}
\f{1}{G^{(d)}_N}=\f{1}{G^{(d+1)}_N}\f{\int
dx^{d+1}\sqrt{g^{(d+1)}}R^{(d+1)}}{\int
dx^{d}\sqrt{g^{(d)}}R^{(d)}}= \f{1}{G^{(d+1)}_N}\int^{\infty}_{\cof}
\f{dz}{z^{d-1}} =\f{1}{(d-2) \,
 \cof^{d-2}}\,  \f{1}{G^{(d+1)}_N}\ .
\Label{newton}
\end{equation}

Since the brane-world theory has gravity coupled to the QFT, we can
consider the Bousso bound  for the $d$ dimensional boundary theory. We then would like to translate the computation of this bound holographically into a calculation from the viewpoint of the bulk  $d+1$ dimensional gravity. In the boundary the calculation would proceed by finding the light-sheets associated with the particular region $\rA$ we want to focus on.  Since in the boundary field theory the  light-sheets bound the region which is relevant for any entropy bound, the corresponding bulk prescription should likewise include no more than this region. A natural expectation  is then that the co-dimension two surface in the boundary has a canonical extension into the bulk spacetime in such a way that the associated bulk light-sheets are anchored on the boundary light-sheets under the appropriate restriction. Of course, there are potentially many surfaces that satisfy this requirement; we will then  pick the one that gives the strongest bound on the bulk entropy. Our claim then amounts to the statement that the dual bulk entropy bounds can be found by extending the boundary light-sheet (which was employed to find the covariant entropy bound in the boundary theory) into the bulk.

One intriguing consequence of this proposal is  that the bulk
results include quantum corrections, while the boundary results do
not, as is familiar in AdS/CFT. Let us see how these quantum
corrections look like in a specific example. We are interested in
the Bousso bound for the spacelike surface $\Sp^{d-2}$, \ie\ a $d-2$
dimensional sphere with the radius $l$ in the $d$ dimensional
brane-world. We choose $\rA$  a submanifold on a time-slice $t=t_0$
 such that $\de \rA=\Sp^{d-2}$. At the
classical level, we obtain the entropy bound
\be S_\rA\leq \f{\area{\de\rA}}{4\, G^{(d)}_N} \ . \ee
In order to take into account the quantum corrections, we extend the
light-sheet from $\de\rA=\Sp^{d-2}$ to that from half of a $d-1$ dimensional
sphere $\ms_\rA$ in the bulk AdS. Then we obtain the quantum
corrected entropy bound
\begin{equation}
S_\rA\leq \f{\area{\ms_\rA}}{4\, G^{(d+1)}_N}=\f{\area{\de\rA}}{4\,
G^{(d)}_N} \,
\left[1-\f{\cof^2}{l^2}\left(\log\left(\f{l}{\cof}\right)+{\rm
const.}\right)\right] \, < \f{\area{\de\rA}}{4\, G^{(d)}_N} \ .
\Label{entros}
\end{equation}
The finite difference between the above quantum and classical
entropy bound is analogous to the Casimir energy.

An alternate way to explain our motivation for considering light-sheets
is to think of the entanglement entropy as being directly related to the
(thermodynamic) entropy computed by the Bousso bound. More precisely,
we would like to claim that {\it the entanglement entropy saturates the Bousso bound}
in the setup of the AdS/CFT correspondence or the related brane-world version.
 While the claim that entanglement entropy is related to light-sheets
is {\it a priori} very surprising, the example of the static AdS background strongly suggests  this interpretation (see also \cite{Hirata:2006jx}).
Similarly, the bulk-boundary relation (so called GKP-W relation
\cite{Gubser:1998bc, Witten:1998qj}) in the AdS/CFT correspondence
leads to the same conclusion. A weaker version of this claim will be
that {\it the entanglement entropy satisfies the Bousso bound}. What
we have argued in the above is summarized as the following proposal
for direct holographic computation of entanglement entropy.

\paragraph{A Covariant Entanglement Entropy Proposal (I):}
Consider the usual AdS/CFT setup in a $d+1$ dimensional
asymptotically AdS spacetime
 $\bulk$ with $d$ dimensional boundary $\bdy$. We will choose the boundary $\bdy$ to be either
   $\R^{1,d-1}$ or $\R\times \Sp^{d-1}$; in the following we usually assume Poincar\'e
    coordinates for  simplicity. As explained earlier, at time $t$, we divide the $d-1$
     dimensional space of the boundary theory into $\rA_t$ and $\rB_t$.
     The boundary $\brA_t$ between these
      domains will play an important role. Note that $\brA_t$ is a $d-2$
       dimensional spacelike surface in $\bdy$.

\begin{figure}
\begin{center}
\includegraphics[width=10cm]{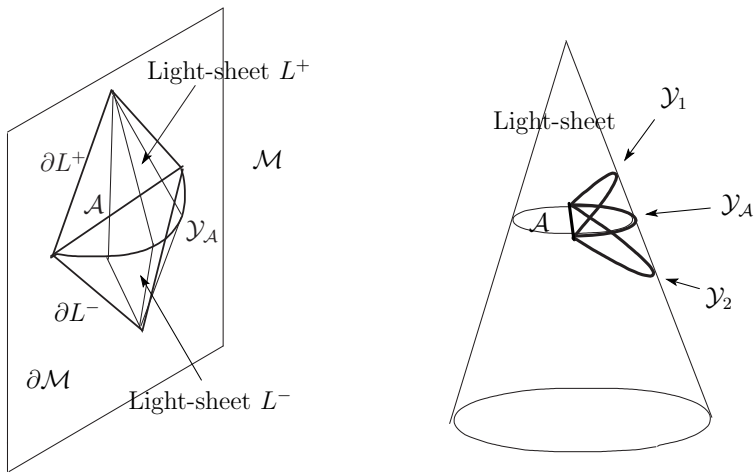}
\caption{A light-sheet construction in  \ads{3}/CFT$_2$.}
\label{lightsheets}
\end{center}
\end{figure}

Now, we can construct the upper and lower light-sheets $\de L_t^{+}$
and $\de L_t^{-}$
 for the spacelike surface $\brA_t$.
 This can be done in a straightforward manner using the conformally flat
   metric on $\bdy$. We then consider extensions $L_t^{\pm}$ of the two light-sheets
   $\de L_t^{\pm}$ into bulk such that they are the light-sheets in $\bulk$ with respect
    to a $d-1$ dimensional spacelike surface $\Lms_t =L_t^{+}\cap L_t^{-}$
    as in the left figure of \fig{lightsheets}.

Given this, we propose that the (possibly time-dependent) entanglement
entropy for the subsystem $\rA_t$ in the
 dual boundary theory is given by
\begin{equation}
S_{\rA_t}(t)=\frac{\min_{\Lms} \(\area{\Lms_t} \)}{4\,
G^{(d+1)}_N} \ . \Label{LmsminA}
\end{equation}
Here $\min_{\Lms} \(\area{\Lms_t} \)$ denotes  the minimum of the
area over the set of $\Lms$ as we vary the form of $L_t^{\pm}$
satisfying the above mentioned conditions with $\de L_t^{\pm}$
fixed. We denote this minimal area surface $\Lms_{\rA_t}^{min}$.
Essentially we then have the analog of \req{minTT},
\begin{equation}
S_{\rA_t}(t) = \frac{\area{\Lms_{\rA_t}^{min}}}{ 4\, G^{(d+1)}_N} \ .
\Label{LmsminB}
\end{equation}
%

\subsection{Expansions of null geodesics}
\label{nexpansion}

As we have already seen, the definition of the
light-sheet involves the expansions of null geodesics. Since this
quantity plays a crucial role in the discussions below, we will
pause to explain its definition and properties (for details
refer to \eg, \cite{Hawking:1973uf, Bousso:2001cf, Ashtekar:2004cn}).

 Given a co-dimension two surface $\ms$ in a spacetime
manifold specified by two constraints
\be
\vp_{1}(x^\nu) = 0 \ , \qquad  \vp_{2}(x^\nu) = 0 \ , \ee
we can define two one-forms
$\nabla_{\! \nu} \, \vp_{i}$, $i=1,2$. Non-degeneracy requires that there be two
linearly independent one-forms and so $\nabla_{\! \nu} \, \vp_{1} + \mu \,
\nabla_{\! \nu} \, \vp_{2}$ has to be a null one-form for two distinct
values of $\mu$. Using this information one can construct two
null-vectors $N_\pm^\mu$ that are orthogonal to the surface of
interest:
\be N_\pm^\mu  =g^{\mu\nu} \, \(\nabla_{\! \nu} \, \vp_{1} + \mu_\pm \,
\nabla_{\! \nu} \, \vp_{2}\)  \ .
\ee
We can fix the null vectors to be normalized such that
\be N_{+}^\mu \, N_{-}^{\nu} \, g_{\mu\nu} =-1 \ . \ee
In terms of $N_\pm^\mu$ and the induced metric $h_{\mu\nu}$ on the surface $\ms$
 we can write down the null extrinsic curvatures:
\be (\chi_\pm)_{\mu\nu} = h_{\ \mu}^\rho \,h_{\ \nu}^\lambda \, \nabla_{\! \rho}
\, (N_\pm)_{\lambda} \ . \ee

The expansion of an orthogonal null geodesic congruence to the
surface is then given by the trace of the null extrinsic
curvature\footnote{Because we are interested in null geodesic
congruences, there is no natural scale associated with the affine
parameter along the congruence.  We can choose to normalize the null
vectors by scaling $N_\pm \to \gamma_\pm\, N_\pm$  ($\gamma_\pm$ are
functions on $\ms$), whilst keeping them tangent to the null
geodesics. In practice, we usually omit the scaling, since we are
typically interested only in the sign of the expansions, and these
scale simply as $\theta_{\pm}\to \gamma_{\pm}\, \theta_{\pm}$.
However, if the rescaling is singular \ie, $\gamma=0$ or
$\gamma=\infty$, such simplification is not possible. This occurs
the case where $\ms$ coincides with an apparent horizon as will be
discussed in \sec{totent}.}
\begin{equation}
\theta_\pm = (\chi_\pm)_{\ \mu}^\mu \ .
\Label{thetadef}
\end{equation}

Physically, the null expansions measure the rate of change of the
area of the co-dimension two surface $\ms$
 propagated along the null vectors. Let us express
 the embedding map from $\ms$ to the spacetime $\bulk$ by
 $X^\mu(\xi^\ap)$, where $\xi^\ap$ denote the coordinates on $\ms$.
 Under an infinitesimal deformation $\delta X^\mu(\xi^\ap)$
orthogonal to $\ms$ with fixed boundary conditions,
the change in the area of $\ms$  is obtained from the value of
 the expansions (see \eg\ \cite{Senovilla:2004dc}):
\be \delta \mbox{Area}\propto
\int_{\ms} \left(\theta_+ \,  N_+^\mu \, \delta X_\mu +\theta_- \,
N_-^\mu \, \delta X_\mu\right) \ ,
\Label{areaform}
\ee
where the proportionality constant is positive. Therefore
determining the sign of the null expansions $\theta_\pm$ is
equivalent to finding whether the area increases or decreases when
we perform an infinitesimal deformation. In addition, \req{areaform}
clearly shows that the surfaces $\ms$ with vanishing null expansions
are extremal surfaces $\Gms$, \ie, saddle points of the area
functional. An explicit proof of this is given in \App{apmin}.

\subsection{The covariant entanglement entropy prescription and extremal surface}
\label{covls}

In \sec{lsmotive} we presented a covariant proposal for calculating
the holographic entanglement entropy based on a light-sheet
construction. Although manifestly covariant, the computation of the
surface $\Lms_\rA$ still involves  first constructing all possible
light-sheets in the bulk $L^\pm$ subject to the appropriate boundary
condition and then minimizing the area of the spacelike co-dimension
two slice $\Lms = L^+ \cap L^-$ over all the possibilities. We now
argue that this procedure can be vastly streamlined to produce a
simple set of partial differential equations for the surface.
Furthermore, in \sec{gkpmotive} we will show that the resulting
prescription follows naturally from a bulk--boundary relation {\it a
la}., GKP-W \cite{Gubser:1998bc, Witten:1998qj} in the AdS/CFT
context.

The covariant construction of \sec{lsmotive} starts from the two
boundary light-sheets, $\de L^+_t$ (future) and $\de L^-_t$ (past),
which are uniquely defined given a subsystem $\rA_t$ in the dual CFT
on $\bdy$ at time $t$. We then pick a co-dimension two spacelike
surface $\Lms_{\rA_t}$  in $\bulk$ whose boundaries coincide with
$\brA_t$ as in \fig{lightsheets}. There are many such surfaces, but
we are only interested in the ones which we can sandwich between the
two light-sheets $L^+$ (future) and $L^-$ (past) in the bulk
spacetime.  The existence of such light-sheets leads to the
constraints for the expansions,
 \be \theta_{\hat{+}}\leq 0\ ,\qquad \theta_{\hat{-}}\leq 0 \ ,
 \Label{expl} \ee
where $\theta_{\hat{+}}$ refers to the expansion along the null
congruence generating $L^+$ and similarly $\theta_{\hat{-}}$ for
$L^-$. The expansions $\theta_{\hat{\pm}}$ are equal to $\theta_\pm$
defined in \sec{nexpansion} up to a  sign.  Along a single
light-sheet, say $L^+$ with $\theta_{\hat{+}}\leq 0$,  small
deformations of the surface $\Lms_{\rA}$ into $\Lms_1$ and $\Lms_2$
(sketched in \fig{lightsheets})  always yield the inequality
$\mbox{Area}(\Lms_1)\leq \mbox{Area}(\Lms_\rA)\leq
\mbox{Area}(\Lms_2)$. Among infinitely many choices of such surfaces $\Lms_{\rA}$, we single out the one whose area becomes the minimum. Of course, there is no minimal surface if we search all surfaces with the same boundary condition due to the Lorentzian  signature. The additional condition \req{expl} of the non-positive  expansions along both light-sheets is crucial for the existence of this minimum.

Now pick a generic surface $\Lms_{\rA_t}$ which is not
necessarily the minimal one, such that
 the null expansions are negative everywhere on
$\Lms_{\rA_t}$. We expect that such a surface reaches in
further than the minimal surface, as can be checked by examining
explicit examples in \sec{examples}. As is clear from the
formula (\ref{areaform}), if we slightly deform the surface
 towards the boundary, its area decreases because
$\theta_{\hat{\pm}}\leq 0$. We will be able to continue this
deformation until both of the expansions become zero. The surface
obtained in this way has the area which is minimum among those
surfaces which allow the light-sheet construction. The validity of
the assumed structure of expansions which allows such a deformation
can be confirmed in an explicit example of AdS$_3$, as shown in the
\fig{lightsheetst} in \sec{stru}. 

The above procedure constructs the surface whose null expansions are
both vanishing. As we have seen in \sec{nexpansion}, this means that
this surface obtained from the minimization procedure
(\ref{LmsminA}) is equivalent to the extremal surface defined by the
stationary point of the area functional. Clearly this argument of
equivalence is rather speculative; we leave a rigorous proof as an
interesting problem for the future.  To summarize, we have obtained the
following proposal:
 \paragraph{Covariant holographic entanglement entropy proposal (II):}

We claim that the holographic entanglement entropy for a region
$\rA$ is given by
\begin{equation}
S_\rA = {\area{\Lms_{ext}}  \over 4 \, G_N^{(d+1)} }, \Label{holeeT}
\end{equation}
where $\Lms_{ext}$ is a co-dimension two surface in $\bulk$ which
has zero null geodesic expansions,  \ie, both $\theta_\pm$ vanish on
$\Lms_{ext}$, and which satisfies  $\p\Lms_{ext} =\brA$. If this
surface is not unique, we choose the one whose area is minimum
 among all such surfaces homotopically equivalent to $\rA$.
 Also by virtue of \req{areaform} and the discussion of \App{apmin},
 we have $\Lms_{ext} = \Gms$. So we can just as well replace $\area{\Lms_{ext}}$ in \req{holeeT}
 by $\area{\Gms}$ without loss of generality.
 Henceforth we will drop the subscript `ext' on $\Lms_{ext}$ and
  simply denote the surface with vanishing null expansions by $\Lms$.

\section{Relations between covariant constructions}
\label{covother}

In the previous section we have motivated a covariant prescription
for calculating the holographic entanglement entropy using
light-sheets, in analogy with the covariant entropy bounds. This
construction involves finding a surface $\Lms_{ext}$ with vanishing null
expansions, which as we discuss, is equivalent to the extremal surface
$\Gms$.  We have also hitherto introduced another natural covariant surface: a
minimal surface on a maximal slice, $\Xms$.

In this section, after we show that the covariant proposal
(\ref{holeeT}) can indeed be also derived from the basic principle
of AdS/CFT, we will
  proceed to discuss the detailed relations between $\Gms(=\Lms_{ext})$ and
  $\Xms$.

\subsection{Equivalence of $\Gms$ and $\Lms$ via variational principles}
\label{gkpmotive}

In the time-dependent setup discussed in \sec{holtimedep}, we can
 directly apply the Lorentzian GKP-W relation  (see \eg,  \cite{Marolf:2004fy}) as
long as the UV limit of the theory becomes conformal. Assuming that
the boundary field theory is in a pure state, we have the
path-integral expressions for the reduced density matrix analogous
to the situation in \sec{tdqft}:
\be [\rho_A(t)]_{\ap
\beta}=\f{\int D\vp~ e^{iS_{sugra}(\vp)}~ \la \beta |\vp(t-\ep)\lb
\la \vp(t+\ep) | \ap \lb} {\int D\vp~
e^{iS_{sugra}(\vp)}} \ ,
\Label{GKPW}
\ee
 The boundary conditions (which will be implemented on a suitable cut-off surface) $\vp=\vp_{\pm}$ are the ones induced from the `indices'
$\vp_{\pm}$ of the density matrix $[\rho_A(t)]_{\vp_+ \vp_-}$. This
is a Lorentzian generalization of the argument in
\cite{Fursaev:2006ih}, where the proposal of \cite{Ryu:2006bv,
Ryu:2006ef} was first proven.

The CFT partition function $Z_n$ in \req{parex} is now
holographically equivalent to the partition function $Z^{sugra}_{n}$
of the supergravity on the dual manifold ${\cal M}_n$ which is
obtained by solving Einstein equations while requiring that it approaches
$\de{\cal{M}}_n$ at the boundary. Since the original
manifold $\de{\cal{M}}_n$ includes the singular surface $\de {\cal
A}$ with a negative deficit angle $2\pi(1-n)$, its holographical
extension ${\cal M}_n$ has the co-dimension two deficit angle surface
$\Gms$.
If we employ the tree level supergravity approximation, the action can be
estimated\footnote{The curvature is delta function localized
along the deficit angle surface. In actual computation, we estimate
this contribution by analytically continuing to the Euclidean
signature. This explains the imaginary factor $i$ in
\req{actiongkp}.} by
\be
\f{i}{16\pi\, G_N^{(d+1)}}\int_{\bulk_n}
\s{-g}\,(R+\Lambda)=\f{1-n}{4\, G_N^{(d+1)}}\int_{\Gms}\s{g}+(\mbox{irrelevant
terms}),
\Label{actiongkp}
\ee
where the irrelevant terms signify those
which cancel between the two terms in \req{entfor}.

In this way, after taking the derivative with respect to $n$ as in \req{entfor},
we obtain the holographic formula
\be
S_\rA=\f{\mbox{Area}(\Gms)}{4\,G^{(d+1)}_N} \ .
\Label{hf} \ee
Moreover, the action principle in the gravity theory instructs us to
single out the extremal surface $\Gms$ among infinitely many choices
of co-dimension two surfaces, that {\it a priori} could be the
extension into the bulk of the region $\rA$ satisfying the required
boundary conditions.   This completes the derivation of the
holographic formula \req{holeeT} of the entanglement entropy in
time-dependent backgrounds. Since a differential geometrical
analysis shows $\Gms = \Lms_{ext}$ (see \App{apmin}), the above
derivation may be viewed as a heuristic proof of our covariant
proposal.

\subsection{Equivalence of $\Xms$ and $\Lms$ on totally geodesic surfaces}
\label{extminmax}

In motivating the existence of a covariant formulation
of holographic entanglement entropy, we argued that one could in
principle choose a preferred slicing of the bulk corresponding to
the maximal area slices and then use the holographic entanglement
entropy proposal of \cite{Ryu:2006bv, Ryu:2006ef}. We will argue
that while the maximal surfaces $\Xms$ don't generically coincide
with $\Gms$ or $\Lms$, there is a
special case wherein this proposal for $\Xms$ is equivalent to the
covariant entanglement entropy proposal for $\Lms$ formulated in
terms of the light-sheets and the expansion along null geodesic
congruences.  The specific restriction on the maximal slices which turns out to
be relevant is the notion of ``totally geodesic submanifold".

To examine this issue, it is useful to recall a few geometric facts
related to
 foliation of spacetimes and extrinsic curvatures.  For a co-dimension one spacelike
 sub-manifold $\Sigma$ in $\bulk$, anchored at some time $t$ in $\bdy$, with
 $\tau^\mu \equiv \(\p_\tau\)^\mu$
being the unit timelike normal to $\Sigma$, we define the induced metric on  $\Sigma$:
\begin{equation}
\gamma_{\mu \nu} = g_{\mu \nu} + \tau_\mu \, \tau_\nu \ ,
\Label{sigmet}
\end{equation}
and extrinsic curvature:
\begin{equation}
K_{\mu \nu} = \gamma^\rho_{\ \mu}\, \gamma^\sigma_{\ \nu} \, \nabla_{\! \rho} \, \tau_\sigma \ .
\Label{sigext}
\end{equation}
Here and in the following, $\nabla_{\! \mu}$ will denote the covariant derivative with respect to the full bulk metric $g_{\mu\nu}$. Now, we can look for a minimal surface $\CS$ in $\Sigma$. For such a putative minimal surface $\CS$, let $s^\mu$ denote the unit spacelike normal to  $\CS$ lying within $\Sigma$, so that $s^\mu \, \tau_\mu =0$ everywhere. Then we can again define the induced metric on the surface $\CS$:
\begin{equation}
h_{\mu\nu} = \gamma_{\mu\nu} - s_\mu \, s_\nu \ ,
\Label{minind}
\end{equation}
and the extrinsic curvature of $\CS$ in $\Sigma$:
\begin{equation}
\pi_{\mu \nu} = h^\rho_{\ \mu} \, h^\sigma_{\ \nu} \, D_\rho\, s_\sigma \ ,
\Label{pidef}
\end{equation}
where $D_\mu$ is the covariant derivative with respect to the metric $\gamma_{\mu \nu}$ on $\Sigma$, which is related to the spacetime covariant derivative by projection from $\CM$:
\begin{equation}
D_\mu \, s_\nu = \gamma^\rho_{\ \mu} \, \gamma^\sigma_{\ \nu} \, \nabla_{\! \rho}\, s_\sigma \ .
\end{equation}
This implies that
\begin{equation}
\pi_{\mu \nu} = h^\rho_{\ \mu}\, h^\sigma_{\ \nu}\, \nabla_{\! \rho}\, s_\sigma \ .
\Label{pidef2}
\end{equation}

We now turn to the question of interest: assuming  that $\ms$ is a minimal surface on the particular slice $\Sigma$ corresponding to the maximal slice of $\CM$, under what conditions does the null geodesic expansion along $N^\mu \propto \tau^\mu \pm s^\mu$ vanish?

Since we assume that $\Sigma$ is a maximal slice, we necessarily have $K_\mu^{\ \mu} = 0$, \ie, the trace of the extrinsic curvature vanishes everywhere on $\Sigma$, which implies that %
\begin{equation}
\nabla_{\! \mu} \, \tau^\mu =0 \ .
\end{equation}
  Similarly, the constraint that $\ms$ is a minimal surface in $\Sigma$ implies that $\pi_\mu^{\ \mu}  =0$, which leads to the identity
\begin{equation}
\nabla_{\! \mu} \, s^\mu = s^\nu \, \tau^\mu \, \nabla_{\! \mu} \, \tau_\nu  \ .
\end{equation}
Having extracted the two relations implied by $\Sigma$ being the maximal slice in $\bulk$ and $\ms$ being the minimal surface in $\Sigma$, we now turn to evaluate the null expansions $\theta_{\pm}$.
Using  \req{thetadef} with $N_\pm^\mu \propto \tau^\mu \pm s^\mu$, we obtain:
\begin{equation}
\theta_\pm  \propto K_\mu^{\ \mu}\pm \pi_\mu^{\ \mu} - s^\nu \, s^\mu \, \nabla_{\! \mu}\, \tau_\nu =  \tau^\nu\, s^\mu\,\nabla_{\! \mu}\, s_\nu
\end{equation}
where the second equality used $K_\mu^{\ \mu}=0$, $\pi_\mu^{\ \mu}=0$ and $s^\mu\,\tau_\mu =0$.
So $\theta_\pm$ will vanish provided we
 have $ \tau^\nu\, s^\mu\,\nabla_{\! \mu}\, s_\nu =0$, which
is equivalent to the condition $K_{\mu\nu} \, s^\mu\,s^\nu =0$.

 Thus we see that for the null geodesic congruence to have vanishing expansion, it does not suffice for the surface $\Sigma$ to be a maximal slice. We must  in addition require that $K_{\mu\nu} \, s^\mu\,s^\nu =0$.
This is satisfied only\footnote{
Note that while $K_{\mu\nu} \, s^\mu\,s^\nu =0$ only picks out the symmetric part of $K_{\mu\nu}$, the antisymmetric part is automatically guaranteed to vanish whenever $\tau^{\mu}$ is hypersurface orthogonal, \ie\ $\tau_{[\mu} \, \nabla_{\! \nu} \, \tau_{\rho]} = 0$, which is the present case.
} when $K_{\mu\nu} =0$ since the vector $s^\mu$
can be taken to be arbitrary. Such a surface is called a totally
geodesic submanifold, and it describes a surface whose geodesics are also
geodesics of the entire spacetime. One can quickly intuit this by
noting that if $s^\mu$ were tangent to a geodesic then
$s^\mu\,\nabla_{\! \mu}\, s_\sigma \propto s_\sigma$, which by
virtue of $s^\mu\,\tau_\mu =0$ will imply the vanishing of
$\theta_\pm$. This leads to the following claim:

\noindent
{\bf Claim:} Assume that a maximal spacelike surface $\Sigma_t$
(anchored at a constant time $t$ on $\bdy$) is totally geodesic.
Then the minimal surface $\Xms$ on $\Sigma_t$ is equivalent to the
surface $\Lms_{ext}$ in the covariant entanglement entropy
prescription of \sec{covls}.
However, if we require $\Sigma_t$ to be totally geodesic for all $t$,
\ie\ if the spacetime $\bulk$ allows a totally geodesic foliation,
then the spacetime must be static.
This is because the condition $K_{\mu\nu}=0$ means that
the hypersurface orthogonal timelike vector
$\tau^\mu$ is in fact a Killing vector.
In this case, the covariant
construction reduces to the minimal surface
prescription \req{minTT}.

Hence we see that the covariant entanglement entropy candidate $\Xms$ reproduces the `correct' prescription $\Gms = \Lms$ for all time only in the trivial case of static bulk geometries.
However, if we relax the requirement of full foliation of $\bulk$ by totally geodesic slices,
 but rather achieve $K_{\mu\nu}=0$ on a single slice, say at $t=0$, then we still have\footnote{
In this case we can easily prove the strong subadditivity of
holographic entanglement entropy \cite{Hirata:2006jx} as in
\cite{Matt} since two minimal surfaces on the same time slice can
intersect with each other if they do so at the boundary of
AdS.} $\Xms_{t=0}=\Lms_{t=0}$. For
example, in a spacetime with time reversal symmetry
$t\leftrightarrow -t$, at time $t=0$ we can compute the entanglement
entropy by using the minimal surface $\Xms_{t=0}$ in $\Sigma_{t=0}$
(in this case the $t=0$ slice).

\section{Consistency checks for time-independent backgrounds}
\label{examples}

Thus far we have kept our discussion at a reasonably abstract level;
we have formulated a clear algorithm for constructing the bulk
surface whose area captures the entanglement entropy associated with
the boundary region $\rA$ in question. A simple consistency check of
our picture is that the covariant proposal should reduce to the
holographic entanglement entropy proposal of \cite{Ryu:2006bv,
Ryu:2006ef} whenever the bulk spacetime is static. To make contact
with that discussion, we examine several examples of asymptotically
AdS static spacetimes. This also allows us to see explicitly the
equivalence between the light-sheet construction $\Lms$ and the
extremal surface proposal $\Gms$, thereby making explicit the
arguments of \sec{lsmotive} and \sec{covls}. Finally we will turn to
an example of a stationary spacetime (rotating BTZ geometry) to
illustrate the shortcomings of the minimal surface on a maximal
slice prescription  $\Xms$ of \sec{holtimedep}.

\subsection{AdS$_{3}$}
\label{3adsex}
First consider the $AdS_3$ geometry described by the Poincar\'e metric
\be
ds^2=\f{-dt^2+dx^2+dz^2}{z^2} \ . \Label{adsth} \ee
We begin by studying the null expansions  for a co-dimension two surface by choosing a particular ansatz and compute the covariant holographic entanglement entropy.

\subsubsection{Expansions of null geodesics}

A general co-dimension two curve $\ms$ in \req{adsth} is described by the
constraint functions
\be \vp_1=t-G(z) \ , \qquad \vp_2=x-F(z) \ .
\Label{constcg}\ee
The normalized null vectors orthogonal to $\ms$ are then given by
the following linear combinations
\begin{equation}
(N_{\pm})^{\mu} = {\cal N} \,  g^{\mu\nu}\, (\nabla_{\! \nu} \,
\vp_1+\mu_{\pm}\nabla_{\! \nu} \, \vp_2) \ ,
\Label{expnull}
\end{equation}
where we defined
\begin{eqnarray}
\mu_{\pm} &=& -\f{G'F'}{1+(F')^2} \pm \f{\s{1+(F')^2-(G')^2}}{1+(F')^2} \ , \nonumber\\
{\cal N} &=& \f{\s{1+(F')^2}}{\s{2}\,z\s{1+(F')^2-(G')^2}} \ .
\Label{expmu}
\end{eqnarray}
As explained earlier, we will ignore the overall normalization
${\cal N}$ of the null vectors in most parts of this paper as we are
only interested in their signs.\footnote{If we rescale $N_+^\mu\to
\gamma \, N_+^\mu$ and $N_-^\mu\to \gamma^{-1}\, N_-^\mu$, the
normalization conditions $N_{+}^\mu\, N_{-\mu}=-1$ and
$N_{\pm}^\mu\, N_{\pm\, \mu}=0$ are unchanged. The geodesic
expansion  scales like the null vectors \ie, $\theta_+\to \gamma\,
\theta_+$ and $\theta_-\to \gamma^{-1}\, \theta_-$. Because we are
interested in the condition $\theta_{\pm} =0$, this non-zero scale
factor is inconsequential except some singular cases where apparent
horizons exist.} Moreover, the induced metric on $\ms$ is
given by\footnote{One can check that the three-metric \req{tindmetr}
is degenerate, as required. For purposes of computing the
expansions, it is more useful to work with this degenerate
three-metric rather than the one-metric on the curve, to ensure the
correct projections of $\nabla_{\! \mu} \, N_\nu$.}
\begin{equation}
h^\mu_{\ \nu}= \f{1}{1+(F')^2-(G')^2}  \, \left(\begin{array}{ccc}
  -(G')^2 & G'F' & G'\\
  -G'F' & (F')^2 & F' \\
   -G' & F' & 1
\end{array}\right)
\Label{tindmetr}
\end{equation}

To compute the expansions \req{thetadef} from $\ms$, we need to calculate the
covariant derivative of the null vectors \req{expnull} projected via \req{tindmetr}.
This yields the expression
\begin{eqnarray}
\theta_{\pm} &=& \f{\mp H\s{1+(F')^2-(G')^2}
-(G')^3+G'(1+(F')^2+zF'F'')-z(F')^2G''-zG''}{\s{2(1+(F')^2)}\, \;
(1+(F')^2-(G')^2)^{3/2}}, \no
\end{eqnarray}
where we defined
\begin{eqnarray}
H \equiv F'(G')^2-F'-(F')^3+z \, F''\ .
\end{eqnarray}

\subsubsection{Extremal surface and holographic entanglement entropy}

While we have written the expression for expansions for a general
curve $\ms$ in \ads{3} parameterized as \req{constcg}, by virtue of
time translation invariance, we expect that the desired extremal
surface (curve)  lies  on a constant $t$ slice. Let us therefore
concentrate on a curve $\ms_0$ in \req{adsth} with no temporal
variation, by requiring $G(z)=0$. Then the expansions for $\ms_0$
are simplified to
\be \theta_+=-\theta_-=\f{-z\, F''(z)+F'(z)+F'(z)^3} {\s{2}\,
(1+F'(z)^2)^{\f{3}{2}}} \ . \Label{nexth}
 \ee
Notice that the expansion in the time direction is vanishing, \ie,
$\theta_+ +\theta_-=0$, because the spacetime is static.

To find the covariant holographic entanglement entropy candidate
$\Lms$ to utilize our proposal \req{holeeT}, we require that both
null expansions vanish. This leads to the equation
\be z \, F''(z)-F'(z)-(F'(z))^3=0 \ , \ee
which determines the requisite surface.
We can easily find the following simple solutions:
\be F(z)=\s{h^2-z^2} \ ,
\Label{extadsth} \ee
where $h$ is an arbitrary non-negative constant. This means that the
half circle $x^2+z^2=h^2$ ($z>0$) is the curve $\Lms_{\rA}$
responsible for the entanglement entropy when we choose the
subsystem $\rA$ to be an interval with length $2h$.

As can be easily verified, this curve also describes
a spacelike geodesic in AdS$_3$ and likewise corresponds to the minimal surface on
the constant $t$ slice. This makes explicit the assertion made
earlier that minimal surfaces on a constant time slice in a static
spacetimes have vanishing null expansions.
Hence we have verified, for the AdS$_3$ example, that $\Gms_{\rA} = \Xms_{\rA} = \Lms_{\rA}$ for any region $\rA$, and given an explicit equation for this surface.  To compute the holographic dual of the entanglement entropy $S_{\rA}$ itself, we need to calculate the proper length along this bulk surface.

The length $L$ of $\Lms_{\rA}$ is found to be
\be L=2h\,  \int^h_{\cof}
\f{dz}{z\,\s{h^2-z^2}}=2\log\f{2h}{\cof}, \Label{lengthth} \ee
where $\cof$ is the lattice spacing corresponding to the UV cut-off. Using the relation
between the central charge of the dual CFT and the bulk Newton's constant
$c=\f{3}{2G^{(3)}_N}$ \cite{Brown:1986nw}, we obtain the expression
in the dual CFT language
\be S_{\rA}=\f{L}{4G^{(3)}_N}=\f{c}{3}\log
\f{2h}{\cof} \ .
\ee
This reproduces the well-known formula in 2D
conformal field theory \cite{Holzhey:1994we, Calabrese:2004eu}.

\subsubsection{Structure of the sign of expansions}
\label{stru}

The signs of expansions of null geodesics are directly related to
the change of the area of a given spacelike surface under an
infinitesimal deformation as the formula \req{areaform} shows.
In this subsection, we discuss how the signs of the
expansions change in explicit examples.

We start with the 3
dimensional flat spacetime $\R^{1,2}$: $ds^2=-dt^2+dx^2+dz^2$. If we
consider the curve $t=$ constant and $x=F(z)$, the null vectors
$N_\pm^\mu$ are given by the formula \req{expnull} with the
modification that ${\cal N}$ of \req{expmu} now becomes ${\cal N} =
\f{\s{1+F'^2}}{\s{2 \, (1+F'^2+G'^2)}}$. The expansions are found to
be
\be \theta_+=-\theta_-=-\f{F''(z)}{\s{2}\, \(1+F'(z)^2\)^{3/2}} \ .
\ee
In the particular case of the circle $x^2+z^2=h^2$, we find
$\theta_+=-\theta_-=\f{1}{\s{2} \, h}>0$ when
 $x\geq 0$. When $x$ is negative we obtain the opposite result.

{}From the formula \req{areaform} and the normalization $(N_+)_\mu
\, (N_-)^\mu=-1$, we find that $\theta_\pm$ measure the increase of
area under the infinitesimal deformations $\delta X^\mu \propto
-N^\mu_\mp$. The signs $\theta_{\hat{+}}\equiv -\theta_+<0$ and
$\theta_{\hat{-}}\equiv \theta_-<0$ for the circle can be intuited
directly as the null vector $N_{\hat{+}}=-N_+$ is ingoing along the
future light-cone and $N_{\hat{-}}=N_-$ is ingoing along the past
light-cone. We thus see that the past and future light-cones
emanating from the circle are examples of light-sheets as we have
explained in \req{expl} (see \fig{boussofig}). Note also that
$\theta_+ - \,\theta_->0$ means that the expansion in the spacelike
direction is positive, which illustrates  the basic fact that the
length of the circle increases as the radius $h$ becomes larger.

Now we move on to the more interesting case  \req{adsth} of AdS$_3$.
The null expansions for static curves are already computed in
\req{nexth}. On the curve defined by the ellipse $x^2+b^2\, z^2=h^2$
for a positive constant $b$, we find that when $x\geq 0$ (for $x<0$
the result has the signs reversed),
\be \theta_+=-\theta_-= \f{b^4 \, (1-b^2)\, z^3}{\s{2}\, \(h^2+b^2\,
(b^2-1)\, z^2\)^{3/2}}\ . \ee
Thus the expansions of ingoing null geodesics
$\theta_{\hat{+}}=-\theta_+$ and $\theta_{\hat{-}}=\theta_-$ are
negative when $b<1 $, \ie, when the curve goes deep into the IR
region, while it becomes positive when $b>1$. Furthermore, the
expansions in AdS$_3$ are vanishing when the curve is a half-circle,
which coincides with the minimal surface. Thus we conclude that the
null geodesic congruences on this ellipse can be used as
light-sheets only when $b\leq 1$.

\begin{figure}
\begin{center}
  \includegraphics[width=10cm]{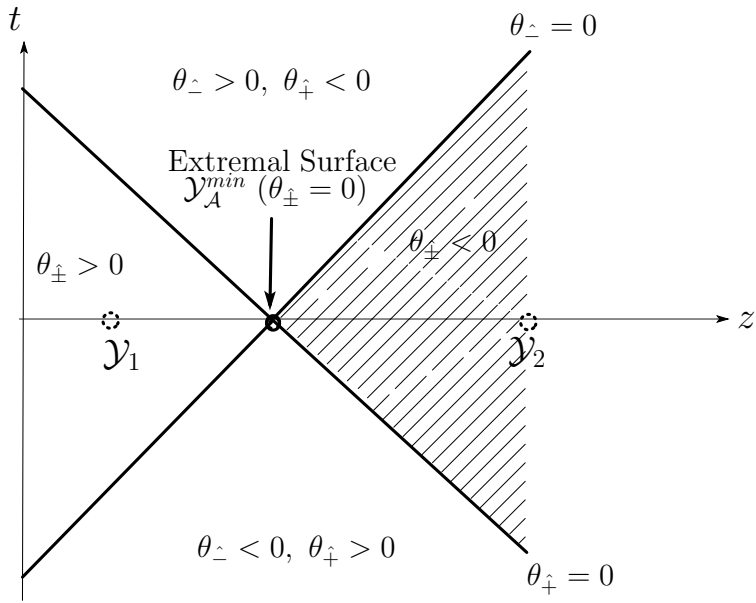}\\
  \caption{The signs of expansions $\theta_{\hat{\pm}}$
  for the ingoing null geodesics in \ads{3}.
  We projected the \ads{3} to the plane $x=0$ assuming a
particular series of curves whose
   null expansions each take the same sign at any points. The shaded region denotes the region where
   two light-sheets exist.} \label{lightsheetst}
\end{center}
\end{figure}

In the \ads{3} background we can notice one more interesting fact:
for any curve on the light-cone, one of the two null expansions is
vanishing, as will also be shown in \sec{Arex}. For example, if we
consider an arbitrary curve on the future light-cone
$t=-\s{x^2+z^2}$, it turns out that $\theta_{\hat{+}}=0$ when $x\geq
0$, while $\theta_{\hat{-}}=0$ when $x<0$. This property can be
easily generalized to higher dimensional AdS spaces. The behavior of
expansions $\theta_{\hat{\pm}}$ in $AdS_3$ is summarized in
\fig{lightsheetst}. 

In this way we observed that in (asymptotically) AdS spacetimes, the
expansions of null geodesics can change their sign at the specific
points in the bulk.  This property clearly plays a crucial
role in our holographic computation of entanglement entropy.

\subsection{Higher dimensional examples: AdS$_{d+1}$}
\label{highdex}

We can repeat the above computations of null expansions
for AdS$_{d+1}$ \req{poincarem}. In these higher dimensional examples,
there are many different choices for the shape of region $\rA$.
Working in Poincar\'e coordinates we can choose an arbitrary region
on the boundary $\R^{1,d-1}$ and in principle figure out the
associated extremal surfaces. For simplicity, we will concentrate on
two  specific examples, where we assume the subsystem $\rA$ in the
dual CFT is given by (i) an infinite strip  and (ii) a spherical
ball in $\R^{1,d-1}$.

\subsubsection{Infinite strip in AdS$_{d+1}$}
\label{inst}

On the boundary of \ads{d+1} in Poincar\'e coordinates we choose the region $\rA$ to be an infinite strip defined as
\begin{equation}
 \rA := \left\{ (t, \vec{ x }) \mid \ t = 0 , \ \  |x_1| \le h,\ \   x_i =
 {\rm arbitrary\; for}\;  i = 2, \ldots, d-1 \right\} \ .
\Label{stripadsd}
\end{equation}
 Here we have singled out one of the spatial coordinates in $\R^{1,d-1}$ called $x_1$ to take values in  finite range. To find the associated extremal surface in the bulk, we choose an ansatz for co-dimension two surface in AdS$_{d+1}$ by
the two constraints \req{constcg}, with a trivial relabeling $x_1 \to x$.  We require that when restricted to the boundary $z\to 0$, the extremal surface  is reduced to the boundary $\brA$ of infinite strip.

The null expansions of this surface can be shown to be  (here $\ti{d}=d-1$)
\begin{eqnarray}
\theta_\pm &=& \f{\mp H\,\s{1+(F')^2-(G')^2} -\ti{d}\,(G')^3+G'\,
(\ti{d} +\ti{d}\, (F')^2+z\, F'\, F'')-z\,(F')^2\,G''-z\,G''}{\s{2
\, \(1+(F')^2\)}\, \; \(1+(F')^2-(G')^2\)^{3/2}} \ , \no
\Label{ddimthetas}
\end{eqnarray}
where we define
\begin{eqnarray}
H=\ti{d}\,F'(G')^2-\ti{d}\,F'-\ti{d}\, (F')^3+z \,F''.
\end{eqnarray}

Again by virtue of the staticity of the background it suffices to consider
only $F(z) \neq 0$ while  $G(z) = 0$.
It is easy to see that the vanishing of both the  null expansions for the surface localized on a constant $t$ slice leads to the known minimal surface
\cite{Ryu:2006bv, Ryu:2006ef},
\be
F'(z) = {z^{\ti{d}} \over
\sqrt{z_*^{2\ti{d}} - z^{2\ti{d}}}} \ ,
\Label{stripms} \ee
where $z_*$ is the maximal $z$ value reached by the surface, given in terms of the width of the region $\rA$ by the relation
\be
z_*  = {\Gamma({1\over 2 \ti{d}} ) \over \sqrt{\pi} \, \Gamma({\ti{d}+1\over 2 \ti{d}} ) } \, h
\Label{stripmszs} \ . \ee
We can obtain the
entanglement entropy from the area of this surface. For details, we
refer the reader to \cite{Ryu:2006ef}.

\subsubsection{3-dimensional ball in AdS$_{5}$}

Our previous examples have focussed on planar symmetry and we now
turn to an example where the region $\rA$ of interest is a ball in
$\R^{d-1} \subset \R^{1,d-1}$ with radius $h$. The region $\rA$ is
given as (for simplicity we choose $t=0$)
\begin{equation}
\rA := \left\{ (t,\vec{x} ) \mid t = 0 , \xi^2 \le h^2 \right\} \ ,
\Label{balldef}
\end{equation}
where $\xi$ is the radial
coordinate of the Poincar\'e metric in the polar coordinates
\be
ds^2=\f{-dt^2+dz^2+d\xi^2+\xi^2 \, d\Omega_{3}^2}{z^2} \ . \ee
An ansatz for surfaces which respect the spherical symmetry is given by
 \be \vp_1=t-G(z),\ \ \ \
\vp_2=\xi-F(z)\ . \Label{constcgg} \ee
Further imposing the staticity inherited from the background
leads to the simplification\footnote{One can evaluate the
expansions  for non-zero $G(z)$ just as easily and check that the
surface given in \req{highdexts} does indeed have vanishing
expansions.} $G(z)=0$. For the particular case of \ads{5} the null
vectors normalized according to our usual convention are then given
by:
\begin{equation}
 N_{\pm}^\mu =\f{z}{\s{2}}\, \( \(\p_t\)^\mu \mp \f{F'}{\s{1+F'^2}}\, \(\p_z\)^\mu \pm
 \f{1}{\s{1+F'^2}} \, \(\p_\xi\)^\mu \) \ .
\end{equation}
One can check that  the induced metric on the surface is given by
\begin{equation}
h_{\mu\nu}=\left(\begin{array}{ccccc}
  0 & 0 & 0 & 0 &0\\
  0 & \f{F'^2}{1+F'^2} & \f{F'}{1+F'^2} & 0 & 0\\
  0 & \f{F'}{1+F'^2} & \f{1}{1+F'^2} & 0 & 0\\
  0 & 0 & 0 & \xi^2 & 0 \\
  0 & 0 & 0 & 0 & \xi^2 \, \sin^2\theta
\end{array}\right).
\Label{tindmetrcir}
\end{equation}
Plugging these expressions  into the formula for the null congruence expansions we find:
\ba \theta_{\pm} &=& \pm {1\over \sqrt{2}} \, {z^4   \over \xi \,
(1+F'(z)^2)^{7/2}} \, (- 9 \, F'(z)^5 \, \xi - 3 \, F'(z)^7\, \xi +
F'(z)^4\, \xi \, z \, F''(z) - 9 \,F'(z)^3\, \xi \no & &
 + 2 \, F'(z)^2 \, \xi \, z \, F''(z) - 3 \,F'(z)\,
  \xi  + \xi \,z \,F'(z) -2 \,z - 6\, z \,F'(z)^2 -6 \,z\, F'(z)^4 - 2
  \,z\,F'(z)^6) \ . \no\ea
One can check these null expansions vanish for the minimal surface
\be F(z) = \sqrt{h^2-z^2}\ . \Label{highdexts}\ee
The entanglement entropy  associated with the region $\rA$ of \req{balldef} can be calculated from the area of this surface. As expected, the surface  \req{highdexts} coincides with the minimal surface of \cite{Ryu:2006ef}. We refer the interested reader to \cite{Ryu:2006ef} for a detailed discussion of the area and comparisons of the holographic entanglement entropy thus obtained to the field theory calculations at weak coupling.

\subsubsection{Area and expansion of surfaces on the light-cone}
\label{Arex}

In \sec{nexpansion} we presented the relation between
the change in the  area of a spacelike surface under a small deformation
and the expansions of the null geodesics. Here we would like to
understand this relation geometrically in the specific example of
AdS$_{d+1}$.

Consider the set-up of the infinite strip region on the boundary as
in \sec{inst} and take a surface which infinitely extends in the
directions  $x^2,x^3,\cdots,x^{d-1}$. Such surfaces can be
described by the ansatz \req{constcg}, with $x\to x_1$. We would
like to  concentrate on the case where the surfaces lie on the
light-cone\footnote{The light-cone in question is the flat space
light-cone by virtue of the Poincar\'e metric \req{poincarem} being
conformally flat.} $t^2=x_1^2+z^2$. These can be  parameterized as
\begin{equation}
x_1= p(s)\, \cos(s)\ , \qquad  z=p(s)\, \sin(s) \ , \qquad t(s)
=h-p(s) , \Label{Infstppar}
\end{equation}
with the boundary condition $p(0)=p(\pi)=h$. 

The area of any of these surfaces given by a particular choice of $p(s)$ is expressed as
\begin{equation}
\area{\Lms} =\int^{\pi-\epsilon_2} _{\epsilon_1}\,
\f{ds}{\sin^{d-1}(s)\, p(s)^{d-2}} \ . \Label{bennst}
\end{equation}
where the boundary condition on the cut-off surface $z=\cof$ is being
implemented through the boundary condition $z(s = \eps_{1,2}) = \cof$.

When $d=2$, the expression \req{bennst} does not depend on the
function $p(s)$ which represents the choice of the curve. This means
that the deformation of any curve on a light-cone in AdS$_3$ does
not change its area (as long as we neglect the UV cut-off). This
nicely agrees with the fact that the expansion along the light-sheet
is vanishing for any curve on it, as mentioned in \sec{stru}. If we
consider the opposite light-cone $t+h=\s{x_1^2+z^2}$, we can find
that on the half circle defined by $x_1^2+z^2=h^2$, $t=0$ and $z>0$,
the null expansions are both vanishing, \ie\ this is an extremal
surface as we noticed in \req{extadsth}. 

On the other hand, in higher dimensions $d>2$, the area becomes
dependent on $p(s)$. Furthermore, we can see the inequality
$\area{\Lms_1} > \area{\Lms_\rA} > \area{\Lms_2}$ where the surfaces are labeled in accord with the conventions of \fig{lightsheets}. This in particular shows that the ingoing expansion along this light-cone $t=-\s{x_1^2+z^2}$ is positive. Thus we can conclude that we cannot regard the light-cone \req{Infstppar} as a light-sheet in $d>2$. This fact can also be confirmed by direct evaluation of the
expansions using  \req{ddimthetas}.

\subsection{BTZ black hole (non-rotating)}
\label{btzbhex}
Our next example will be one which is
not globally static, but one which has a horizon and a static patch
extending out to the boundary.  Consider the BTZ black hole, with a
mass proportional to $m$, in the Poincar\'e coordinates
\cite{Banados:1992wn}, \cite{Aharony:1999ti}
\be
ds^2=-(r^2-m)\, dt^2+\f{dr^2}{(r^2-m)}+r^2\, dx^2 \ .
\Label{btzmet}
\ee
We will pick the region $\rA$ on the boundary $\R^{1,1}$ with coordinates $(t,x)$ to be at a constant $t$ slice and a finite interval in $x$ with $|x| \le h$. One can again take as an ansatz for the extremal surface \req{constcg} and compute the expansions to derive the differential equations for the functions $G(z)$ and $F(z)$. It is however simpler to exploit the fact that the extremal surfaces in \ads{3} are spacelike geodesics on a constant $t$ slice and find the relevant surface  directly.

\begin{figure}[htbp]
\begin{center}
\includegraphics[width=6.5in]{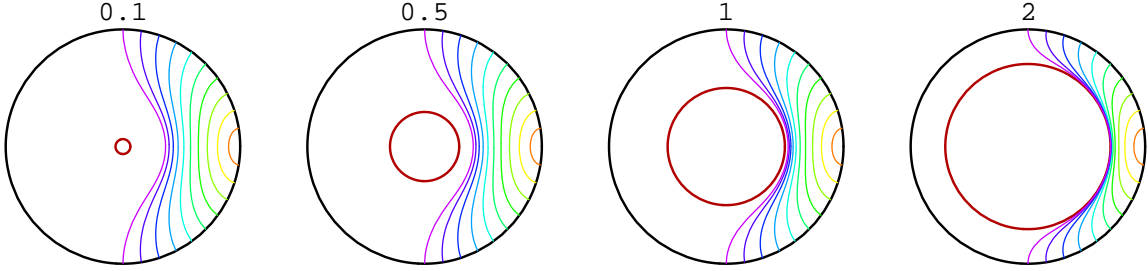}
\caption{Minimal surface in BTZ (in this 3-d case a geodesic)
plotted on $r -x$ slice of the bulk; the radial
 coordinate $r$ is compactified using $\tan^{-1}$ function,
 the thick outer circle represents the global AdS boundary,
 and the thick (red) inner circle the horizon radius,
  $\s{m} = 0.1,0.5,1,2$, as labeled.} \label{mpBTZgeodsA}
\end{center}
\end{figure}

Therefore we would like to find the spacelike geodesics of the form $t=$ constant and $r=r(x)$ in order to calculate the entanglement entropy. The conservation equation resulting from the the fact that $\p_x$ is a Killing field leads to a
constant Hamiltonian:
\be \f{dr}{dx}=r\s{\(r^2-m\)\, \(\f{r^2}{r^2_{*}}-1\)} \ .\ee
where $r_*$ is determined by the fact $|x| \le h$:
\be
2\, h=\int^\infty_{r_*}
\f{dr}{r\s{(r^2-m)(r^2/r^2_{*}-1)}}=\f{1}{\s{m}}\log\f{r_*+\s{m}}{r_*-\s{m}} \ .
\ee
For future  use we also record the exact relation between $x$ and $r$
\ba x&=&-\f{1}{2\s{m}}\log\left(\f{-2r_*\s{m(r^2-m)(r^2-r_*^2)}
-2mr_*^2+r^2r_*^2+mr^2}{r^2(r^2_*-m)}\right)\no
&=&\f{1}{2\s{m}}\log\left(\f{r_*+\s{m}}{r_*-\s{m}}\right)-\f{r_*}{2r^2}+\cdots \ea
The spacelike geodesics in BTZ for compact $x$ are plotted on
constant $t$ slices in \fig{mpBTZgeodsA} for various values of $m$.

Finally, the length $L$ of the geodesics in the BTZ spacetime is given as
 \ba
L&=&2\int^{r_{\infty}}_{r_*}\f{r\, dr}{r_*\, \s{(r^2-m)\, (r^2/r^2_{*}-1)}}
\no
&=&2 \, \log(2\, r_\infty)
-\log(r^2_*-m) =2\log(2\, r_\infty)+\log\f{\sinh^2(\s{m}\, h)}{m} \ .
\Label{entv} \ea
where we introduced the UV cut-off at $r=r_{\infty}$. This is related to the lattice spacing defined in \req{lengthth} via $r_{\infty}=\f{1}{\cof}$.

For large $m$ we find that the regularized length of the geodesic is given by
\be
L_{reg}=  L-2\, \log(2\, r_\infty)\simeq
2\, \s{m}\, h \ ,\ee
which can be interpreted as the  length of a part of the
horizon.\footnote{When $m$ is very small, we find $L_{reg}\sim
\f{ml^2}{12}+\log\f{l^2}{4}$.}

Using the relation between the mass and the inverse
temperature $\beta=\f{2\pi}{\s{m}}$
\cite{Banados:1992wn,Aharony:1999ti},
we finally obtain the entanglement entropy
computed holographically \cite{Ryu:2006bv, Ryu:2006ef} from the BTZ
black hole:
 \be S_\rA =\f{L}{4\, G^{(3)}_N}=
\f{c}{3}\, \log\left(\f{\beta}{\pi\, \cof}\sinh\f{2\, \pi\,  h}{\beta}\right) \ ,
\Label{finitet} \ee
where $c$ is again the central charge of the
dual 2D CFT. The result \req{finitet} agrees perfectly with the
known result in the 2D CFT at finite temperature \cite{Calabrese:2004eu}.

\subsection{Star in AdS$_5$}

Our final example of a static spacetime is the
5-dimensional AdS radiation star background\footnote{These
considerations can be generalized of course to any static,
spherically symmetric, asymptotically AdS spacetime.} considered in
\cite{Hubeny:2006yu},
\be ds^2=-f(r)\, dt^2+h(r)\, dr^2+r^2\, d\Omega_3^2\ ,
\Label{metstar}
\ee
where the function $h(r)$ is given in terms of the mass $M(r)$ of
the star within radius $r$ by
\be h(r)=\left[r^2+1-\f{8\,
G^{(5)}_N}{3\,\pi}\f{M(r)}{r^2}\right]^{-1} \ ,
\Label{hstar}
\ee
and the mass density $\rho(r)$, defined by $T_{tt}=\rho(r)\,f(r)$, is related to the mass function by $M(r) \propto \int_0^r \rho(\rb) \, \rb^3 \, d \rb$.  (For further details, see \cite{Hubeny:2006yu}.)

We consider the entanglement entropy defined by dividing the $\Sp^3$
into two hemispheres $\rA$ and $\rB$. The minimal surface for
$S_{\rA}$ is clearly given by the largest two-sphere $\brA$ times
the radial direction $r$. Thus its area is given by
\be \mbox{Area}=4\, \pi\,
\int^\infty_0 dr \, r^2\, \s{h(r)} \ .
\Label{estar} \ee
 We are interested in  the difference $\Delta S_{\rA}$ between the entanglement entropy for the region $\rA$ in the star geometry \req{metstar},\req{hstar} and in pure \ads{5}.
 This difference will capture the excess entanglement by virtue of the state of the boundary theory being an excited state of the CFT, and in a sense provide a measure of how many degrees of freedom are excited (and entangled) in the region in question. One can check that $\Delta S_{\rA}$ is finite and positive; the finite increase of the entanglement entropy clearly represents the degrees of freedom of the matter which composes the star.

When $M(r)$ is very small we  approximate the increase in entanglement entropy  (measured with respect to pure AdS or the CFT vacuum) by
\be \Delta S_{\rA}=\f{\Delta \mbox{Area}}{4\, G^{(5)}_N}\simeq
\f{4}{3}\, \int^\infty_0 dr\,
\f{M(r)}{\left(1+ r^2\right)^{\f{3}{2}}}>0 \ . \ee
%

\subsection{Stationary spacetimes: the rotating BTZ geometry}

Our final example of a spacetime with a timelike Killing field
(outside ergo-regions) will be a rotating black hole spacetime. We
will use this example to illustrate the inadequacy of the min-max
proposal of \sec{holtimedep}, providing a more robust confirmation
of our light-sheet construction discussed in \sec{covls}.

\subsubsection{The holographic computation of entanglement entropy}
We consider 3 dimensional Kerr-AdS solution (\ie, rotating BTZ black hole) and
compute the holographic entanglement entropy for a finite interval on the boundary.
In this example, as we will see one can no longer assume any constant time slice on
which the extremal curve lives.

The metric is given by
\be
ds^2=-\f{(r^2-r_{+}^2)\,(r^2-r_{-}^2)}{r^2}dt^2+\f{r^2}{(r^2-r_{+}^2)\,(r^2-r_{-}^2)}dr^2
+r^2\, \left(dx+\f{r_+r_-}{r^2}dt \right)^2,
\Label{btzrot}
\ee
where the coordinate $x$ is compactified as $x\sim x+l$ and we
assume $r_{+}\geq r_-$. The mass $M$ and angular momentum $J$ of
this black hole becomes
\be 8\,G^{(3)} M= r_+^2+r_-^2 \ , \qquad J=\f{r_+r_-}{4\, G^{(3)} }
\ .   \ee
If we set $r_-=0$, then the angular momentum becomes zero and the
black hole (\ref{btzrot}) becomes identical the static example
\req{btzmet} discussed in \sec{btzbhex} by setting
$m=r_+^2$.

This rotating black hole background (\ref{btzrot}) is dual to a 1+1
dimensional CFT on a circle at finite temperature $\beta^{-1}$ with
a potential $\Omega$ for the momentum. The radius of the circle is
defined to be $l$ and we assume that the system is at a very high
temperature ($\beta \ll l$). The potential $\Omega$ is conjugate to
the angular momentum of the rotating black hole.

The temperature and the potential in the dual CFT  are found from
the relations
\be
\beta_\pm \equiv \beta\, (1\pm \Omega) = \f{2\,\pi\,  l}{\Dr_\pm} \ , \qquad
\Dr_\pm \equiv r_+ \pm r_-\ .
\Label{pote} \ee
The dual CFT is then described by the density matrix
\be \rho=e^{-\beta \,H +\beta\, \Omega \, P } \ ,
\Label{adenst} \ee
where $H$ and $P$ are the Hamiltonian and the momentum of the CFT. Equivalently we can
regard $\beta_\pm=\beta\,(1\pm \Omega)$ as the inverse temperatures for the left and
right-moving modes.

To obtain the geodesics explicitly, it is convenient to  remember that all BTZ
black holes are locally equivalent to the pure AdS$_3$.  Explicitly, this map is given
by (\cf, \cite{Carlip:1994gc})
\ba && w_{\pm}=\s{\f{r^2-r_+^2}{r^2-r_-^2}}\;\; e^{ \(x\pm t\)\,\Dr_\pm }\equiv X\pm T, \no &&
z=\s{\f{r_+^2-r_-^2}{r^2-r_-^2}}\; e^{x\,r_{+}+ t\,r_{-}  } \ .
\Label{cordw} \ea
This maps the metric \req{btzrot} to the Poincar\'e metric
\be ds^2= \f{dw_{+}dw_{-}+dz^2}{z^2} \ .
\Label{poinc}\ee
We know that the spacelike geodesics in pure \ads{3}
\req{poinc} are given by the half circles of the form $(X-X_{*})^2+z^2=h^2$ on a
constant $T$ slice and their boosts $w_{\pm}\to \gamma^{\pm 1 }\,w_{\pm}$. Indeed,
by mapping these geodesics in pure AdS$_3$ into the rotating black hole, we can
obtain the  relevant extremal surface. Note that despite the spacetime being just
stationary, the extremal surface $\Gms = \Lms_{ext}$ is indeed given by spacelike geodesics.

Thus we can assume that a series of spacelike geodesics in  AdS$_3$
are all situated on some spacelike hypersurface
\be
\gamma \, w_{+}-\gamma^{-1}\,  w_{-}=\mbox{const.}
\Label{linee}\ee
 Since we are
considering the subsystem $\rA$ which is an interval at a fixed time
$t_0$, the value of $t$ should be the same at the two endpoints of the geodesic. If we define the value of $x$ at the endpoints by $x_1$ and $x_2$, this requirement leads to the constraint
\be \gamma^2 \,
e^{\(x_1+t_0\) \Dr_+  }-e^{\(x_1-t_0\) \Dr_- }
=\gamma^2\,
e^{\(x_2+t_0\) \Dr_+ }-e^{\(x_2-t_0\)\Dr_- } \ .
\Label{gamrel} \ee

The geodesic length in  AdS$_3$ (\ref{poinc}) leads to the
holographic entanglement entropy $S_{\cal A}=\f{c}{3}\log \f{\Delta
x }{\cof}$ when the length of the interval $\rA$ on the boundary is $\Delta x$
as we have seen in \sec{3adsex}.
The UV cut-off $z=\cof$ is mapped to the cut-off in the
background \req{btzrot} via
\be
\cof_{1,2}=
\f{\s{r_{+}^2-r_{-}^2}}{r_{\infty}}\, e^{r_{+}\, x_{1,2}+
r_{-}\,  t_0}\ , \ee
 where $\cof_{1,2}$ denote the cut-off at each of
the two endpoints in \req{poinc}. Further, the UV cut-off $r_{\infty}$ in
\req{btzrot} can be identified with the cut-off (\ie, the lattice
spacing) $\cof$ in the dual CFT via $  r_{\infty}=1/\cof$.
The length of the interval $\D x$  is easily found to be
\be (\Delta
x)^2=\Delta w_{+}\, \Delta
w_{-}=\left(e^{\Dr_+(x_1+t_0)}-e^{\Dr_+ (x_2+t_0)}\right)\,
\left(e^{\Dr_- (x_1-t_0)}-e^{\Dr_- (x_2-t_0)}\right).
\ee

Putting these together we obtain the holographic entanglement
entropy in the rotating BTZ geometry to be
\ba S_{{\cal A}}&=&\f{c}{6}\, \log\f{(\Delta
x)^2}{\cof_1\, \cof_2} \no
&=&
\f{c}{6}\, \log \left[\f{\beta_+\,\beta_-}{\pi^2\, \cof^2}\sinh\left(\f{\pi
\, \Delta l}{\beta_+}\right)\, \sinh\left(\f{\pi \, \Delta
l}{\beta_-}\right) \right]\ ,
\Label{rotent} \ea
where $\Delta l=(x_1-x_2)$ is the length of the interval in the dual CFT. The final answer is manifestly time-independent as required. Further, if we set
$\Omega=0$, then the above result reduces to the non-rotating BTZ answer \req{finitet}.

\subsubsection{CFT and  left-right asymmetric ensembles}

We would like to compare the holographic result \req{rotent} with
the entanglement entropy calculated directly from two dimensional
CFT in the ensemble \req{adenst}. This can be done by exploiting the
fact that the value $\Tr{\rho_\rA^n}$ for the reduced density matrix
$\rho_\rA$ for the subsystem $\rA$ is equal to the two point
function of twist operators whose conformal dimension is
$\Delta_{n}=\f{c}{24}(n-\f{1}{n})$ as shown in
\cite{Calabrese:2004eu}.

For a  CFT  defined on a 2 dimensional non-compact plane (Euclidean) and a region $\rA$ whose boundaries are at  $u_1$ and $u_2$, one can show that
\be
\Tr{\rho^n_\rA} =\left(\f{|u_1-u_2|}{\cof}\right)^{-\f{c}{6}\,(n-\f{1}{n})}\ , \ee
where $\cof$ is the UV cut-off in the CFT. This leads to the well-known formula
of the entanglement entropy at zero temperature
\be S_\rA =-\f{\p}{\p n}\, \log \Tr{\rho^n_\rA}\biggr|_{n=1}=\f{c}{3}\, \log
\f{|u_1-u_2|}{\cof} \ .
\Label{entfl} \ee

To derive the result at finite $\beta$ and $\Omega$ described by \req{adenst}, we need to periodically identify the (Euclidean) two dimensional manifold on which the CFT is defined. The total partition function of this system is given by
\be
Z_1=\Tr{ e^{-\beta\, H+i\, \beta\,\Omega_E \, P }}\ ,
\Label{pattw} \ee
 where we defined $\Omega_E=-i\,\Omega$. For the Euclidean CFT we will take $\Omega_E$ to be real as is conventional. This is achieved by the following conformal map
\be
w'=\f{\beta\, (1-i\,\Omega_E)}{2\pi}\, \log w\ . \ee
Notice that the new coordinate $w'$ satisfies the periodicity $w'\sim w'+i\,\beta\,
(1-i\Omega_E)$, in agreement with \req{pattw}.  Performing the conformal transformation,  we find
\be
\Tr{\rho^n_\rA} =\left[\f{\beta^2(1+\Omega^2_E)}{\pi^2\, \cof^2}\, \sinh\left(\f{\pi\, \D l}{\beta\, (1+i\,\Omega_{E})}\right)
\, \sinh\left(\f{\pi\,\D l}{\beta\, (1-i\, \Omega_E)}\right)\right]^{-\f{c}{12}\,(n-\f{1}{n})} \ ,
\Label{cftnpart}
\ee
 where we have set $\Delta l=\f{\beta(1-i\Omega_E)}{2\pi}\log
\f{u_{1}}{u_2}$, which is the length of the interval $\rA$ in
the $w'$ coordinate. After differentiating with respect to $n$ as
in (\ref{entfl}) and remembering the relation $\Omega_E=-i\Omega$,
this precisely agrees with \req{rotent}.
It is also intriguing to notice that the expression factorizes into the left and right moving contributions: $S_\rA =S^{L}_\rA+S^{R}_\rA$, suggesting a left-right decoupling in the two dimensional CFT.

\subsubsection{Comments on the min-max construction}
The prime reason for focusing on the rotating BTZ geometry is that
it clarifies some of the arguments regarding the min-max proposal
and the associated surface $\Xms$. While we motivated the existence
of a covariant construction using $\Xms$, a minimal surface on a
maximal slice, in \sec{holtimedep}, we subsequently argued that this
prescription doesn't agree with the light-sheet construction of
\sec{covls}. In fact, we claimed in \sec{extminmax} that the
surfaces $\Xms$ and $\Gms$(=$\Lms_{Ext})$ generically agree only
when the spacetime admits a totally geodesic foliation.

The rotating BTZ black hole has a Killing field $(\p_t)^\mu$ which is timelike outside the ergo-regions, but is not hyper-surface orthogonal.\footnote{A necessary and sufficient condition for  a vector field $\xi^\mu$ to be hypersurface orthogonal is $\xi_{[\mu}\nabla_{\!\nu}\,\xi_{\rho]} =0$. It is easy to check that $\(\p_t\)^\mu$ doesn't satisfy this condition in the metric \req{btzrot}.} As a result, while it is true that surfaces of constant $t$ are maximal, \ie, have $K^{\mu}_{\ \mu}= 0$, they do not contain the extremal surface $\Gms$. This is also clear from the fact that constant $t$ surfaces are not everywhere spacelike. From our explicit construction of the geodesic \req{cordw} and \req{linee} it is apparent that the geodesic moves in $t$ despite being pinned on the boundary at $t =t_0$ at both ends of the interval $\rA$.

By an explicit CFT computation we have confirmed that the covariant holographic entanglement entropy obtained from the surface $\Gms$ is indeed the correct one.
While {\it a priori} it was plausible that the surface $\Xms$ provided the covariant
generalization of the holographic entanglement entropy prescription, this example makes it manifest that light-sheets or extremal surfaces are crucial to capture the correct measure of entanglement. This example should therefore be viewed as a strong support for our covariant proposal.

\section{Entanglement entropy and time-dependence}
\label{timedep}
One of the motivations behind covariantizing the holographic
entanglement entropy proposal was to be able to address the question
of entanglement entropy in genuine time-dependent states. We will
now turn to applying our proposal to geometries with explicit
time-dependence. By virtue of the AdS/CFT duality these spacetimes
will correspond to states in the CFT with non-trivial time
evolution.  However, we do not always have an explicit CFT
description of the state in question. While this hinders direct
comparison of the results on time variation of the entanglement
entropy from the geometric perspective with field theory, it
nevertheless provides an interesting qualitative picture (which
could be made quantitative once the dictionary between states in the
field theory and geometry becomes more explicit).

\subsection{Vaidya-AdS spacetimes}
\label{adsvaidya}
One of the most important examples in time-dependent gravitational
backgrounds will be the black hole formation process via a collapse
of some massive object. As a simplest such example, we would like
to study the Vaidya background which describes the time-dependent
process of a collapse of an idealized radiating star (\cf, \cite{Stephani:2003tm}).
The metric of $d+1$ dimensional Vaidya-AdS spacetime is given in Poincar\'e coordinates as
\be
ds^2=-\left(r^2-\f{m(v)}{r^{d-2}}\right)\, dv^2+2\, dv\,
dr+r^2\, \sum_{i=1}^{d-1}\, dx_i^2\ , \ee
and in global coordinates by
\be
ds^2=-\left(r^2+1-\f{m(v)}{r^{d-2}}\right)\, dv^2+2\, dv \, dr+r^2\,
d\Omega_{d-1}^2 \ . \ee
If we assume that the function $m(v)$ does not depend on the
(light-cone) time $v$, then the background is exactly the same as
the Schwarzschild-AdS black hole solution after a coordinate
transformation. In this sense the Vaidya metric is a simple example
of black hole with a time-dependent mass or temperature.

The property of null geodesics in AdS Vaidya background has been studied
in \cite{Hubeny:2006yu} from the view point of AdS/CFT
correspondence. The authors were interested in using the geodesics to compute singularities of boundary correlation functions. It was argued that the geodesic structure (which clearly probes the spacetime geometry)  can be read off from the correlation function and thus a map was provided between geometric information in the spacetime and the natural observables of the field theory. In particular, it was shown how the field theory correlation functions could be used to ascertain the formation of a horizon in the bulk spacetime.

Given that null geodesics can be used to decipher the map between field theory observables and geometry, a natural question is whether there is some more information to be gained from studying other geometric structures -- spacelike geodesics or surfaces.  We expect this to be generally the case, because in certain cases, such as in spacetimes with null circular orbits, spacelike geodesics probe more easily further into the bulk than null geodesics.  Furthermore, null geodesics are manifestly insensitive to conformal rescaling of the spacetime, which is not the case for the spacelike ones.
Motivated by these ideas we wish to ask whether the entanglement entropy of the boundary theory can be used as a non-local probe of the bulk geometry.

Hence in the following we wish to calculate a time-dependent entanglement entropy in
the Vaidya-AdS background. We will specifically focus on the 3-dimensional Vaidya-AdS  metric
\be  ds^2=-f(r,v)\, dv^2+2\, dv\,  dr+r^2 \, dx^2 \ ,\ \ \ \
f(r,v)\equiv r^2-m(v),
 \Label{vaidyam} \ee
for simplicity. The coordinate $x$ can be either non-compact (Poincar\'e coordinate) or compact (global coordinate). When $m(v)$ is a constant $m$, this
background is same as the BTZ black hole \req{btzmet}, which can be confirmed using the coordinate transformation
\be
v=t+\f{1}{2\s{m}}\, \log\left(\f{r-\s{m}}{r+\s{m}}\right)\simeq
t-\f{1}{r}-\f{m}{3\, r^3}+\cdots \ ,\ee
where we have also recorded the large $r$ expansion for future use. In the metric \req{vaidyam}, the only  non-zero component of the energy-momentum tensor (defined by the Einstein's equation $T_{\mu \nu}= R_{\mu \nu} -\frac{1}{2} \, R \, g_{\mu\nu} + \Lambda \, g_{\mu \nu}$) is
\be T_{vv}=\f{1}{2\,r}\f{dm(v)}{dv}\ .
\Label{curvav} \ee
By imposing the null energy condition \ie, $T_{\mu\nu}N^\mu
N^\nu\geq 0$ for any null vector $N^\mu$, we find that the time-dependent mass $m(v)$
always increases as the time $v$ evolves
\be \f{dm(v)}{dv}\geq
0 \ .\Label{nullenergy}\ee
Below we would like to see how the entanglement entropy computed
holographically changes under this time-evolution.

\subsection{Extremal surface in Vaidya-AdS}
In order to compute the holographic entanglement entropy, we need to
find the minimal surface and then compute its area. The advantage of
our example of the 3 dimensional Vaidya-AdS  spacetime is that the
minimal surface is the same as the spacelike geodesic. We can
express the general geodesic by using (the non-affine)
parameterization
 \be \vp_1=r-r(x)=0,\qquad  \vp_2=v-v(x)=0 \ .
\Label{gline} \ee

We define the subsystem ${\cal A}_v$ at time $v$ by the region
$-h\leq x\leq h$ so that it always has the width $2h$. In the dual
gravity side, this leads to the following boundary condition along the
geodesic:
\be r(h)=r(-h)=r_{\infty}\ ,\qquad v(h)=v(-h)=v \ ,
\Label{boundconv} \ee
where $r_{\infty}\to\infty $ is the UV cut-off which is inversely related to the lattice spacing $\cof$ \ie, $r_{\infty}=1/\cof$. Note that we can require $r(x)=r(-x)$ and $v(x)=v(-x)$ due to the reflection symmetry of the background.

We would like to calculate the length $L$ of this geodesic
\be
L=\int^{h}_{-h} dx \,\s{r^2+2\,r'\,v'-f(r,v)\, v'^2}
\ ,
\Label{lengthv} \ee
where the derivative with respect to $x$ is denoted by the prime $'$. This length functional being independent of $x$, we have a conserved quantity
\be \f{r^4}{r_*^2}=r^2+2\, r'\, v'-f(r,v)\, v'^2\ , \Label{ham}\ee
where $r_*$ is a constant.
In addition, we get two equations of motion for $r$ and $v$  from
the action principle. As usual, only one of them is independent of
the previous conservation equation \req{ham}. It is given by
\be
r^2-r^2\, (v')^2-r\, v''+2\, v'\, r'=0\ .  \Label{eomvat} \ee

Thus we have to solve these ODEs (\ref{ham}) and (\ref{eomvat}) in
order to find the geodesics. For a generic $m(v)$, it is unfortunately not easy to
find an analytical solution. To obtain an explicit example, we
performed a numerical analysis in the specific case smoothly interpolating between pure AdS and BTZ,
\be f(r,v) = r^2 - {m_0+1 \over 2} \, \tanh{v \over v_s} - {m_0-1
\over 2} \ . \Label{Vadf} \ee
Roughly speaking, this corresponds to a null shell of characteristic thickness
 $v_s$ collapsing to form a BTZ black hole of mass $m_0$ at time $v=0$.
We can numerically integrate to find the spacelike geodesics in this
geometry.  For definiteness, for the result shown below, we chose
$v_s = 1$ and $m_0=1$.   (Note that in 3 dimensions, unlike in the
higher dimensional analogs, the horizon starts at finite $v$; in
this case $v=0$.)
\begin{figure}[htbp]
\begin{center}
\includegraphics[width=4.5in]{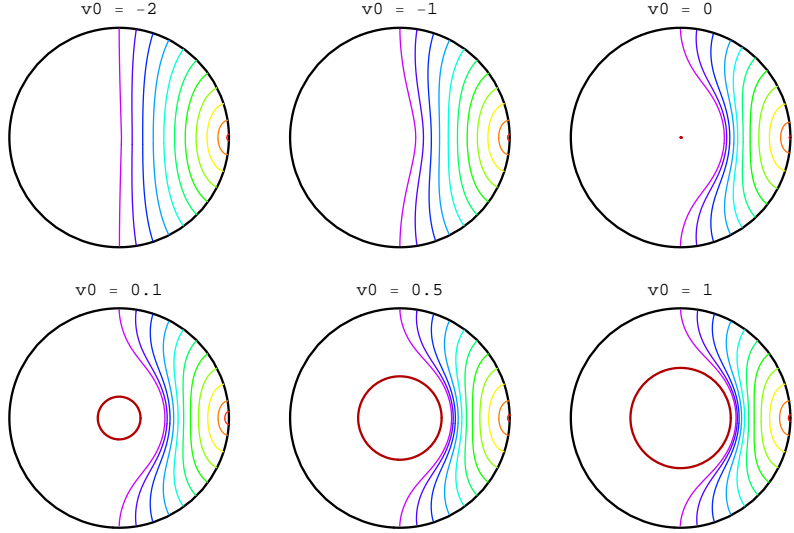}
\caption{Minimal surface in Vaidya-AdS (in this 3-d case a geodesic)
projected onto $r -  x$ slice of the bulk ($x$ is compact);
the radial coordinate $r$ is compactified using $\tan^{-1}$
function, the thick outer circle represents the global AdS boundary,
and the thick (red) inner circle the horizon radius at the value of
$v = v_0$ reached by the geodesic at minimum radius, as labeled.}
\label{mpvaidyageodsA}
\end{center}
\end{figure}
\fig{mpvaidyageodsA} shows several plots (snapshots for different times
$v(\rmin) \equiv v_0$ as labeled) of a series of extremal surfaces.\footnote{The surfaces will now vary in time as well; here we show just the
$r -x$ behaviour.  In fact, it is easy to confirm that the extremal surface $\Gms$ cannot coincide with the minimal surface on a maximal slice $\Xms$. This follows simply from the fact that the geodesics anchored at constant $v$ on the boundary do not all lie on a single spacelike surface in the bulk Vaidya-AdS spacetime.} As the horizon grows with increasing $v$, the
plots look similar to different size static BTZ black holes.

\subsection{Null expansions in Vaidya-AdS}

Having seen the behaviour of the spacelike geodesics which give us
the requisite minimal surface, we next study the null expansions for
the curve defined by \req{gline} in the three dimensional Vaidya-AdS
background \req{vaidyam}. Consider a generic curve parameterized as
in \req{gline} which is not necessarily a geodesic. Its two
orthogonal null vectors are given by
\begin{equation}
N^\mu_{\pm} = \CN \, \( \mu_{\pm}\, \(\p_v\)^\mu +(1 + \mu_{\pm}\,
f(r,v))\, \(\p_r\)^\mu - {1\over r^2}\, \(v' + \mu_{\pm} \, r'\) \,
\(\p_x\)^\mu \)  \ , \Label{vaidnull}
\end{equation}
where we have defined
\ba &&{\cal{N}}=\f{1}{\s{2}}\, \s{\f{r^2\, f(r,v)+r'^2}{r^2+2\,r'\,v'-v'^2\,
f(r,v)}}\ ,\no &&\mu_{\pm}=-\f{r^2+r'\, v'\, \mp r\, \s{r^2+2\,
r'\, v'-v'^2\,  f(r,v)}}{r^2\,f(r,v)+r'^2} \ . \ea
The expansions for these null vectors are then found to be
 \ba
\theta_+ +\theta_- &=& -\f{\Theta_{1}}{\s{2}\, \s{r^2\, f(r,v)+r'^2}\,
\(r^2+2\,r'\,v'-f(r,v)\, v'^2\)^{3/2}} \ ,\cr
\theta_+ -\theta_-  &=& \f{\Theta_2}{\s{2}\, \s{r^2\, f(r,v)+r'^2}\,
\(r^2+2\, r'\, v'-f(r,v)\, v'^2\)} \ , \Label{expvaa} \ea
where we have defined
\ba  \Theta_1 &=& - 2\,  r^2\, r'' + 2\, r'\, v' \, r^2 \p_r f+ 2
\,r^2 \,f \, v'' + r^2 \,v'^2 \, \p_v f - 2\,f\,r \, r'\,v' + 2\,
r\, r'^2 \no &&\qquad+ 3\,v'^2\, r'^2 \,\p_r f + 2 \,r'^2\, v'' - r'
\,v'^3 \,f\,\p_r f - 2 \,r'\, r''\, v' + r'\,v'^3 \,\p_v f, \no
\Theta_2 &=& 2\,r^2 \,f + 2\,r\, r'\,v'\,\p_r f -2\, r\, r'' - r\,
f\, v'^2\,\p_r f + r\, v'^2 \,\p_v f + 4 \,r'^2 \ . \ea
After some algebra we can show that both null expansions
$\theta_{\pm}$ are vanishing iff the equations of motion for the
geodesic \req{ham} and \req{eomvat} are satisfied. This justifies our assertion
in the previous sub-section that the extremal surface in question is given by a
spacelike geodesic in Vaidya-AdS.

\subsection{Time-dependent entanglement entropy}

Having obtained the extremal surface $\Gms$ for the Vaidya-AdS geometry, we can compute the entanglement entropy of the region $\rA$ using the area of $\Gms$.
In particular, we would now like to return to the original question about
time-dependence of the entanglement entropy. If we assume that the
time-dependence of the mass function $m(v)$ in \req{vaidyam} is very weak, $m'(v) \ll  1$, then we can use the adiabatic approximation. First we compute the
entropy for the static three dimensional AdS black hole (\ie\ BTZ) and
then treat the mass as a function of time $v$.

In the BTZ black hole background, the length of the geodesic is
given by the formula \req{entv}. The adiabatic approximation allows
us to regard $m$ as a time-dependent function $m(v)$, so that the
finite part of the geodesics length, denoted by $L_{reg}(v)$,
becomes
\be
L_{reg}(v) = L(v)-2\log(2\,r_{\infty})=\log\f{\sinh^2(\s{m(v)}\, h)}{m(v)} \ .
\Label{entvv} \ee
When the mass is very small $m(v)\ll 1$, \req{entvv} reduces to a regularized proper length as a function of (light-cone) time $v$:
\be L_{reg}(v)\simeq 2\log
h+\f{h^2}{3}m(v) \ . \Label{btzenta} \ee

Now let us recall the monotonicity property \req{nullenergy}. If
we combine it with the expression \req{entvv}, we can show that the
entanglement entropy increases in the adiabatic approximation. Hence assuming that the matter undergoing collapse to form the black hole satisfies the null energy condition, it is clear from the adiabatic approximation that the entanglement entropy
$\D S_\rA(v) \propto L_{reg}(v)$ increases in the process of a gravitational collapse.

\begin{figure}[htbp]
\begin{center}
\includegraphics[width=6.5in]{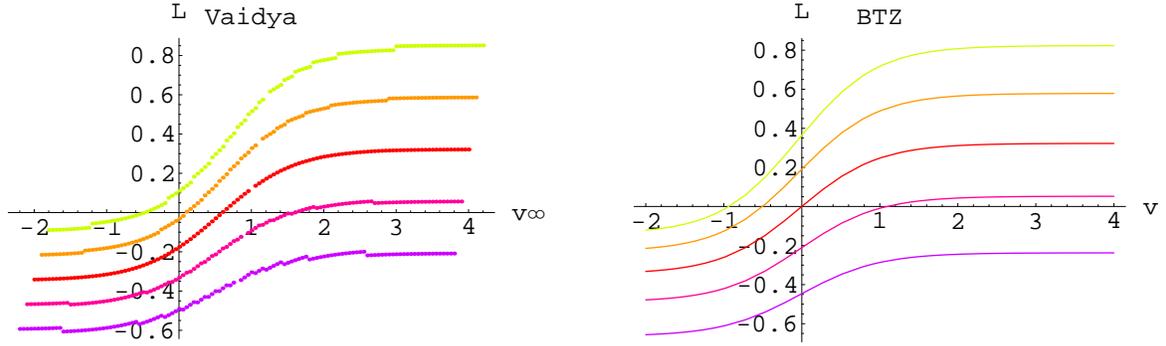}
\caption{{\bf left:} Regularised proper length $L_{reg}$ as a function of
the boundary $v_{\infty}$, for several regions,
$\pho=0.8,0.9,1,1.1,1.2$ in the Vaidya-AdS spacetime (\ref{Vadf}). {\bf
right:} the corresponding prediction in BTZ from (\ref{entvv}) . }
\label{LvarVad}
\end{center}
\end{figure}

\begin{figure}[htbp]
\begin{center}
\includegraphics[width=4.5in]{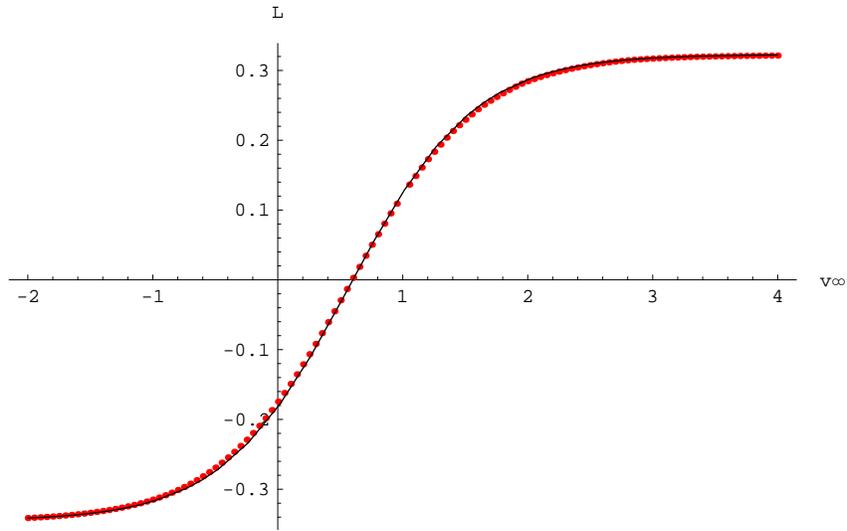}
\caption{ Regularised proper length $L_{reg}$ as a function of the
boundary $v_{\infty}$, for the particular region $\pho=1$ in the
Vaidya spacetime (\ref{Vadf}) (red dots) and the corresponding
prediction in BTZ from (\ref{entvv}), with a shifted $v$ value
(black curve). } \label{LvarfitVad}
\end{center}
\end{figure}

This claim can also be checked by a direct numerical analysis as
shown in \fig{LvarVad} for the profile \req{Vadf}. Not
only is it apparent that the proper length increases monotonically with time,
but we can also see that the adiabatic formula (\ref{entvv}) is actually quite
accurate for $v_s=1, m=1$. Note that there is a slight offset in the
$v$-values between the two plots; presumably this is because of the
dynamics, in particular the identification of the $v$ values.
However, if we shift the $v$ value appropriately and overlay the two
plots, the fit is almost perfect as shown in \fig{LvarfitVad}.

%

Below we will see that the monotonicity property can be proven via a
direct perturbative analysis and furthermore that it is related to
 the second law of the black hole thermodynamics.

\subsection{Perturbative proof of entropy increase}

Consider the change of the area functional when the surface is
deformed slightly. The infinitesimal shift of the $d-1$ dimensional
spacelike surface $\Gms$ is described by the deviation $\delta
X^\mu$. In general we find
\ba \delta \mbox{Area} &=&\delta
\int_{\Gms} d\xi^{d} \, \s{\mbox{det} \, g_{\ap\beta}}\no
&=& \int_{\Gms}
d\xi^{d}\, \delta X^\nu \, \Pi_\nu+\int_{\de \Gms}\, \s{g}\,g^{\ap\beta}\,g_{\mu\nu}\,\f{\de X^\mu}{\de \xi^\ap}\,\delta X^\nu\ , \Label{deviationa} \ea
where $g_{\ap\beta}=g_{\mu\nu} \, \f{\de X^\mu}{\de \xi^\ap} \, \f{\de X^\nu}{\de \xi^\beta}$
is the induced metric on the surface. $\beta$ in the final expression is orthogonal to the
submanifold $\de \Gms$ and $\Pi_\nu$ is defined such that the equation of motion for this
variational problem is given by $\Pi_\nu=0$.

This clearly shows that the area of extremal surface does not change
under any infinitesimal deformation provided we keep the same
boundary condition or the surface $\Gms$ is closed. However, since we
are interested in changing the boundary condition, corresponding to
the time-evolution, the final term in \req{deviationa}, which comes
from the boundary contribution via the partial integration, plays an
important role.

Let us now concentrate on the specific case of the three dimensional Vaidya-AdS spacetime and assume that $\Gms_{v}$ is the extremal surface at the asymptotic time $v$ as in \req{boundconv}. The equation of motion vanishes on shell by
definition; so only the boundary term contributes and it can be written as:
\be
\int_{\de\Gms}\s{g}\, g^{\ap\beta}\, g_{\mu\nu}\, \f{\de X^\mu}{\de
\xi^\ap}\delta X^\nu=2\, \delta v_0\, r\,
\left(1-f(r,v_0)\f{dv}{dr}\right)\biggr|_{r=r_{\infty}} \ ,
\Label{boundc}\ee
where the right hand side should be evaluated on the boundary with the cut-off $r=r_{\infty}$. The factor of two in \req{boundc} arises due to the two endpoints $x=\pm h$. To derive the above result, we set $\xi=r$ and use the fact
$g=g_{rr}\simeq \f{1}{r^2}$ and the deviation $\delta X^\mu=\delta
v_0\, \(\p_v\)^\mu$.  As a consequence, we obtain the following expression for the
time-dependence of the geodesic length for any choice of $m(v)$:
 \be \f{dL(v)}{d v}=-2\, r^3\,
\left[\left(\f{dv}{dr}\right)-\f{1}{r^2}\right]\biggr|_{r=r_{\infty}} \ .
\Label{timeevo} \ee
 Here we have used that the fact that in the UV
limit $r=r_{\infty}\to \infty$, the leading behavior of the relation
between $r$ and $v$ becomes $v\simeq \mbox{constant}-\f{1}{r}$. This
result \req{timeevo} shows a remarkable fact that the
time-dependence of the entanglement entropy only depends on the
asymptotic form of the function $v=v(r)$.

To evaluate \req{timeevo} explicitly, let us perform a perturbative
analysis by assuming that the time-dependent mass $m(v)$ is very
small and by keeping only its leading perturbation. The details of
this computation are described in the \App{apvaidya}. The
upshot is that the asymptotic expansion of $\f{dv}{dr}$ is found using (\ref{expvr}),
to be
\be \f{dv}{dr}\simeq
\f{1}{r^2}-\f{h^2 \, m'(v_0)}{6\,r^3}+\CO(r^{-4}) \ . \ee
Plugging this into \req{boundc}, we finally find the time-dependence of the
geodesic length
\be \f{dL(v)}{dv}=\f{h^2 \, m'(v_0)}{3}\geq
0.\Label{monot}\ee
This precisely agrees with the one obtained from
an adiabatic argument \req{finrt}. Notice that this is non-negative
when we impose the null energy condition \req{nullenergy}.

In this way we have confirmed the monotonicity property of the
entanglement entropy in the process of a gravitational collapse. It
would be an interesting problem to prove this for any general
function $m(v)$. We leave this for future investigation.

\subsection{Relation to the second law of black hole thermodynamics}
\label{totent}

Up to now we have used holography to examine the entanglement
entropy for a subsystem $\rA_v$ at a time $v$ in a two dimensional
theory. It is interesting to consider the limit where the subsystem
approaches the total space. In this limit,  it turns out that the
extremal surface $\Gms$ covers the whole apparent horizon, as we
will explain below. Thus the finite part of the holographic
entanglement entropy is dominated by the contribution from the area
of apparent horizon. The analogous result for static AdS black holes
has been already obtained in \cite{Ryu:2006bv, Ryu:2006ef}. When
$\rA_v$ finally coincides with the total system, the end points $\de
\rA_v$ annihilate with each other and the extremal surface becomes
the closed surface defined by the apparent horizon at time $v_*$,
where $v_*$ is the limiting value of the coordinate $v$ on the
extremal surface toward IR region. 

\begin{figure}
\begin{center}
  \includegraphics[width=10cm]{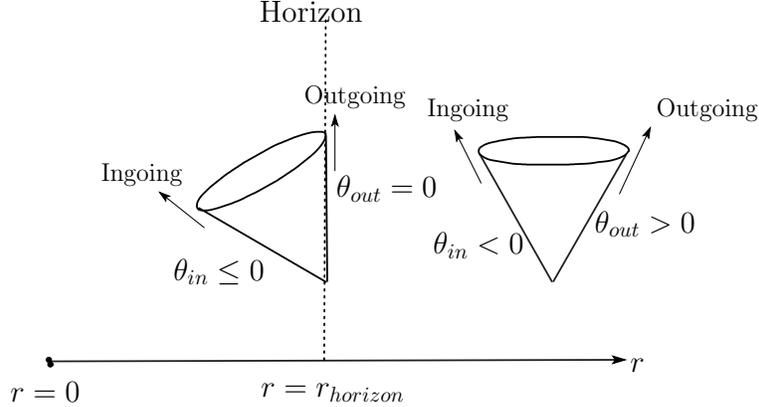}\\
  \caption{The behavior of the null geodesics near an apparent horizon.}\label{apparenth}
\end{center}
\end{figure}

The (future) apparent horizon is defined by the boundary of a
(future) trapped surface \cite{Hawking:1973uf}. In other words, on
the apparent horizon the expansion $\theta_{out}$ of the outgoing
future-directed null geodesics is vanishing, while the other expansion
$\theta_{in}$ of the ingoing null geodesics is non-positive (see
\fig{apparenth}),
 \be \theta_{out}=0\ , \qquad \theta_{in}\leq 0 \ .
\Label{aph}\ee
%

Let us find an apparent horizon in the Vaidya metric \req{vaidyam}.
Consider the particular class of co-dimension two surfaces defined by
$r=$ constant and $v=$ constant. Then the null expansions can be read from
\req{expvaa} as follows:
\be \theta_{in}=-\theta_+=-\f{\s{r^2-m(v)}}{\s{2}\, r},\ \ \ \ \
\theta_{out}=-\theta_-=\f{\s{r^2-m(v)}}{\s{2}\, r}\ .
\Label{norwr}\ee
One might naively think the expansions of null geodesics are both
vanishing at $r=\s{m(v)}$. However, this is actually not true
because we have not normalized the null vectors $N_{\pm}$
\req{vaidnull} such that they satisfy the geodesic equations
$N_{\pm}^\mu\nabla_{\! \mu} N_{\pm}^\nu=0$. The correct null vectors
are given by
\be N_{in}^\mu= -\(\p_r\)^\mu \ , \qquad
N_{out}^\mu=2\, \gamma(r,v)\, \(\p_v\)^\mu + f(r,v) \, \gamma(r,v) \, \(\p_r\)^\mu
\ ,
\ee
where $\gamma(r,v)$ is a positive function determined as a solution
to $2 \, \de_v \gamma(r,v)+\de_r\gamma(r,v)+2 \, r \, \gamma(r,v)=0$ which is
smooth at $r=\s{m(v)}$. The corresponding expansions of the null
geodesic congruences then become
\be \theta_{in}=-\f{1}{r}<0 \ , \qquad
\theta_{out}=\f{f(r,v) \, \gamma(r,v)}{r}\ .
\ee
With these correct normalizations we find that the condition
\req{aph} for an apparent horizon is satisfied at $f(r,v) =0 $. Thus
we can conclude that $r=\s{m(v)}$ is an apparent horizon in the
3-dimensional Vaidya-AdS metric.  While in  general in time-dependent
backgrounds the apparent horizon does not coincide with the event
horizon, we are guaranteed that event horizon  always lies outside the apparent horizon \cite{Hawking:1973uf}.

In the above example, the formula \req{expvaa} did not give the
correct sign of the expansions, as the rescaling needed to satisfy
the geodesic equation is singular. Since this occurs because
$f(r,v)=r^2-m(v)$ vanishes on the apparent horizon, for generic
curves which do not reach the apparent horizon this problem does not
appear and we can read off the correct sign of expansion from
\req{expvaa}.

Let us now return  to the reason for the extremal surface $\Gms$ to almost wrap the apparent horizon when the subsystem is taken to be as large as the total system. Finding the  extremal surface is equivalent to solving for the vanishing null expansion given by \req{expvaa};  the apparent horizon provides  a solution to this criterion, as is
manifest from \req{norwr}. Thus we can conclude that the limit of the subsystem engulfing the entire system,  the extremal surface appears to coincide with a
spatial section of the apparent horizon\footnote{Strictly speaking, as we have argued above, the apparent horizon is not a minimal surface. Moreover, the full apparent horizon is  a bulk co-dimension one surface. We will interpret the fact that the extremal surface $\Gms$ dips down almost all the way to the location of the apparent horizon and wraps the spatial section before returning back to the boundary to signify that the area of the apparent horizon plays an important role in computing the entanglement entropy in the limit of the subsytem $\rA$ approaching the full system $\bdys$.} (this fact can also be observed nicely in \fig{mpvaidyageodsA}).

Therefore we can argue that the total entropy $S_{tot}(t)=-\mbox{Tr}
\, \rho(t)\log\rho(t)$ in the dual time-dependent theory is given by
the area of the apparent horizon at $t=v_*$. The time-dependence of
$S_{tot}$ means that the evolution of the system is non-unitary;
this is the usual issue of evolution of a density matrix.\footnote{ One could
try to interpret this as unitary evolution in a tensor product
theory, where the second Hilbert space is hidden behind the horizon,
as in the eternal AdS black hole \cite{Maldacena:2001kr}. Of course,
in the dynamical situation we do not have exact thermal periodicity
and this would imply that the `shadow CFT' lives on a shifted locus
in the complex time plane.} It is
also interesting to note that for reproducing a physical quantity,
the apparent horizon, which is defined using local quantities, is
more crucial than the event horizon, whose definition is rather
global.

The second law of black hole thermodynamics tells us that the area
of apparent horizon  always increases under any physical process
which satisfies the appropriate energy condition \cite{Hawking:1973uf} (also \cf,
\cite{Ashtekar:2004cn}). This can be shown explicitly from the
condition \req{aph} and the basic formula \req{areaform}, which
guarantees that the area increases under an infinitesimal
deformation $\delta X^\mu$ along the evolution of the apparent
horizon $\delta X^\mu\propto N_{out}^\mu-N_{in}^\mu$. On the other
hand, we can derive this second law of the apparent horizon area
from the monotonicity property \req{monot} by taking the mentioned
limit of the minimal surface. In this way the two concepts are
naturally connected with each other. Notice that on both sides the
monotonicity stems from the positive energy condition (with assumption of 
homogeneity in the dual field theory).

We note in passing that in the limit of the subsystem $\rA$ engulfing the system,
the temporal evolution of the extremal surface $\Gms$ is captured by the behaviour of {\em dynamical horizons}. A {\em dynamical horizon} is defined to be a smooth, co-dimension one spacelike submanifold of the spacetime, which can be foliated by a family of closed spacelike surfaces, such that the leaves of the foliation have one null expansion vanishing and the other null expansion being strictly negative \cite{Ashtekar:2003hk,Ashtekar:2004cn} as in \req{aph}. It is tempting to infer from this that the results proved for the area increase of dynamical horizons can be ported to the present situation and in particular used to establish a ``second law of entanglement entropy'' from holographic considerations.

\section{Other examples of time-dependent backgrounds}
\label{wholebub}

\subsection{Wormholes in AdS and entanglement entropy}

 Consider the (entanglement) entropy $S_{tot}=-\mbox{Tr} \, \rho_{tot} \log \rho_{tot}$ for the total system as in \sec{totent}. It is vanishing if the system is in a pure state. When it is non-vanishing, it is usually interpreted as the
thermal entropy and correspondingly its AdS dual spacetime is
expected to have an event horizon. In such examples the total
entropy $S_{tot}$ is dual to the Bekenstein-Hawking entropy of the
black hole in question. In Lorentzian geometries such as the eternal
Schwarzschild-AdS geometry,  we can equivalently regard the entropy
as arising from the entanglement between the total system and
another identical system hidden behind the horizon as in
\cite{Maldacena:2001kr} (\cf, also ~\cite{Freivogel:2005qh} for
other examples). A recent discussion of issues relevant to this
context can be found in \cite{Balasubramanian:2007qv}.

In this section we would like to point out an example which has a
non-zero total entropy $S_{tot}$ and its origin seems to be
different from the example mentioned above. In particular, the example we have in mind is an Euclidean spacetime with no event horizons. These are the
Euclidean AdS wormholes discussed in \cite{Maldacena:2004rf}.
\begin{figure}
\begin{center}
\includegraphics[width=10cm]{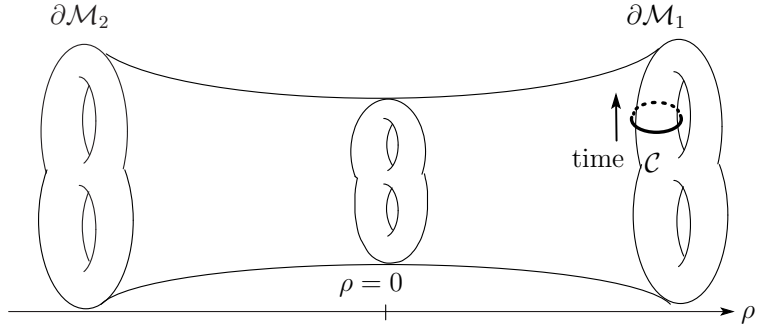}\\
\caption{The AdS wormhole geometry and the topologically non-trivial cycle ${\cal C}$ on the boundary.}\label{wormhole}
\end{center}
\end{figure}
They are obtained by considering the hyperbolic slices of Euclidean
AdS$_{d+1}(=H_{d+1})$
\begin{equation}\label{MM}
ds_{H_{d+1}}=d\rho^2+\cosh^2\rho\ ds^2_{H_{d}} \ ,
\end{equation}
and by taking a quotient of $H_d$ by a discrete group $\Gamma$ to generate a compact manifold. We mainly consider the case $d=2$, because in this case the background is perturbatively stable \cite{Maldacena:2004rf}.  Also the dual two dimensional CFT is well-defined on a background of negative curvature $H_{2}/\Gamma$; a Riemann surface. The two boundaries $\bdy_1$ and $\bdy_2$ are given by the two limits $\rho\to \infty$ and $\rho\to -\infty$. After the quotient by the
Fuchsian group $\Gamma$, the two boundaries become the same Riemann surface
with genus $g\geq 2$ (see \fig{wormhole}).

Such a solution leads us to a puzzle\footnote{There is another
possibility that the path-integral over infinitely many such
geometries cure the problem as discussed in \cite{Maldacena:2004rf}.
Here we are assuming that each of perturbatively stable
asymptotically AdS solutions should have its dual CFT interpretation
before summing over the geometries.} immediately as pointed out in
\cite{Maldacena:2004rf}. From the CFT side, we expect that the two
CFTs on the two boundaries are decoupled from each other. Thus all
correlation functions between them should be vanishing. However,
from the gravity side, there are non-trivial correlations since the
two boundaries are connected through the bulk.

Here we would like to point out a possible resolution to this
problem. Our claim is that the CFT$_1$ on $\bdy_1$ and the
CFT$_2$ on $\bdy_2$ are actually entangled with each other despite
the absence of an event horizon. To check CFT$_1$ and CFT$_2$ are
indeed entangled, we need to compute the entanglement entropy $S_1$
for the total system of CFT$_1$. This should coincide with the entanglement entropy
$S_2$ for the CFT$_2$ (we expect that the total  system CFT$_1$  $\cup$
CFT$_2$ to be in a pure state).

When we choose a Euclidean time-direction locally in the two dimensional space
$\bdy_1$, the total system  (at a specific time)  in CFT$_1$ is defined by a
circle ${\cal C}$ in $\bdy_1$, which is topologically non-trivial.
This setup can be regarded as a higher genus generalization of the
computation at a finite temperature using Euclidean BTZ black hole
done in \cite{Ryu:2006ef, Ryu:2006bv}.

Let us define the circle ${\cal C}_{min}$ in the Riemann surface to
be cycle with minimal length among those which are homotopic to
${\cal C}$. Then the minimal surface which is relevant  to the
holographic computation of $S_1$ turns out to be the circle ${\cal
C}_{min}$ at the throat $\rho=0$. This can be understood as follows;
see \fig{wormhole} below. We first consider the entropy $S_{\rA}$
assuming that $\rA$ is a submanifold of ${\cal C}$. Then we can
easily find the minimal surface whose end point at $\rho=\infty$
coincides with $\brA$. As we gradually increase the size of $\rA$,
the minimal surface anchored  on one boundary dips deeper into the
bulk. In the limit $\rA \to {\cal C}$ the two end points of $\brA$
 annihilate and the minimal surface gets localized at the throat. So the maximum entropy
  is given by the area of the neck. It is clear from the geometric picture that
$S_1=S_2$, since both are measured by the area of the throat.

In this way we find that the entanglement entropy $S_1$ between
CFT$_1$ and CFT$_2$ is given by
\be S_1=S_2=\f{\mbox{Area}({\cal
C})}{4\,G^{(3)}_{N}}>0. \ee
As this is clearly non-vanishing due to the throat connecting the two boundaries, we can conclude that the two CFTs are entangled with each other. Interestingly, the existence of such a minimal surface at the throat also plays the crucial role when we present a generic definition of wormhole as discussed in \cite{Hochberg:1997wp}.

The above definition of entanglement entropy between two CFTs only
depends on the topological class of the cycle ${\cal C}$. Thus we
can define $2g$ different entropies when $\bdy_1=\bdy_2$ is a genus
$g$ Riemann surface. We would also like to stress that the genus one version of the above calculation is equivalent to the ordinary Euclidean computation of the Bekenstein-Hawking entropy of black holes.

In the above discussion, we have concentrated on Euclidean wormholes. In
the Lorentzian case, the topological censorship \cite{Galloway:1999br}
(with assumptions about energy conditions) guarantees that disconnected
boundaries  are separated from each other by event horizons. This is a
simple consequence of null geodesic convergence following from Raychaudhuri's equation;
essentially if two disconnected boundaries were in causal communication then null geodesics
which are initially contracting will have to re-expand, violating the null convergence condition.
If we allow the presence of some exotic matter so that the Lorentzian wormholes exist,
the above computation of the entanglement entropy in wormhole geometries can be equally
applied to the Lorentzian case.

\begin{figure}
\begin{center}
  \includegraphics[width=10cm]{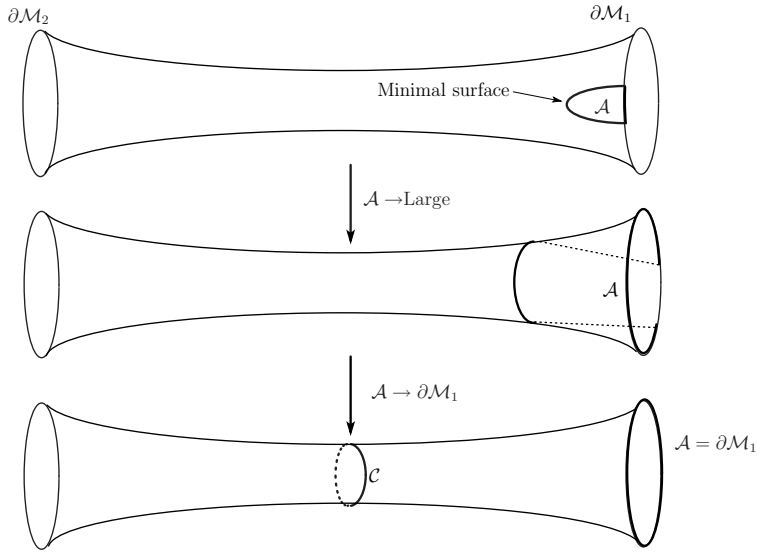}\\
  \caption{The computation of entanglement entropy in the
presence of two boundaries in the AdS wormhole geometry at fixed
value of the Euclidean time.}\label{wormhole2}
\end{center}
\end{figure}

\subsection{Entanglement entropy of the AdS bubble}

Our final example of a time-dependent asymptotically AdS background
is the AdS bubble solution \cite{Birmingham:2002st,
Balasubramanian:2002am}
\begin{equation}
ds^2=f(r) \, d\chi^2+\f{dr^2}{f(r)}+r^2 \, \(-d\tau^2+\cosh^2 \tau
\, d\Omega_2^2\), \Label{bubble}
\end{equation}
where $f(r)=1+ r^2-\f{r_0^2}{r^2}$. This is obtained via a double
Wick rotation of the Schwarzschild-AdS black hole in the global
coordinates. This solution represents a background where a bubble of
nothing shrinks from infinite size to a minimum value $r_+^2$
\req{bubparams} and subsequently re-expands out as the time evolves
from $\tau=-\infty$ to $\tau=\infty$. If we consider a double Wick
rotation of the  (planar) Schwarzschild-AdS black hole in Poincar\'e
coordinates, we obtain the static AdS bubble (or AdS soliton), whose
entanglement entropy was computed in \cite{Nishioka:2006gr} and a
quantitative comparison with the dual Yang-Mills has been made
successfully.

The coordinate $\chi$ in \req{bubble} is compactified and the
smoothness of this solution requires the periodicity $\chi\sim
\chi+\Delta\chi$, where $\Delta\chi$ is given by
\be
\Delta\chi=\f{2\pi \, r_+}{2\,r_+^2+1 } \ , \qquad
r^2_+\equiv
\f{1}{2}\, \left(\s{1+4\, r_0^2}-1\right) \ .
\Label{bubparams}
 \ee
One can show that this solution is asymptotically AdS as
$r\to\infty$. The important point is that the time $t$ in the
asymptotically AdS global coordinate is different from $\tau$ in
\req{bubble}. They are related via \cite{Balasubramanian:2002am}:
 \be \tan t=\f{r}{\s{r^2+1}}\, \f{\sinh \tau}{\cosh \chi}.
\ee
The boundary of the metric \req{bubble} is dS$_3$ $\times$ $\Sp^1_\chi$,
with $\tau$ being the deSitter time coordinate.

Now we are interested in the entanglement entropy $S_{\rA}$ at fixed
time $t=t_0$ in the boundary theory on dS$_3$ $\times$ $\Sp^1_\chi$.
The radius of $\Sp^2 \subset$ dS$_3$ has a time varying radius
$\sim\cosh \tau$. We define the subsystem $\rA$ such that its
boundary $\brA$ is $T^2=\Sp^1_{\chi}\times \Sp^1$, where the second
$\Sp^1$ is the equator of the $\Sp^2$. Then the extremal surface
will be given by the two dimensional surface defined by
$g(\tau,\chi,r)=0$ times the $\Sp^1$.  To explicitly determine the
function $g$, one needs  to solve a complicated set of partial
differential equations.

If we assume\footnote{Recall that we are working in units where the
AdS radius is set to unity.} $r_0 \ll 1 $ (\ie, $\Delta\chi\sim 2\pi
\, r_0 \ll 1$), then the condition $t=$ constant is approximated by
$\tau=$ constant. To avoid solving the differential equations, we
consider a minimal surface on the time slice defined by $t={\rm const}.$
as a further approximation.\footnote{Clearly this surface does not
coincide with $\Lms$ in our covariant construction except
$\tau=t=0$. However, we believe that we can obtain a qualitative
behaviour of the time-dependent entanglement entropy using this
approximation.} Under this approximation we can easily find
the entanglement entropy:
\begin{equation}
S_{\rA}(t)=\f{\mbox{Area}(\Gms)}{4\, G^{(5)}_N}\simeq
\f{2\pi}{4G^{(5)}_N}\, \cosh \tau \int^\infty_{r_0} r \,
dr\int^{2\pi r_0}_0 d\chi=\f{\pi^2 r_0}{2G^{(5)}_N}\,
(r^2_{\infty}-r^2_0)\cosh \tau ,
\end{equation}
where $r_{\infty}$ is the UV cut-off. We find that the entropy is
proportional to $\cosh \tau\simeq \f{1}{\cos t}$. This is
consistent with the known area law of the entanglement entropy
because Area$(\de \rA)\propto \cosh \tau$. Note that the finite term
has a minus sign. This is because the emergence of the bubble means
the disappearance of degrees of freedom as discussed in \cite{Nishioka:2006gr}
about the static AdS bubble example. It would be interesting to understand this
 time-dependent entanglement entropy from the dual field theory side and to find
 the explicit extremal surface for this geometry.

\section{Discussion }
\label{discuss}

\hspace{.5cm} In this paper, we have presented the covariant
holographic formula \req{holeeT} (or equivalently \req{LmsminB}) of
the entanglement entropy (or von-Neumann entropy) in AdS/CFT
correspondence within the supergravity approximation. We propose that
it is simply given by the area of the extremal surface in a given
asymptotically AdS background. This allows us to calculate
entanglement entropy of dual (conformal) field theories even in
time-dependent backgrounds. This is a natural generalization of the
previously proposed holographic formula for static AdS backgrounds
\cite{Ryu:2006ef, Ryu:2006bv}.

Our covariant holographic proposal claims that the
entanglement entropy $S_{\rA}$ for the subsystem $\rA$ is equal to
the area of a certain bulk surface $\ms$, which is anchored at the
boundary $\brA$,  in Planck units as in the Bekenstein-Hawking
formula. Our main conclusion is that the surface $\ms$ is given
by the extremal surface $\Gms$, which is an extremum of the area functional.
We argued that $\Gms$ is equivalent to the surface $\Lms$, which we motivated from the covariant entropy bound.  In particular, $\Lms$ is defined to be the minimal area surface
among the family of co-dimension two bulk surfaces satisfying the requisite boundary conditions with the additional constraint that they support two light-sheets \ie, the null geodesic congruences directed toward the boundary have non-positive expansion.
More constructively, $\Lms$ corresponds to the surface with vanishing null expansions.
We gave an argument which supports our claim  $\Gms=\Lms$; a rigorous proof is left as an intriguing problem for the future. We also pointed out  another potential candidate for a covariantly defined surface, a minimal surface on a maximal time-slice, $\Xms$, which reduces to the minimal surface in static spacetimes.
We showed that $\Xms$ coincides with $\Gms$ when the bulk spacetime is foliated by totally geodesic spacelike surfaces and argued that generically $\Xms$ doesn't capture the holographic entanglement entropy of a specified boundary region.

We argued that our covariant proposal can be derived naturally
using the light-sheet construction and thus is closely related to
the covariant entropy bound (Bousso bound) \cite{Bousso:1999xy}. At
first sight, this relation is rather surprising, since the entropy
bound is usually associated with the thermodynamic entropy while
the entanglement entropy has a different origin. It strongly
supports the historical idea that the entanglement entropy is connected with a
microscopic origin of the gravitational entropy with quantum
corrections \cite{Bombelli:1986rw, Srednicki:1993im} (see also
\cite{Hawking:2000da, Emparan:2006ni} and references therein). We leave further exploration of this relation as an important open problem. It would  also be interesting to generalize the covariant holographic formula beyond the
supergravity approximation assumed in the above discussion (see \cite{Fursaev:2006ih} for  recent progress in the Euclidean case).

We believe that deeper understanding of the entanglement entropy
will provide crucial insights into the nature of the holographic relation
between quantum gravity and its dual
non-gravitational lower dimensional theory. One reason for this stems
from the universality of the definition of this quantity for any system
 described by the laws of quantum mechanics. We
can deal equally well with diverse systems such as spin chains,
 quantum Hall liquids, gauge theories, matrix models, and even
cosmological models from this viewpoint. For each, we can then examine if
a holographic dual exists, and then exploit the covariant construction above to compute the entanglement entropy. The second reason is that the
entanglement entropy is holographically described by a basic
geometrical quantity, namely the area of a
well-defined co-dimension two surface, as we have discussed extensively.
 Given a specific bulk geometry, which is described by some particular CFT state, we can calculate the proper areas of the requisite surfaces which we have conjectured to correspond to the entanglement entropy for that state.

Conversely, given a specific state of the boundary theory, we can ask how much information is encoded in the entanglement entropy.  In particular, for a system  in a given state admitting a gravitational dual, if we know the entanglement entropy for all subsystems $\rA$ of the boundary, can we decode the full geometry of the gravitational dual corresponding to that state, at least at the supergravity level?  Even though we leave the actual metric extraction for future work, we believe that most, if not all, of the metric information can indeed be extracted from the entanglement entropy data by a suitable inversion technique.

An analogous problem has been discussed in \cite{Hubeny:2006yu}, where the singularities in the CFT correlators, the so-called bulk-cone singularities, were used to distinguish different geometries.  The basic idea is that the bulk-cone singularities occur for correlators whose operator insertions are connected by a null geodesic through the bulk spacetime; and since bulk geodesics are determined by the bulk geometry, knowing the endpoints of null geodesics allows us to extract a large amount of information about the bulk geometry.  Using this technique, \cite{Hammersley:2006cp} has numerically demonstrated metric extraction for a class of static, spherically symmetric bulk spacetimes.

However, null geodesics have their limitations.  They are insensitive to conformal rescaling of the metric, and they probe only the part of the bulk which allows their endpoints to remain pinned at the boundary.  Hence the metric extraction of \cite{Hammersley:2006cp} does not probe the bulk past null circular orbits.
In this respect, a spacelike geodesic, or more generally a spacelike surface, would bypass both of these shortcomings. This has been confirmed in \cite{Hammersley:2007ab}.  Not only are spacelike geodesics sensitive to the conformal factor, but also they probe deeper into the bulk while remaining pinned at the boundary.  This is demonstrated in \fig{mpBTZgeodsA}, where metric extraction would be allowed all the way down to the horizon.  Moreover, when the bulk is 4 or higher dimensional, the co-dimension two surfaces are likewise higher-dimensional, and therefore may be expected to contain a larger amount of information than geodesics which are only one-dimensional quantities.

The computation of the entanglement entropy in a time-dependent
background is in general a very hard question due to technical
complications. However, our holographic formula allows us to solve this
problem simply, provided the system under consideration has a holographic dual.
In this paper, we examined
several examples of time-dependent backgrounds. First we considered
the 3-dimensional AdS-Vaidya background, which is dual to a time-dependent
background of a 2-dimensional CFT. There we found that the entanglement entropy
computed holographically increases under time evolution. We have
also seen that this is closely related to the second law of the
black hole thermodynamics. This result suggests the expected monotonicity
property; namely that given the null energy condition,
in any gravitational collapse the entanglement
entropy always increases. It would
be interesting to  see if
monotonicity is preserved once we take into account quantum corrections
described by the Hawking radiation on the
gravity side.

Another example we discussed concerns wormholes in AdS. Even though
  the two dual CFTs on the two disconnected
 boundaries look decoupled from each other, there are
 non-vanishing correlation functions from the bulk gravity viewpoint \cite{Maldacena:2004rf}.
 We proposed a possible resolution to this puzzle
 by showing that the entanglement entropy between the two CFTs is
 actually non-vanishing. This confirms that they are quantum mechanically entangled.

Since the concept of entanglement entropy is well-defined in any
time-dependent system, it provides a very useful physical quantity
to analyze in a quantum system which is far from the equilibrium,
where we cannot define the usual thermodynamical quantities. At the
same time, it is an important quantity bearing on quantum phase
transitions of various low dimensional systems at zero temperature.
Therefore our results can be regarded as a first step toward the
analysis of condensed matter physics using the AdS/CFT
correspondence (see \eg, \cite{Herzog:2007ij} for other recent
interesting approach).

\section*{Acknowledgements}
We would like to thank O.~Aharony, R.~Azeyanagi, D.~Eardley, D.~Garfinkle, M.~Headrick, T.~Hirata, G.~Horowitz, D.~Marolf, T.~Nishioka, S.~Ross,  S.~Shenker, T.~Shiromizu, E.~Silverstein, L.~Susskind, S.~Yamaguchi for extremely useful discussions. We would like to thank the organisers of the Indian Strings Meeting 2006 for hospitality during the initial stages of this project. VH and MR would in addition  like to thank the KITP, Santa Barbara for hospitality during the course of this project.

\appendix
\section{Covariant construction of causally-motivated surface $\Cms$}
\label{3dspl}

Above we have discussed three distinct constructions as candidate covariant duals of the entanglement entropy, namely the surfaces $\Gms$, $\Xms$, and $\Lms$.   All of these require solving an extremization problem.  However, in \sec{covEEprev} we have also mentioned an alternate construction, $\Cms$, which may be computationally simpler to find.  This is because the requisite co-dimension one surface on which we define $\Cms$ is constructed purely based on causal relations and therefore does not require \eg\ solving for geodesics (though in practice, in many examples it is quite easy to find this by using null geodesics).

The causal covariant construction $\Cms$ will be achieved by a series of steps:
\begin{enumerate}
\item
Starting with the spatial region $\rA_t \subset \bdys_t \in \bdy$, construct
its domain of dependence
$D_t \subset \bdy$.  This is the set of all boundary points $q$
through which all causal boundary curves $\curt q$ necessarily intersect $\rA_t$,
\begin{equation}
D_t = \set{q \in \bdy \st \forall \ \curt q \in \bdy,  \set{\curt q \cap A_t}
\ne \emptyset }
\end{equation}
\item
Construct the bulk ``causal wedge" $C_t$ of the boundary
region $D_t$.  This is defined as the set of bulk
points $p$ from which there exists both a future-directed and a
past-directed causal curve, $\curf p$ and $\curp p$, which intersects\footnote{
For purposes of this definition
we treat $\bdy$ as a subset of $\bulk$, so that a bulk
curve can ``intersect'' (\ie, terminate on) the boundary.
This is motivated by using the usual ``cut-off" surface instead of the actual boundary.
More technically, we want the ideal points associated with the TIP or
 TIF of the requisite curve through $p$ to lie in $D_t$.
} $D_t$.
\begin{equation}
C_t = \set{p \in \bulk \st
\exists \ \curf p ,  \set{\curf p \cap D_t} \ne \emptyset
\ \ {\rm and} \ \
\exists \ \curp p ,  \set{\curp p \cap D_t} \ne \emptyset  }
\end{equation}
\item
Let $B_t$ be the boundary (in the bulk) of $C_t$.
In some simple cases, this is constructed from the
future and past bulk light-cones from the past and future tip of $D_t$.
\begin{equation}
B_t = \set{ p \in \bulk \, \backslash \, \bdy \, \cap \, p \in \p C_t}
= \p C_t \, \backslash \, D_t
\end{equation}
\item
Finally, consider the set of all spacelike surfaces
lying in $B_t$ and pinned at $\brA_t$.
From these spacelike surfaces, take one with the
minimal area.\footnote{There may in general be more
than one such surface, but we are ultimately interested
in the area of such a surface, and this value is unique.
The fact that the area is bounded from above is tied to the
fact that we are looking for a surface which has only one
dimension less than $B_t$; lower dimensional surfaces could
achieve arbitrarily high area by ``crumpling". See also footnote \ref{v3addition}.}  We denote this maximal
surface by $\Cms_t$.
So we have $\Cms_t \in B_t \in \bulk , \ \p \Cms_t = \brA_t  $.
\end{enumerate}

\begin{figure}[htbp]
\begin{center}
\includegraphics[width=6in]{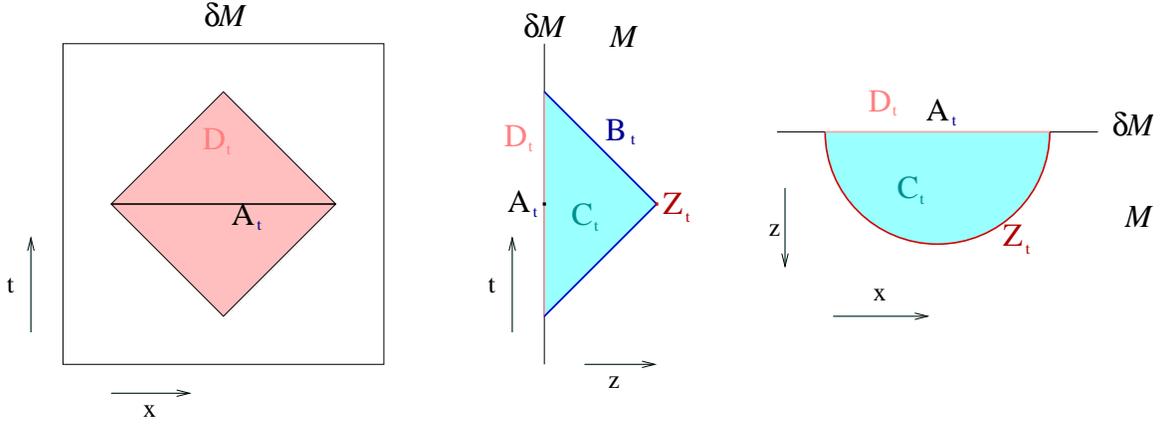}
\caption{Sketch of the proposed construction of the desired bounding
surface $\Cms_t$.} \label{EntConstrFig}
\end{center}
\end{figure}
In \fig{EntConstrFig} we indicate the construction of $\Cms$
more explicitly by sketching various 2-dimensional slices of the spacetime, as labeled.
\begin{table}[htdp]
\caption{Dimensionality of the various regions discussed.}
\begin{center}
\begin{tabular}{|c|c|c|}
\hline
region & dimensionality & bulk/bdy \\
\hline
$\bulk$  & $d+1$ & $\bulk$ \\
$\bdy$ & $d$ & $\bdy$ \\
$\rA_t$ & $d-1$ & $\bdy$ \\
$\brA_t$ & $d-2$ & $\bdy$ \\
$D_t$ & $d$ & $\bdy$ \\
$C_t$ & $d+1$ & $\bulk$ \\
$B_t$ & $d$ & $\bulk$ \\
$\Cms_t$ & $d-1$ & $\bulk$ \\
$\Gms, \Xms, \Lms$ & $d-1$ & $\bulk$ \\
\hline
\end{tabular}
\end{center}
\label{dimensionality}
\end{table}%
In Table~\ref{dimensionality} we show the dimensionality
 of the various regions discussed and specify whether they
  lie in the bulk or in the boundary.  Recall that the
  $\Cms_t$ is a co-dimension two surface in $\bulk$, as is the
   boundary region $\rA_t$.
Furthermore, this construction does not depend on a choice
of coordinates, but only on physically meaningful quantities:
causal relations in the spacetime and proper ``area" of a given
 spacelike surface.  This ensures that we can apply the same
 construction for time dependent bulk geometries just as easily.

\subsection{Discrepancy in AdS$_{d+1}$ for $d\ge3$}

Now let us consider whether the construction $\Cms$ provides a viable candidate for the dual of the entanglement entropy. In order for the area of $\Cms$ to be equal to the entanglement entropy for general states, a minimal requirement is that $\Cms$ reduces to the correct minimal surface for static spacetimes.  Therefore we wish to check whether in any static spacetime, $\Cms$ coincides with $\Gms$ (which, as we argued above, automatically coincides with $\Xms$ and $\Lms$ for all static spacetimes).

We can find an easy counter-example, even for pure AdS, in more than three dimensions for non-spherical regions.  For simplicity, let us consider the infinite strip in AdS$_4$, in Poincar\'e coordinates.  The bulk metric is
$ds^2 = {1 \over z^2} \, \( -dt^2 + dz^2 + dx^2 + dy^2 \)$, and let the region $\rA$ on the boundary be an infinite strip extended along the $y$ direction; $\{t=0, x \in (-h,h)\}$.  The minimal surface is given by \req{stripms}, with $x(z)$ given by $\ti{d}=2$ and smeared over all $y$.  On the other hand, the causal construction of $\Cms$ outlined above is determined by past/future directed null geodesics at constant $y$, from $\{z=0,x=0,t=\pm h\}$ into the bulk.  Since these are insensitive to the conformal factor of the bulk metric, they behave just as in flat spacetime; the maximal area surface, lying on the intersection of the future and past light-cones from the tips of $D_0$, is given simply by the half-circle $z^2+x^2=h^2$, uniformly smeared in the $y$-direction.
\begin{figure}[htbp]
\begin{center}
\includegraphics[width=2.5in]{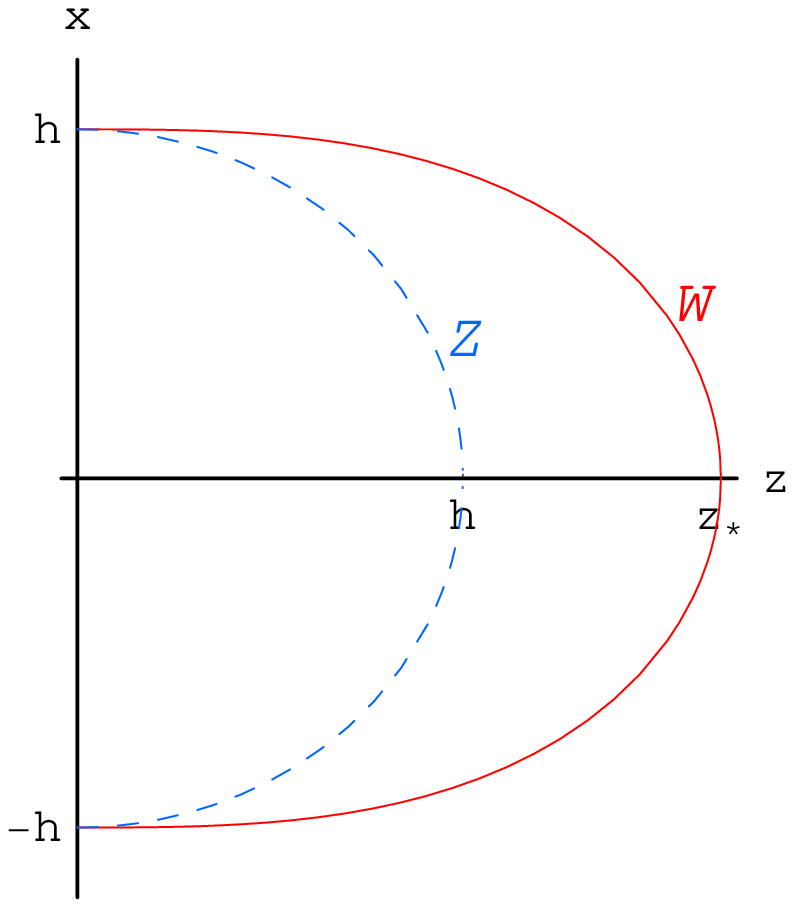}
\caption{A constant-$y$ cross-section of the two surfaces $\Cms$
and $\Gms$ for infinite strip of width $2h$ in AdS$_4$.  This example
demonstrates that $\Cms \ne \Gms  = \Xms = \Lms$.}
\label{ZneW4d}
\end{center}
\end{figure}
We can easily check that this surface $\Cms$ does {\it not} coincide with the minimal surface $\Gms$ since $\Cms$ does not satisfy\footnote{
Note however that, remarkably, for a circular region $\rA$, the two surfaces $\Cms$ and $\Gms$ would coincide exactly.
}  \req{stripms}.  In particular, \fig{ZneW4d} demonstrates the difference between the two surfaces.  Moreover, we can easily check that the area of $\Cms$ is much larger than the area of $\Gms$ (since $\Cms$ lies closer to the boundary where the warp factor diverges), so the former cannot yield the entanglement entropy.

The above discrepancy gets exacerbated in higher dimensions, where
as $d$ increases, the solution to \req{stripms} becomes more
separated from the curve $\Cms$, given by
 $z^2+x^2=h^2$, which is independent of $d$.  In fact, in the more physically interesting
 case of AdS$_5$, we have an infinite discrepancy between the area of $\Cms$ and that of $\Gms$.
 Here, we can actually compare our results directly with a free Yang-Mills calculation, and check
 explicitly which surface yields a better estimate of the entanglement entropy.
In particular, the entropy density corresponding to $\Gms$, which coincides with the minimal surface
considered in \cite{Ryu:2006ef} (see Eq.(7.6) of that paper), is given by
\begin{equation}
S_{\Gms}=\f{1}{4\,G^{(5)}_N}\, \left(\f{1}{\cof^2}-0.32\,\f{1}{4h^2}\right) \ ,
\Label{areaW}
\end{equation}
where $z=\cof$ is the usual UV cut-off. Note that we are quoting here the result for the entropy density and the AdS radius is set to unity.
On the other hand, the entropy density associated with $\Cms$ can be easily computed to be
\begin{equation}
S_{\Cms}=
\f{1}{4\, G^{(5)}_N}\, \left(\f{1}{\cof^2}-2\, \f{1}{4h^2}+\f{1}{h^2}\,
\log\f{2h}{\cof}\right) \ .
\Label{areaZ}
\end{equation}
Note that apart from the standard $1/\cof^2$ divergence, $S_{\Cms}$ also suffers from a
 logarithmic divergence (related to the conformal anomaly),
 so that the discrepancy in the areas of $\Cms$ and $\Gms$ is actually
 infinite, though the leading divergence is the same.

Now, to compare these gravity results with the Yang-Mills results,
we consider the direct free Yang-Mills computation.  This leads to
the following entropy density when we expressed it in terms of AdS
quantities \cite{Ryu:2006ef}:
\begin{equation}
S_{YM}=\f{1}{4\, G^{(5)}_N}\left(({\rm const})\cdot\f{1}{\cof^2}-0.49\cdot\f{1}{4h^2}\right) \ .
\Label{areaYM}
\end{equation}
Of course, we do not expect the free Yang-Mills result to agree
quantitatively with the AdS gravity computation, since the latter
corresponds to the strongly coupled gauge theory. Also we cannot
directly compare the divergent term since the UV cutoff in
Yang-Mills calculation is not necessarily equal to the one in the
AdS side. However, we do expect that the finite part of entropy
should agree with each other semi-quantitatively as evidenced by the
famous $4/3$ entropy factor for the black D3-branes.\footnote{ In
fact, it's amusing to note that the ratio of the coefficients of the
leading finite terms in the entanglement entropy expressions
\req{areaYM} and \req{areaW} also lies very close to $4/3$. }
Comparing \req{areaYM} with \req{areaW} and \req{areaZ}, we
immediately see that the extremal surface $\Gms$ yields a much
better approximation of the entanglement entropy for the free
Yang-Mills system than $\Cms$.  We expect this to remain true even
at strong coupling.

\subsection{3-Dimensional static bulk geometries}

Above we have seen that for AdS$_{d+1}$ with $d\ge3$, the surface $\Cms$ does not necessarily coincide with the requisite minimal surface for static spacetimes.  This {\it a priori} rules it out as a candidate covariant dual of entanglement entropy in time-dependent scenarios as well.
However, we may still ask whether in 3 dimensions the $\Cms$ construction works better.
After all, the dual field theory lives in 2 dimensions, so we would expect many special properties.  Indeed, from the geometrical point of view, 3-dimensional bulk is special: since $\rA$ is 1-dimensional, $B_t$ is always described simply by  a light-cone.
Moreover, performing the above check for $d=2$, we find that $\Cms = \Gms$, since the surface $z^2+x^2=h^2$ satisfies \req{stripms}.

In fact, slightly less trivially, we can likewise check by explicit calculation that in global AdS$_3$, with $ds^2 = - (r^2 + 1) \, dt^2 + {dr^2 \over r^2 + 1} + r^2 \, d\ph^2$ and
$\rA = \set{ (t,\ph) \st t=0 \, , \ \ph \in (-\pho,\pho) }$, both $\Cms$ and $\Gms$ are given by
\begin{equation}
r^2 (\ph) = { \cos^2 \pho  \over \sin^2 \pho \, \cos^2 \ph - \cos^2 \pho \, \sin^2 \ph } \ .
\Label{FAdS}
\end{equation}
Similarly, in BTZ, with
$ds^2 = - (r^2 - \rh^2) \, dt^2 + {dr^2 \over r^2 - \rh^2} + r^2 \, d\ph^2$,
$\Cms$ and $\Gms$ likewise coincide and are given by
\begin{equation}
r^2 (\ph) = \rh^2 \
{ \cosh^2 ( \rh \, \pho)  \over \sinh^2
  ( \rh \, \pho) \, \cosh^2  ( \rh \, \ph) - \cosh^2
   ( \rh \, \pho) \, \sinh^2  ( \rh \, \ph) } \ .
\Label{FBTZ}
\end{equation}

Let us therefore ask whether this agreement holds in general.  Specifically, consider a metric for a general static, spherically symmetric 3-dimensional spacetime,
\begin{equation}
ds^2 = -f(r) \, dt^2 + h(r) \, dr^2 + r^2 \, d\ph^2
\Label{genmet}
\end{equation}
which we take to be asymptotically AdS
($f(r) \to r^2$ and $h(r) \to 1/r^2$ as $r \to \infty$).
Let the boundary region $\rA$ be the same as above,
$\rA = \set{t=0 \, , \ \ph \in (-\pho,\pho) }$.
We want to ask whether $\Cms$ and $\Gms$ coincide in this general static case.
Note that $\Gms$ is simply a spacelike geodesic anchored at $\ph=\pm\pho$, whereas $\Cms$ is the projection to $t=0$ of a null geodesic congruence from the tip of $D_0$.

The effective potential for geodesics with energy $E$ and angular momentum $L$, defined by
$\rd^2 + \Veff(r) = 0$, is given by
\begin{equation}
\Veff(r) = { 1 \over h(r)} \, \[ -\kappa - {E^2 \over f(r)} + {L^2 \over r^2} \]
\Label{genVeff}
\end{equation}
where $\kappa=0$ for null geodesics and $\kappa=1$ for spacelike geodesics.
Since $\Gms $ corresponds to the spacelike
geodesic at constant $t$, pinned at $\ph = \pm \pho$, we have
 $\kappa=1$, $E=0$, which fixes the relation
  between $L$ and $\pho$.  Also, the minimum radius
   reached $\rmin$ is easy to find from $\Veff(\rmin)=0$;  we simply have $\rmin = L$.
Expressing $\Gms $  as $\ph(r)$, we then obtain
\begin{equation}
\ph(\rb) = \pm L \, \int_L^{\rb} \sqrt{{h(r) \over r^2 - L^2}} \, {1 \over r} \, dr \ .
\Label{Mgen}
\end{equation}
To construct $\Cms $,
we consider a null geodesic
 congruence labeled by $\l \equiv L/E$ (and we choose parameterization such that $E=1$).  Then $t$ and $\ph$ along the $\l$ geodesic, written in terms of $r$, are given by
\begin{equation}
t_{\l}(\rb) = \pho -  \int_{\rb}^{\infty}
  \sqrt{{h(r)  \over r^2 - \l^2 \, f(r) }} \, {r \over \sqrt{f(r)}} \, dr \ ,
\Label{trn}
\end{equation}
\begin{equation}
\ph_{\l}(\rb) = \pm \l \, \int_{\rb}^{\infty}
  \sqrt{{h(r)  \over r^2 - \l^2 \, f(r) }} \, {\sqrt{f(r)} \over r} \, dr \ .
\Label{phirn}
\end{equation}
Now, let $\ro(\l)$ be the value of $r$ along the $\l$ geodesic at which $t_{\l}$ reaches zero, $t_{\l}(\ro(\l)) = 0$.  Then $\Cms $ is given by $\ph_{\l}(\ro(\l))$, which as written is a parametric curve parameterized by $\l$, but out of which $\l$ should be eliminated to compare directly with \ref{Mgen}.

To make progress, consider the change in $\ph_{\l}$ as we vary $\ro$ (\ie, as we vary $\l$):
\begin{equation}
{\d \ph_{\l}(\ro(\l)) \over \d \ro(\l)} =
{{\p \ph_{\l}(\ro(\l)) \over \p \l } \over {\p \ro(\l) \over \p \l}}
= {\p_{\l} \ph_{\l}(\ro(\l)) \over \ro'(\l) }
\Label{philvar}
\end{equation}
which we can find using the generalized Leibnitz rule.  We want to compare the resulting expression with the corresponding variation for the spacelike geodesic $\Gms $,
\begin{equation}
{d \ph(r) \over dr} = {L \over r} \, \sqrt{{h(r) \over r^2 - L^2}}
\qquad {\rm at} \ r = \ro(\l) \ .
  \Label{phisvar}
  \end{equation}
We then obtain a long integral equation, which we can simplify (eliminate the integral) by observing that $\p_{\l} t_{\l}(\ro(\l)) = 0$, which follows from the definition of $\ro(\l)$; we can again use the generalized Leibnitz rule to write this explicitly.

Hence the assumption that $\Cms=\Gms$ reduces to the much simpler equation, which we wish to verify/falsify for general $f(r)$, and for all $\l$:
\begin{equation}
\ro^2(\l) - \l^2 \,  f(\ro(\l) ) = L^2 \ .
\Label{roLrel}
\end{equation}
While it is straightforward to check that this mysterious relation does hold for the cases discussed above of AdS and BTZ, as consistency demands, it is less trivial to check it for general $f(r)$.
Resorting to numerical analysis, we find that unfortunately \req{roLrel} is not satisfied for arbitrary $f(r)$ (although $\Cms$ is typically well-approximated by $\Gms$).  Hence we conclude that even in 3 dimensions, $\Cms \ne \Gms$.

To understand better why this is the case, consider the particular point on the surfaces $\Cms$ and $\Gms$ corresponding to $\ph=0$, namely when $r$ reaches its minimal value.  For the spacelike geodesic $\Gms$, this is simply $L$; whereas for $\Cms$, it corresponds to $\ro(\l=0) \equiv \ro$.  Therefore a simple way to check that $\Cms \ne \Gms$ is to show that in general $\ro \ne L$.
We can extract the relation between $\ro$ and $L$ by writing $\pho$ using the spacelike geodesic \req{Mgen} and the null geodesic \req{trn} with $\l=0$:
\begin{equation}
\pho
= \int_{\ro}^{\infty}  \sqrt{{h(r)  \over f(r) }} \, dr
= \int_{L}^{\infty}  \sqrt{{h(r)  \over r^2 \, \({r^2 \over L^2} - 1 \) }} \, dr  \ .
\Label{phiorel}
\end{equation}
But this relation clearly indicates that whereas for any fixed $\pho$, $L$ depends only on $h(r)$ (since the spacelike geodesic at constant $t$ cannot be sensitive to $f(r)$), $\ro$ depends on both $h(r)$ and $f(r)$ -- so $\ro$ cannot coincide with $L$ for arbitrary $f(r)$.  This provides a proof that $\Cms \ne \Gms$ for general 3-dimensional static spherically symmetric spacetimes.

\subsection{Use of $\Cms$ to bound the entanglement entropy}

Above, we have described the construction $\Cms$ and argued that it does
not in general coincide with $\Xms$, $\Lms$, or $\Gms$, even in static backgrounds.
Hence, although  $\Cms$ is based only on causal relations and therefore carries a
certain appeal due to its simplicity, we may well ask what is it useful for.

In the context of entanglement entropy, we propose that computing
$\Cms$ is useful (if simpler than computing $\Gms$, $\Lms$, or
$\Xms$) because it provides a bound on the entanglement entropy. In
particular, we expect that the area of $\Cms$ is larger than (or
equal to) the areas of $\Gms$ and $\Lms$, at least for ``sensible"
spacetimes. If the spacetime is static, this is clearly true by
definition because the correct surface $\Gms=\Lms$ is the minimal
area surface. In more general case, one \req{LmsminB} of our
covariant constructions, implies this speculation, though we cannot
offer a general proof.

Imagine the situation where we want to find the minimal surface
$\Gms=\Lms$ for a complicated choice of the subsystem $\rA$ in order to
compute the holographic entanglement entropy in a static higher
dimensional spacetime. In this case, it seems almost impossible to
find the minimal surface analytically because we need to solve a
partial differential equation with a generic initial condition.
 However, to find null
geodesics will be much more tractable. Then we can find a definite
and useful bound for the entanglement entropy by employing the
construction of $\Cms$.

\section{Null expansions and extremal surfaces}
\label{apmin}

\subsection{Definition of extrinsic curvature}

Consider a $d$-dimensional spacelike submanifold $\ms$ in a
$D$-dimensional spacetime $\bulk$ with Lorentzian signature.
The coordinates of $\bulk$ are denoted by $x^\mu$ and those of the submanifold $\ms$ are $\xi^\a$.

We define the extrinsic curvature $K^{(m)}_{\mu\nu}$ as follows.
There are $D-d$ vectors  $n^{(m)}_\mu$ $(m=1,2,...,D-d)$ on $\ms$ which are
 orthogonal to $\ms$. The extrinsic curvature
is defined by
\be \nabla_{\! \mu}n^{(m)}_\nu=K^{(m)}_{\mu\nu}. \Label{extaap} \ee
In particular, if we choose
a coordinate system adapted to ${\cal S}$ so that
$x^\mu=(\xi^\ap,y^l)$, where $l=1,2,...,D-d$ labels the directions
normal to $\ms$, then we obtain
\be K^{(m)}_{\ap\beta}=-\Gamma^l_{\ap\beta}\, n^{(m)}_{l}\ . \ee
Picking  any two tangent vectors $u^\mu, v^\mu \in T\ms$, we
can equivalently define the extrinsic curvature by
\be K^{(m)}_{\mu\nu}u^\mu v^\nu=(u^\mu\nabla_{\! \mu} \, v^\nu)\,
n^{(m)}_\nu \ . \ee
%

\subsection{Extremal surfaces}
We define an extremal surface by the saddle point of the area
functional
\be \mbox{Area}(\ms)=\int_{\ms}
(d\xi)^{d}\s{\det{g}}, \ee
where $g_{\ap\beta}$ is the induced
metric and is written in terms of the total spacetime metric
$g_{\mu\nu}$ as $g_{\ap\beta}=g_{\mu\nu}\f{\de X^\mu}{\de
\xi^\ap}\f{\de X^\nu}{\de \xi^\beta}$.

 After a little algebra, the equation of motion can be rewritten as
\be \Pi^\mu_{\ap\beta}\, g^{\ap\beta}=0 \ , \ee
where $\Pi^\mu_{\ap\beta}$ is defined by
\be \Pi^\mu_{\ap\beta}=\de_\ap\de_\beta
X^\mu+\Gamma^\mu_{\nu\lambda}\de_\ap X^\nu \de_\beta X^\lambda
-\Gamma^\gamma_{\ap\beta}\de_\gamma X^\mu\ . \ee
It is possible to show that $\Pi^\mu_{\ap\beta}$ is orthogonal
to $T\ms$ \ie, $\Pi^\mu_{\ap\beta}\, \de_\gamma X_\mu=0$. Thus the
only independent components are $D-d$ vectors $n^{(m)}_\mu
\Pi^\mu_{\ap\beta}$.

Let us choose the specific coordinate system such that
$X^\mu=(\xi^\ap,y^l)$ as before. Then it is easy to see
\be n^{(m)}_\mu \,\Pi^\mu_{\ap\beta}=\Gamma^l_{\ap\beta}\,
n^{(m)}_{l}=-K^{(m)}_{\ap\beta}\ . \ee
Thus we find that the extremal surface condition is
equivalent to the vanishing of the trace of the extrinsic curvature
\be -g^{\ap\beta}K^{(m)}_{\ap\beta} =g^{\ap\beta}n^{(m)}_\mu
\Pi^\mu_{\ap\beta} =0\ .\ee
In a generic coordinate frame, this is expressed as
\be g^{\mu\nu}K^{(m)}_{\mu\nu}=0 \ .\ee
%

\subsection{Expansions of null geodesics: relation to extremal surfaces}

Consider a co-dimension two spacelike surface ${\cal S}$. There are
two independent normal vectors at each point on ${\cal S}$. We can
choose them to be lightlike and call them $N_{+}^\mu$ and
$N_{-}^\mu$. They are normalized such that $N_{+}^\mu N_{+
\mu}=N_{-}^\mu N_{- \mu}=0$ and $N_{+}^\mu N_{-\mu}=-1$. We can
define the extrinsic curvatures $K^{(\pm)}_{\mu\nu}$
 for these vectors.

The two null expansions $\theta_{\pm}$  are defined by
 \be \theta_{\pm}=g^{\mu\nu}K^{(\pm)}_{\mu\nu}. \ee
It is clear from the above definition of the extrinsic curvature
that when $g^{\mu\nu}K^{(\pm)}_{\mu\nu}=0$ (\ie, ${\cal S}$ is a
extremal surface), both of the expansions are zero $\theta_{\pm}=0$.

Also from this definition we can find that when  $N^\mu$ is a null
Killing vector,
\ie, $\nabla_{\! \mu} \, N_\nu + \nabla_{\! \nu} \, N_\mu = 0$,
the null expansion $\theta$ is obviously vanishing.

\section{Details of perturbative analysis in Vaidya-AdS background}
\label{apvaidya}

We perform a perturbative analysis by only keeping the linear order
about the mass $m(v)$ in the Vaidya.

We consider the extremal surface (or equivalently a geodesic) in the
background \req{vaidyam} assuming that $m(v)$ is very small. At the
boundary $r=r_{\infty}\to \infty$, the two end points of the
geodesic are given by $(v,r,x)=(v_0,r_{\infty},\pm h)$. We assume
the following profile
 \ba && r(x)=\f{1}{\s{h^2-x^2}}+s(x), \no &&
v(x)=v_0-\s{h^2-x^2}+u(x),\ \ \ \ \ \ (-h\leq x\leq h) .
\Label{sudefs}
\ea

After we plug this into (\ref{ham}) and (\ref{eomvat}), we obtain
the differential equations for the perturbation $s(x)$ and $u(x)$ at
linear order:
\ba &&
2x(h^2-x^2)^{3/2}s'(x)-2\s{h^2-x^2}(2x^2+h^2)s(x)+x^2(h^2-x^2)m[v(x)]
+2\ep h=0 \ , \no  \label{vek} \\ &&
(h^2-4x^2)s(x)+2x(h^2-x^2)s'(x)-(h^2-x^2)u''(x)=0 \ ,  \Label{vvvk} \ea
where, we have defined the very small quantity $\ep$ by
\be h=\f{1}{r_*}+\ep\ . \Label{vekk} \ee
 (Recall that
$h=\f{1}{r_*}$ if $m=0$.)

Integrating (\ref{vek}), we obtain
 \be
s(x)=-\f{x}{(h^2-x^2)^\f{3}{2}}\left[-\f{\ep h}{x}+\int^x_0
dy\f{m[v(y)]}{2}(h^2-y^2)\right]. \label{sole} \ee
Clearly we find $s(0)=\f{\ep}{h^2}$ and this is consistent with \req{vekk}. Notice
also the property $s(x)=s(-x)$. To make sense of our perturbative
argument, we need to require that $s(x)$ does not include the
singular term $\sim (h-x)^{-3/2}$ when we take the limit $x\to h$.
This determines the value of $\ep$ as follows:\footnote{When $m$ is a
constant this leads to $\ep=\f{m}{3}h^2$, which is consistent with
our previous analysis in the BTZ geometry.}
\be \ep=\f{1}{2}\int^h_0 dy~
m[v(y)]\, (h^2-y^2)\ . \Label{epcom} \ee
The other function $u(x)$ can be found by integrating \req{vvvk}
twice by using \req{sole}.

After some analysis we can show that the UV cut-off $r=r_{\infty}$
is related to the UV cut-off $x=h-\delta$ of $x$ via
\be
\delta=\left(1+\f{\ep}{h}\right)\f{1}{2h\,r_{\infty}^2}.
\Label{cutoffl} \ee

The total geodesic length $L$ is then found to be
\be
L=\int^{h-\delta}_{-(h-\delta)} dx \, \f{r(x)^2}{r_*}
=(h-\ep)\,\int^{h-\delta}_{-(h-\delta)}dx\,
\left(\f{1}{h^2-x^2}+\f{2s(x)}{\s{h^2-x^2}}\right)\ . \Label{geol} \ee

\paragraph{Explicit Calculations of Geodesic Length:} Consider the following
specific approximation to the time-dependent mass:
\be m(v)=m(v_0)+m'(v_0)(v-v_0).  \Label{profli} \ee
This is true when the time-dependence is small and  is exact
when the mass is linear function of the time $v$. Further  we assume
the mass itself is also very small. Under these conditions we obtain
from \req{epcom}
\be \ep=\f{1}{2}\int^h_0 dx
(m(v_0)-m'(v_0)\s{h^2-x^2})(h^2-x^2)=\f{h^3}{3}m(v_0)-\f{3\pi
h^4}{32}m'(v_0) \ . \ee

With the specific profile \req{profli}, we can integrate \req{sole}
and \req{geol} analytically. After a somewhat lengthy computation we
find
\be  L(v_0) = \log
\left(\f{2h}{\delta}\right)+\f{2}{3}h^2m(v_0)-\f{5\pi}{32}h^3
m'(v_0). \Label{finaf} \ee
Substituting the relation \req{cutoffl} into
\req{finaf}, we obtain the final expression
\be L(v_0)=2\log
(2hr_{\infty})+\f{1}{3}h^2m(v_0)-\f{\pi}{16}h^3 m'(v_0).  \ee
The finite part of the geodesic length after we subtract the
universal divergent piece $2\log (2h\, r_{\infty})$ is now given by
\be
L_{reg}=\f{1}{3}h^2m(v_0)-\f{\pi}{16}h^3 m'(v_0)  \ . \ee
 In the case of the linear profile (or when $m''(v_0)$ is small enough) we find
the geodesic length at generic time $v$
\ba
L_{reg}(v)&=&\f{1}{3}h^2[m(v_0)+m'(v_0)(v-v_0)]-\f{\pi}{16}h^3
m'(v_0) \no &\simeq& m(v-3\pi h/16)\ ,  \ea
 and its time derivative is given by
 \be \f{d}{dv}L_{reg}(v)=\f{1}{3}h^2 m'(v) \ . \Label{finrt}
\ee

\paragraph{Explicit form of $s(x)$ and $u(x)$:}
Under the assumption \req{profli}, we can find the following
explicit solutions for the functions $s(x)$ and $u(x)$ introduced in \req{sudefs}:
\ba  s(x) & =& s \f{2h^2-x^2}{6\s{h^2-x^2}}\,  m(v_0)\no && +
\f{-9\pi h^5+(30h^2x^2-12x^4)\s{h^2-x^2}+18h^4x \arctan
\left(\f{x}{\s{h^2-x^2}}\right)}{96(h^2-x^2)^\f{3}{2}}m'(v_0)\ ,\no
u(x)&=&\f{x^2\s{h^2-x^2}}{6}m(v_0)+\Bigl[\f{x^2(4x^2-9h^2)}{48}
-\f{3}{32}h^4\log h^2 \no &&
+\f{1}{32\s{h^2-x^2}}\left(6h^4x\arctan\left(\f{x}{\s{h^2-x^2}}\right)
-3\pi h^5\right)\Bigr]m'(v_0). \ea

\paragraph{Asymptotic Expansion of $r(x)$ and $v(x)$:}From the previous explicit
 expression of $s(x)$ and $u(x)$, the
asymptotic expansion of $r(x)$ in the limit $x\to h$ is given by
\ba
&& r(x)\simeq \left(\f{1}{\s{2h}}+\f{h^{\f32}}{6\s{2}}m
-\f{3h^{\f52}\pi}{64\s{2}}m'\right)(h-x)^{-\f{1}{2}}\no && \ \
+\left(\f{1}{4\s{2}h^{\f32}}+\f{3\s{h}}{8\s{2}}m
-\f{9h^{\f32}\pi}{256\s{2}}m'\right)
(h-x)^{\f{1}{2}}-\f{h}{5}m'(h-x)+\CO((h-x)^{\f{3}{2}}) \ .\no
\Label{expansionr} \ea
Notice that the coefficient of
$(h-x)^{-\f{1}{2}}$ in (\ref{expansionr}) is the same as $
\s{\f{r_*}{2}}\simeq \f{1}{\s{2h}}(1+\f{\ep}{2h})$.

On the other hand, the asymptotic expansion of $v(x)$ in the limit
$x\to h$ becomes
\ba v(x) &\simeq&
v_0-h^4\left(\f{7}{24}+\f{9}{96}\log h^2\right)m'+\left(-\s{2h}
+\f{\s{2}}{6}h^{\f{5}{2}}m -\f{9\s{2}\pi}{192}
h^{\f{7}{2}}m'\right)(h-x)^{\f{1}{2}}\no && \ \ \ \
+\f{h^3}{6}m'(h-x)+\CO((h-x)^2) \ .  \ea
Finally, the asymptotic relation between $r(x)$ and $v(x)$ can be
shown with some effort to be:
\be v(x)=v_{0}-\f{h^4m'(v_0)}{48}(14+9\log
h)-\f{1}{r}+\f{h^2m'(v_0)}{12r^2}+\CO(r^{-3}). \Label{expvr} \ee The
first two terms represent the constant contribution $v\equiv v(h)$
as considered in \req{boundconv}.



\providecommand{\href}[2]{#2}\begingroup\raggedright\endgroup


\end{document}